\documentclass[aps,showpacs,twocolumn,prd,preprintnumbers,amsmath,amssymb,letterpaper]{revtex4}

\newcommand{\beqn}{\begin{eqnarray}}
\newcommand{\eeqn}{\end{eqnarray}}

\def\beqar{\begin{eqnarray}}
\def\eeqar{\end{eqnarray}}
\newcommand{\llabel}[1]{\label{#1}}              

\newcommand{\labeq}[2]{ \begin{equation} \llabel{#1}
{#2}
\end{equation}}

\newcommand{\beq}{\begin{equation}}
\newcommand{\eeq}{\end{equation}}
\usepackage{graphicx}
\usepackage{epsf}
\usepackage{amsmath}
\usepackage{graphics,epsfig,placeins,subfigure,wrapfig}
\begin{document}
\title{Head-on collisions of binary white dwarf--neutron stars: \\ Simulations in full general relativity}
\author{Vasileios Paschalidis${}^1$, Zachariah Etienne${}^1$, Yuk Tung Liu${}^1$, and
Stuart L. Shapiro${}^{1,2}$}                           
%
\affiliation{
${}^1$ Department of Physics, University of Illinois at Urbana-Champaign,
Urbana, IL 61801 \\
${}^2$ Department of Astronomy and NCSA, University of Illinois at Urbana-Champaign,
Urbana, IL 61801 
}
%
\begin{abstract}

We simulate head-on collisions from rest at large separation of binary white dwarf -- neutron stars (WDNSs)
in full general relativity. Our study serves as 
a prelude to our analysis of the circular binary WDNS problem.
We focus on compact binaries whose total mass exceeds the maximum mass that a cold degenerate star can support, 
and our goal is to determine the fate of such systems.
A fully general relativistic hydrodynamic computation of a realistic 
WDNS head-on collision is prohibitive due to the large
 range of dynamical time scales and length scales involved. 
For this reason, we construct an equation of state 
(EOS) which captures the main physical features of NSs while, 
at the same time, scales down the size of WDs. 
We call these scaled-down WD models ``pseudo-WDs (pWDs)''. Using these pWDs, 
we can study these systems 
via a sequence of simulations where the size of the pWD gradually
 increases toward the realistic case.
We perform two sets of simulations; One set
studies the effects of the NS mass on the final outcome, when the pWD is kept fixed.
The other set studies the effect of the pWD compaction on the final outcome, 
when the pWD mass and the NS are kept fixed.  All
 simulations show that after the collision,  14\%-18\% of the initial
 total rest mass escapes to infinity. 
All remnant masses still exceed the maximum rest mass that our
 {\it cold} EOS can support ($1.92M_\odot$), but  no case leads to prompt collapse to a black hole. 
This outcome arises because the final configurations are {\it hot}. 
All cases settle into spherical, quasiequilibrium
 configurations consisting of a cold NS core surrounded by a hot mantle, 
 resembling Thorne-Zytkow objects. 
Extrapolating our results to realistic WD compactions, we predict
that the likely outcome of a head-on collision of a realistic, 
massive WDNS system will be the formation of a quasiequilibrium Thorne-Zytkow-like object. 

\end{abstract}


\maketitle
%

\section{Introduction}
\label{sec:introduction}

The inspiral and merger of compact binaries represent some of the 
most promising sources of gravitational waves (GWs) for detection by
ground-based laser interferometers like LIGO \cite{LIGO1,LIGO2}, VIRGO \cite{VIRGO1,VIRGO2}, 
GEO \cite{GEO}, TAMA \cite{TAMA1,TAMA2} and AIGO \cite{AIGO}, as well as by
 the proposed space-based interferometers LISA \cite{LISA} and DECIGO \cite{DECIGO}. 
Extracting physical information from gravitational radiation emitted 
by compact binaries and determining their ultimate fate 
requires careful modeling of these systems in full general relativity
 (see \cite{BSBook} and references therein).
Most effort to date has 
focused on modeling black hole--black hole (BHBH) binaries  (see \cite{2010CQGra..27k4004H} and references therein),
and neutron star--neutron star (NSNS) binaries (see \cite{DuezNSNSReview} for a review),
with some recent general relativistic work on black hole--neutron star (BHNS) binaries
\cite{Rantsiou08, Loffler06, Faber, Faber06, Shibata06, Shibata07, Shibata08,Yamamoto08, 
Etienne08a, Etienne08, Duez08,2009PhRvD..79d4030S,2009PhRvD..79l4018K,2010AAS...21530001M,
2010arXiv1006.2839C,2010CQGra..27k4106D,2010arXiv1007.4160P,2010arXiv1007.4203F,2010arXiv1008.1460K}. 

In this work we consider white dwarf--neutron star (WDNS) binaries in full general relativity. 
WDNS binaries are promising sources of low-frequency GWs for LISA and DECIGO and, 
as we argued in \cite{WDNS_PAPERI}, possibly also high-frequency GWs
for LIGO, VIRGO, GEO, TAMA and AIGO, if the remnant ultimately collapses to form a 
black hole.

Like NSNS binaries, WDNS binaries are known to exist. In \cite{WDNS_PAPERI} 
we compiled tables with 20 observed WDNS binaries and their orbital properties. 
The NS masses in these systems range between $1.26M_\odot$ and 
$2.08M_\odot$, and their distribution is centered around $1.5M_\odot$.
On the other hand, the WD masses in these systems range between $0.125M_\odot$
and $1.3M_\odot$, and their distribution is centered around $0.6M_\odot$. 
Finally, 18 of these observed WDNS binaries have total mass greater 
than $1.65M_\odot$, of which 8 have a WD component with mass greater 
than $0.8M_\odot$,  and 5 have total mass greater than $2.2M_\odot$. 
This is interesting because the expected TOV limiting mass for a cold,
 degenerate gas ranges between $1.65M_\odot$ and $2.2M_\odot$ 
\cite{2000ApJ...528L..29B,2004ApJ...610..941M,CooST94, APR,
1993PhRvL..70..379L,1988PhRvC..38.1010W,1975BAAS....7..240P,
BetheJohnson,Pandha1971}, depending on the equation of state (EOS) and degree of rotation,
and one of the main goals of this work is 
to determine whether a WDNS merger can lead to prompt collapse to a black hole.

Population synthesis calculations by Nelemans et al. \cite{Nelemans01} show
 that there are about $2.2\times 10^{6}$ WDNS binaries in our Galaxy, 
and that they have a merger rate of $1.4\times 10^{-4} \rm yr^{-1}$. 
Furthermore, Nelemans et al. find that after a year of integration, 
LISA should be able to detect $128$ WDNS pre-merger binaries and, after
 considering the contribution of the double WD background GW noise, 
resolve $38$ of these. On the other hand, calculations by Cooray 
\cite{Cooray2004} give much more conservative numbers of resolvable WDNS binaries.
 In particular, Cooray finds that the number of LISA-resolvable WDNS binaries 
ranges between $1$--$10$, using a WDNS merger rate between
 $10^{-6} \rm yr^{-1}$--$10^{-5} \rm yr^{-1}$. Finally, recent work
 by Thompson et al. \cite{TThompson09} suggests that the lower limit on 
the merger rate in the Milky Way, at 95\% confidence, is 
$2.5\times 10^{-5}\rm yr^{-1}$. Thompson et al. also suggest that the
 merger rate in the local universe is $\sim 0.5 -  1 \times 10^4 \rm Gpc^{-3} yr^{-1}$.
 Therefore, leaving some uncertainties aside, all recent work on population 
synthesis suggests that LISA should be able to detect a few WDNS
 pre-mergers per year.

We note here that Newtonian work on binaries with a WD component
has been performed analytically in
\cite{Rappaport82,Rappaport83,Verbunt88,Podsiadlowski92,Marsh,WDNS_PAPERI} 
and via Newtonian hydrodynamic simulations in
\cite{Benz1990, RasioShapiro95, Segretain1997,
  Guerrero2004, Yoon2007, Dan08}.

In \cite{WDNS_PAPERI} we focused on WDNS binaries that have spiraled  sufficiently 
close that they reach the termination point for
equilibrium configurations. This is the Roche limit for WDNSs, 
at which point the WD fills its Roche lobe and 
may experience one of at least two possible fates: i) stable mass transfer (SMT) 
from the WD across the inner Lagrange point onto the 
NS, or ii) tidal disruption  of the WD by the NS via unstable mass transfer (UMT). 

We also studied the key parameters that determine whether a system
will undergo SMT or UMT and found that, for a given NS mass, there
exists a critical mass ratio $q_{\rm crit} \approx 2/3$ which separates the UMT
and SMT regimes.  If the mass ratio $q=M_{\rm WD}/M_{\rm NS}$ of a
WDNS system is such that $q>q_{\rm crit}$, the WD quickly overfills its
Roche lobe, and the binary will ultimately undergo UMT. In the opposite
case, $q<q_{\rm crit}$, the system will undergo SMT. 
We showed that a
quasistationary treatment is adequate to follow the evolution of an SMT
binary during this secular phase and calculated the gravitational
waveforms. We also pointed out that WDNS observations suggest that there are candidates
residing in both regimes.

In the case of tidal disruption (UMT), by contrast, 
the system will evolve on a hydrodynamical (orbital) time scale. 
In this scenario the NS may plunge into the WD and spiral 
into the center of the star, forming a quasiequilibrium
 configuration that resembles a
 Thorne-Zytkow object (TZO) \cite{ThorneZytkow77};
alternatively, the NS may be the receptacle of massive debris from the
 disrupted WD.

Depending on the details of the EOS, a cold degenerate 
gas can support 
a maximum gravitational 
mass between $1.65M_\odot$ and $2.2M_\odot$ 
against catastrophic collapse, if it is not rotating (the TOV limit).  
It can support $20\%$ more mass, if it is rotating uniformly at the mass-shedding
limit (a ``supramassive NS'' \cite{CooST94}), and about $50\%$ 
more mass, if it rotates differentially (a ``hypermassive NS'' 
\cite{CooST94,2000ApJ...528L..29B,2004ApJ...610..941M}). 
If the total mass of the 
merged WDNS exceeds the maximum mass supportable by a cold EOS, delayed collapse to a black hole is inevitable after the remnant cools.
However, the ultimate fate of the merged WDNS 
depends on the initial mass of the cold progenitor stars, the 
degree of mass and angular momentum 
loss during the WD disruption and binary merger phases, the angular momentum 
profile of the WDNS remnant and the
extent to which the disrupted debris is heated by shocks as 
it settles onto the NS and forms an extended, massive mantle.
These are issues that require a hydrodynamic simulation to 
resolve. Moreover, ascertaining whether or not the neutron star ultimately
undergoes a catastrophic collapse (either prompt or delayed) to a black hole 
requires that such a simulation be performed 
in full general relativity.  In fact, even the 
final fate of the NS in the alternative scenario in which there is 
a long epoch of SMT may also lead to catastrophic collapse, 
if the neutron star mass is close to the neutron star maximum mass, and
this scenario too will require a general relativistic hydrodynamic simulation to track.

In this paper we employ the Illinois adaptive mesh refinement (AMR)
relativistic hydrodynamics code \cite{Etienne08,Illinois_new_mhd} to
perform our first simulations of these alternative scenarios. 
In particular, we study the head-on
 collision from rest at large separation of a massive WD and a NS as a prelude to
our investigation of the circular binary problem, which we will report in a future work. 
We focus on compact objects whose total mass exceeds
the maximum mass supportable by a cold EOS to determine whether such a
 collision leads to prompt
collapse of the remnant, or a hot gaseous mantle composed of WD debris
surrounding a central NS -- a Thorne-Zytkow-like object (TZlO).

The vast range of dynamical time scales and length scales involved 
in this problem make fully general relativistic simulations extremely challenging. 
For example, a near-Chandrasekhar-mass WD has a radius 
$R_{\rm WD}\simeq 10^3 \rm km$ and dynamical time scale of about 1s. 
On the other hand a typical NS has a radius of order $R_{\rm NS} \simeq 10 \rm km$ 
and dynamical time scale of about 1ms. Therefore, there is a difference of 
several orders of  magnitude both in length scales and time scales. 
Current numerical relativity techniques and available computational 
resources make such calculations prohibitive. 
For this reason, we tackle this problem using a different strategy. 

In particular, we construct a piecewise polytropic 
EOS which captures the main physical features of a NS while, at the same time,
 scales down the size of the WD. We call these scaled-down WDs ``pseudo-WDs (pWDs)''. 
We perform a sequence of simulations where we change the EOS
 so that the pWDs have the same mass ($0.98M_\odot$) but different compactions, 
while the compaction and mass of the NS involved remain practically unchanged. 
In other words, while keeping the masses of the binary components and
 the NS radius fixed, we adjust the ratio of the radius of
 the pWD to that of the NS so that it varies from 5:1 to 20:1 and
then use our results to predict the outcome of the realistic case. 
The common feature among all versions of the piecewise EOS we employ is that the
 maximum NS mass always is $1.8M_\odot$ and the maximum WD mass always is
 $1.43M_\odot$, i.e., the Chandrasekhar mass. 

In addition to studying the effects of the pWD compaction, we also study
the effects of the NS mass. We consider NSs with masses $1.4M_\odot$
$1.5M_\odot$ and $1.6M_\odot$.

All simulations that we perform show that after the collision,  
14\%-18\% of the initial total rest mass escapes to infinity. The remnant
 mass in all cases exceeds the maximum
 rest mass that our {\it cold} EOS can support ($1.92M_\odot$), but we find 
that no case leads to prompt collapse to a black hole. This outcome arises because the 
final configurations are {\it hot}. All our cases settle into a
 spherical quasiequilibrium configuration consisting of a cold NS core
 surrounded by a hot mantle. Hence, all remnants are TZlOs. 
Extrapolating our results to realistic WD compactions, we predict
that the likely outcome of a head-on collision from rest at large separation
 of a realistic massive WDNS system will be the formation of a quasiequilibrium TZlO.

This paper is organized as follows. In Section \ref{sec:timescales} we 
review the time scales and length scales involved in a
 WDNS merger and discuss why this problem presents such
 a computational challenge. In Section~\ref{sec:pWD} we introduce the 
EOS adopted for our computations and describe our pWD models. 
Section ~\ref{sec:Init_data} outlines our method for
 preparing initial data, 
and Section~\ref{sec:evolutions} summarizes our methods for evolving 
the gravitational and matter fields. 
We present the results of our fully relativistic hydrodynamic simulations
 in Section~\ref{sec:results}, and conclude in Section~\ref{sec:summary} with
 a summary of our main findings. Throughout we use geometrized units,
where $G=c=1$.

%
%

\section{Computational Challenge}
\label{sec:timescales}

Simulating a WDNS merger in full
 general relativity is a difficult computational task. In this section we sketch in 
quantitative terms exactly why this is so.

There are three fundamental time scales and length scales involved in
 the WDNS merger that must be resolved. The relevant time scales are the 
dynamical time scale of each component of the binary and the
orbital period; the relevant length scales are the NS and WD radii
 and their orbital separation. 

Resolving the WD length scale and dynamical time scale is
 necessary in order to assess what happens to the WD at merger. 
Merger occurs on the orbital time scale, so this time scale must also be
resolved.  Resolving the NS dynamical time scale will enable us to
assess whether the NS promptly collapses and forms black hole, or
remains inside the remnant WD, settling into a TZlO.

The dynamical time scale of the NS, $t_{\rm d,NS}$, is given by
\labeq{tdynNS}{
t_{\rm d,NS} = \sqrt{\frac{R_{\rm NS}^3}{M_{\rm NS}}},
}
where $R_{\rm NS}$, and $M_{\rm NS}$ are the characteristic NS radius and mass respectively.

Similarly, the dynamical time scale of the WD, $t_{\rm d,WD}$, is given by
\labeq{tdynWD}{
t_{\rm d,WD} = \sqrt{\frac{R_{\rm WD}^3}{M_{\rm WD}}}, 
}
where $R_{\rm WD}$, and $M_{\rm WD}$ are the characteristic WD radius and mass respectively.
Finally, the orbital time scale, $T$, is given by
\labeq{torb}{
T = 2\pi\sqrt{\frac{A^3}{M_{\rm T}}},
}
where $M_{\rm T} = M_{\rm NS}+M_{\rm WD}$ is the total mass, and $A$
is the orbital separation. Note that, at this separation, the head-on collision 
time scale is
\labeq{tcoll}{
T_{\rm coll}= \frac{\pi}{2\sqrt{2}}\sqrt{\frac{A^3}{M_{\rm T}}}
           = \frac{T}{4\sqrt{2}},
}
assuming the stars free-fall from rest as Newtonian point masses,
and hence it is roughly the same order of magnitude as $T$.

By use of Eqs.~\eqref{tdynNS} and~\eqref{tdynWD} the 
ratio of the WD time scale to the NS time scale is
\labeq{twdotns}{
\frac{t_{\rm d,WD}}{t_{\rm d,NS}} = \sqrt{\frac{R_{\rm WD}^3}{q R_{\rm NS}^3}}, 
}
where 
\labeq{q}{
q = \frac{M_{\rm WD}}{M_{\rm NS}}
}
is the binary mass ratio. 

For $M_{\rm WD} = 1M_\odot$ and using a cold degenerate 
electron EOS one finds $R_{\rm WD} \approx 5000\rm km$ \cite{Shapiro,WDNS_PAPERI}. 
On the other hand, typical NS masses and radii are
 $M_{\rm NS} = 1.5M_\odot$, $R_{\rm NS} \approx 10\rm km$.
 Hence, in realistic scenarios the ratio of the WD
 size to the NS size is 
\labeq{RwdoRns}{
\frac{R_{\rm WD}}{R_{\rm NS}} \sim 500,
}
and from Eq.~\eqref{twdotns} the ratio of the dynamical time scales is
\labeq{}{
\frac{t_{\rm d,WD}}{t_{\rm d,NS}} \sim 10^4.
}

At the Roche limit, $A$ is typically a few (two to five)
 times the WD radius \cite{WDNS_PAPERI}. Using, $A \sim 2 R_{\rm WD}$,
 and the values for the masses and 
radii used above, the orbital time scale becomes
\labeq{Totdns}{
\frac{T}{t_{\rm d,NS}} = 2\pi \sqrt{\frac{A^3}{(1+q) R_{\rm NS}^3 }} \sim 10^{5}.
}

It is thus clear that there is a vast range of 
length scales and time scales involved in this problem. 
The only way to simulate the WDNS merger 
is by exploiting the power of adaptive mesh refinement, so
 that resolution is high only where required. However, even this
 does not suffice to tackle the timescale problem as we explain below. 

Given that all current numerical 
relativity schemes for evolving both the spacetime and the
 fluid are explicit, there are strong limitations imposed
 on the size of the timestep 
by the Courant-Friedrich-Lewy condition
\labeq{CFL}{
\frac{\Delta t}{\Delta x}=\lambda < C ,
}
where $\lambda$ is the Courant number and $C$ a constant of order unity
 that depends on the integration scheme employed. 
If one uses AMR, this implies that the size of the timestep has to be
different in regions of different mesh size $\Delta x$. 
 If we resolve the stars adequately, the mesh size will be much smaller
near the NS than near the WD,
because in typical scenarios 
the NS is 500 times smaller than the WD. Eq.~\eqref{CFL}
 then implies that the smallest timestep must be in
 the domain of the NS. In particular, if the NS
 is covered by $N_{\rm NS}=2R_{\rm NS}/\Delta x_{\rm NS}$ 
grid zones and the WD is covered by $N_{\rm WD}=2R_{\rm WD}/\Delta x_{\rm WD}$ 
grid zones, then from Eq.~\eqref{CFL} we have
\labeq{dtwdodtns}{
\frac{\Delta t_{\rm WD}}{\Delta t_{\rm NS}} = 
                   \frac{\Delta x_{\rm WD}}{\Delta x_{\rm NS}}
                 = \frac{N_{\rm NS}}{N_{\rm WD}} \frac{R_{\rm WD}}{R_{\rm NS}}, 
}
where $\Delta t_{\rm NS},\Delta x_{\rm NS}$ and
 $\Delta t_{\rm WD},\Delta x_{\rm WD}$ denote the timestep
 and mesh size in the vicinity of the NS and WD, respectively. 

To assess the potential formation of a black hole
 requires at least $N_{\rm NS}=50$ grid zones across the NS, in order that a $2M_\odot$ 
BH (which is a probable mass a BH would have after the merger of
 a WDNS system of total mass of about $2.5M_\odot$) would be 
covered by at least 20 grid zones. Even if a BH does not form, covering the NS
with 50 grid zones is necessary to reliably model the NS 
and maintain small hamiltonian and momentum constraint violations. 
Resolving the WD requires about $N_{\rm WD}=30$ grid
 zones to reliably model the star. 
If we combine Eq.~\eqref{RwdoRns} and Eq.~\eqref{dtwdodtns},  we obtain
\labeq{}{
\frac{\Delta t_{\rm WD}}{\Delta t_{\rm NS}} \approx 833. 
}
This means that for one timestep in the vicinity of the WD we would 
have to take about 833 timesteps in the vicinity of the NS. 
Even evolving the system 
for only one WD dynamical time scale would require millions of 
timesteps in the vicinity of the NS. 
This shows how difficult it is
 to resolve both the WD and the NS at the same time.

 However, what renders the computation of the WDNS merger in full GR prohibitive
is that a realistic merger takes place on an orbital time scale,
 which is equivalent to $10^{5}$ NS dynamical time scales 
(see Eq.~\eqref{Totdns}).

\begin{figure}[t]
\centering
\includegraphics[width=0.495\textwidth,angle=0]{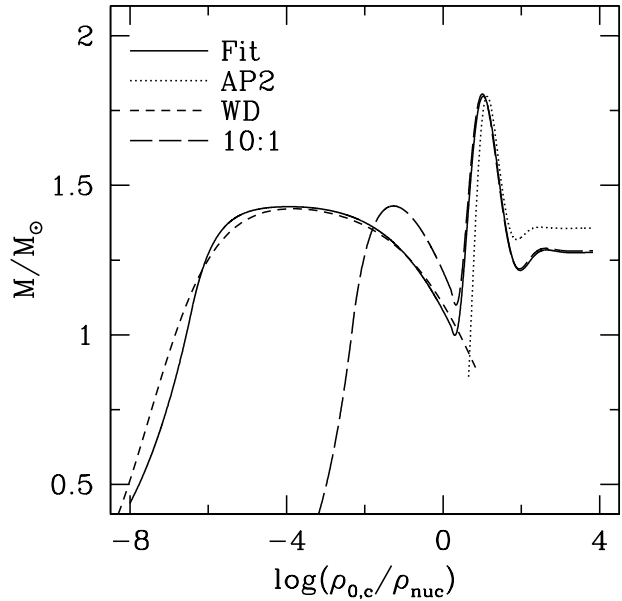}
\caption{Mass (M) -- central rest-mass density ($\rho_{_{0,c}}$) relationship of single TOV stars
for various cold EOSs. In the plot
 $\rho_{\rm nuc} = 2\times 10^{14} g/cm^3$ is the nuclear density.
  Plotted are the Chandrasekhar electron-degenerate EOS for mean molecular
weight per electron $\mu_e=2$ (labeled as WD), 
 the AP2 version of the Akmal-Pandharipande-Ravenhall EOS 
\cite{APR,ReadLackey2009}, a polytropic approximation of 
these realistic EOSs using EOS \eqref{EOS}  (labeled as Fit)
 and a version of EOS \eqref{EOS} where the ratio of the isotropic 
radius of a $0.98M_\odot$ pWD to the isotropic radius of a $1.5M_\odot$ 
NS is reduced to $\approx 10$ (labeled as 10:1). The parameters of these EOSs
 are listed in Table~\ref{tab:eosparams}.
\label{fig:eos}
}
\centering
\end{figure}

To make this quantitative, let us compare the
 orbital time scale with a typical timestep in the vicinity of the NS.
Using $N_{\rm NS}$ grid zones across the NS
 and combining Eq. \eqref{tdynNS} and Eq. \eqref{CFL} we find
\labeq{tdynNSodt}{
\frac{t_{\rm d,NS}}{\Delta t_{\rm NS}} = 
\frac{N_{\rm NS}}{2\lambda}\sqrt{\frac{R_{\rm NS}}{M_{\rm NS}}} \sim 100,
}
where in the last step we used a typical value
 for the Courant number $\lambda = 0.4$ and the 
values for $M_{\rm NS},R_{\rm NS},N_{\rm NS}$ we cited above.
Combining Eq.~\eqref{Totdns}  and Eq.~\eqref{tdynNSodt} we obtain
\labeq{}{
\frac{T}{\Delta t_{\rm NS}} \sim 10^{7}.
}
Hence, a realistic WDNS simulation would require a minimum of
$10^{7}$ timesteps in the vicinity of the NS. In fact this number 
of timesteps is an underestimate because extracting GWs would require 
a few orbits and the final system would settle in equilibrium or collapse 
within a few orbital time scales after merger. 
As a result, a dynamical, fully general relativistic 
hydrodynamics WDNS calculation would require of order $10^{8}$ timesteps.

We can give an estimate of how long such a simulation would be based
on high resolution ($192^3$ grid points in the innermost refinement level)
 benchmark runs we performed for a WDNS system
with a $1.5M_\odot$-NS and a $1.0M_\odot$-WD at separation of about $2.7R_{\rm WD}$
(close to the Roche limit),
which has an orbital period of about $1.4\times 10^6M$, where $M=2.5M_\odot$.
Using 256 cores on the Ranger cluster of the Texas Advanced Computing Center 
we found that the Illinois GR hydrodynamics code advances about $6M$ per hour. Thus,
the entire simulation (of about 10 orbital periods) would require about 264
years of pure computational time.

Realistic WDNS simulations are beyond the capabilities of current computational
resources and numerical relativity techniques. For this reason, we will
 tackle the problem of WDNS mergers and head-on collisions 
adopting an alternate strategy. We carefully construct an
 EOS which mimics a realistic cold NS EOS and, at the same time, 
scales down the size of the WD to make such a calculation feasible. 
Using sequences of these systems, where the WD
 size gradually increases, we can extrapolate 
our results to the realistic case.   
These scaled down WDs or pseudo-WDs are
 the subject of the following section. 

%
%



\begin{center}
\begin{table*}
\caption{Parameters for the piecewise
 polytropic EOS \eqref{EOS} used in generating different stellar models. The first column 
corresponds to the name of the EOS. 
An EOS named $\rm M:N$ corresponds to a version of \eqref{EOS} for
 which the mass -- radius relationship of TOV stars is such
 that the ratio of the isotropic radius of a $0.98M_\odot$ pWD 
to that of a $1.5M_\odot$ NS is $\rm M:N$
 (see Fig.~\ref{fig:models}). 
The EOS named AP2 is the same as the AP2 EOS defined in \cite{ReadLackey2009}. 
Finally, the EOS named Fit is an approximate fit to the
 Chandrasekhar EOS (for $\mu_e=2$) joined onto the AP2 EOS (see Fig.~\ref{fig:eos}).
Here $\rho_{\rm nuc} = 1.48494\times 10^{-4} {\rm km}^{-2}$ and
$\kappa_3$ is given in geometrized units.}
\begin{tabular}{ccccccc}\hline\hline
\multicolumn{1}{p{2.0cm}}{\hspace{0.1 cm} EOS Name } & 
\multicolumn{1}{p{2.0cm}}{\hspace{0.75 cm} $\kappa_3$ \quad}  &
\multicolumn{1}{p{2.0cm}}{\hspace{0.7 cm} $n_1$ \quad}  &
\multicolumn{1}{p{2.0cm}}{\hspace{0.7 cm} $n_2$ \quad}  &
\multicolumn{1}{p{2.0cm}}{\hspace{0.7 cm} $n_3$ \quad}  &
\multicolumn{1}{p{2.0cm}}{\hspace{0.1 cm} $\log(\rho_1/\rho_{\rm nuc})$ \quad}  &
\multicolumn{1}{p{2.0cm}}{\hspace{0.1 cm} $\log(\rho_2/\rho_{\rm nuc})$ \quad}  \\  \hline 
20:1    	       &    5064.2599    &  1.56128  &  2.98418  & 0.714286      &  -3.17219    & 0.180473        \\  \hline
10:1    	       &    4993.0688    &  1.51515  &  2.96971  & 0.714286      &  -2.26862    & 0.208502        \\  \hline
5:1    	               &    6123.5567    &  1.51883  &  2.94291  & 0.699301      &  -1.2909     & 0.267623        \\  \hline
AP2    	               &   145414.043    &  0.60864  &  0.49652  & 0.514139      &  0.398915    & 0.698922        \\  \hline
Fit    	               &    4458.0491    &  2.       &  2.96736  & 0.716         &  -6.39356    & 0.208502        \\  \hline\hline 
\end{tabular}
\label{tab:eosparams}
\end{table*}
\end{center}

\section{Pseudo-white dwarfs}
\label{sec:pWD}

In this section we introduce our EOS and describe resulting models for
 pWDs. Our EOS is the following 6-parameter piecewise polytropic 
EOS  
\labeq{EOS}{
\frac{P}{\rho_0} = 
\left\{
\begin{array}{ll}
\kappa_1 \rho_0^{1/n_1},  &  \rho_0 \leqslant  \rho_1  \\
& \\
\kappa_2 \rho_0^{1/n_2},  &  \rho_1<\rho_0 \leqslant  \rho_2 \ ,\\
& \\
\kappa_3 \rho_0^{1/n_3},  &  \rho_0 > \rho_2  
\end{array}
\right.
}
where $P$ is the pressure, $\rho_0$ is the rest-mass density
 and $\kappa_1, \kappa_2, \kappa_3, n_1 , n_2, n_3, \rho_1, \rho_2$ are the 
parameters of the EOS. Note that the parameters are 8 in number but
continuity requires that the following conditions be true
\labeq{EOScond}{
\kappa_1=\kappa_2 \rho_1^{1/n_2-1/n_1}, \ \ 
\kappa_2=\kappa_3 \rho_2^{1/n_3-1/n_2}.
}
As a result, the adopted EOS \eqref{EOS} has 6 free parameters and
 throughout this paper 
we use $\kappa_3, n_1 , n_2, n_3, \rho_1, \rho_2$ to specify an EOS. 
Note also that we use the polytropic indices $n_i$ 
instead of the adiabatic indices $\Gamma_i = 1+1/n_i$.

The freedom of our multi-parameter EOS enables us to capture
 the same characteristic curves and turning points on a TOV mass-central density plot 
as for a cold-degenerate realistic EOS (see \cite{Shapiro}),
as shown in Fig.~\ref{fig:eos}. The figure shows that EOS \eqref{EOS} 
can provide a reasonable approximation to the mass-central density
 relation of realistic compact objects, 
 exhibiting both stable ($dM/d\rho_{0,c}>0$) and unstable
 ($dM/d\rho_{0,c}<0$) branches for both WDs and NSs.

Furthermore, EOS \eqref{EOS} allows us to adjust the
 size of a pWD of given mass, thereby shifting the pWD branch
 to smaller radii (see Fig.~\ref{fig:models}), while
 keeping the NS masses and radii approximately unchanged
 for $M_{\rm NS} \geqslant 1.3 M_\odot$. 
The shifted branches in Fig.~\ref{fig:models} correspond to
 the stars that we call pseudo-WDs in this work.

Finally, note that all versions of EOS \eqref{EOS} considered
 in this work have been carefully constructed so that the maximum gravitational
 mass of a NS is $1.8M_\odot$, i.e., the same as that
 which for the AP2 version of the Akmal-Pandharipande-Ravenhall 
(APR) EOS \cite{APR,ReadLackey2009}, and the maximum gravitational mass
 of a pWD is $1.43M_\odot$, i.e., the maximum mass of a
 TOV WD obeying the Chandrasekhar EOS for $\mu_e = 2$.
In addition, EOS \eqref{EOS}
is constructed to preserve the shape of the $M$ vs $\rho_{0,c}$ curve, yielding
both stable and unstable branches and turning points appropriately.


%
%

\section{Initial Data}
\label{sec:Init_data}

In this section we present the basic formalism for generating valid general
relativistic initial data for the head-on collision in a
given pWDNS system.

\subsection{Gravitational Field Equations} \label{formalism_cts}

We begin by writing the spacetime metric in the standard 3+1 form \cite{ADM3plus1}
\begin{equation}\label{ADMmetric}
ds^2 = - \alpha^2 dt^2 + \gamma_{ij}(dx^i +\beta^i dt)
	(dx^j +\beta^j dt),
\end{equation}
where $\alpha$ is the lapse function, $\beta^i$ the 
shift vector and $\gamma_{ij}$ the three-metric on
 spacelike hypersurfaces of constant time $t$. 
Throughout the paper Latin indices run from 1 to 3, whereas Greek
indices run from 0 to 3. 

We conformally decompose the three-metric $\gamma_{ij}$ as
\begin{equation} \label{conflatmet}
\gamma_{ij} = \Psi^4 f_{ij},
\end{equation}
where $\Psi$ is the conformal factor and $f_{ij}$ the conformal metric. 
We adopt the standard approximation of a conformally flat spacetime, so that 
$f_{ij} = \delta_{ij}$ in Cartesian coordinates.

Since we are interested in head-on collisions between compact objects,
we assume that initially the stars begin to accelerate towards each other
starting from rest. As a result the extrinsic curvature is
 initially  zero and the momentum constraints
 are identically satisfied \cite{BSReview}. 
Hence, we need only prepare initial data for $\Psi$.

Under the aforementioned assumptions the only equation 
we have to solve is the Hamiltonian constraint, which becomes
\begin{equation} \label{ham2}
\nabla^2 \Psi =
	- 2\pi\Psi^5 \rho,
\end{equation}
where $\nabla^2$ is the flat Laplacian operator.
Here the source term $\rho$ is defined as
\begin{equation} \label{rho}
\rho \equiv n^\alpha n^\beta T_{\alpha\beta},
\end{equation}
and where $n^{\alpha}$ is the normal vector to a $t = {\rm constant}$ slice, and
$T_{\alpha\beta}$ is the stress-energy tensor of the matter.

The gauge is chosen so that the initial slice is maximal, i.e., $K=0$
and $\partial_t K=0$, and the shift vector is set equal to zero. 
Using the assumption of maximal slicing it is
 straightforward to derive an equation for the lapse
\cite{Baum98_NSNS,BSBook}
\begin{equation} \label{lap2}
\nabla^2\tilde\alpha = 2 \pi \tilde\alpha \Psi^4 (\rho + 2S),
\end{equation}
where 
\labeq{alphatilde}{
\tilde\alpha = \alpha\Psi,
}
and the source term $S$ is defined as
\begin{equation} \label{S}
S \equiv \gamma^{ij} T_{ij}.
\end{equation}

Equations \eqref{ham2} and \eqref{lap2} are
 elliptic and hence have to be supplemented with outer boundary conditions.
Following \cite{Baum98_NSNS}, we impose 1/r 
fall-off conditions on $\alpha-1$ and $\Psi-1$ at the outer boundary.

\subsection{Matter fields}

To model the matter, a perfect fluid stress-energy tensor is assumed:
\begin{equation}
T^{\alpha\beta} = (\rho_0 + \rho_i + P) u^\alpha u^\beta + P g^{\alpha\beta},
\end{equation}
where $g^{\alpha\beta}$ is the inverse of
 the four-metric and $\rho_0, \rho_i, P, u^\alpha$ are the rest-mass density, 
internal energy density, pressure, and four-velocity of the fluid respectively.

Since the initial configuration is assumed to be 
 at rest in the center of mass frame,
the initial fluid four velocity is given by
\begin{equation}
u^\alpha = u^t(1,0,0,0)
\end{equation}
or 
\begin{equation}
u^\alpha = \alpha u^t n^\alpha.
\end{equation}

A straightforward calculation shows that the source term $\rho$ 
in Eq.~(\ref{rho}) can then be written as
\begin{equation} \label{rho2}
\rho 	= \rho_0 + \rho_i,
\end{equation}
and $S$ in Eq.~(\ref{S}) as
\begin{equation} \label{S2}
S 	=  3P.
\end{equation}

\subsection{Computational methods}

To solve the elliptic equations \eqref{ham2} and 
\eqref{lap2} we developed a fixed-mesh-refinement (FMR)
 finite difference code based 
on the Portable, Extensible Toolkit for Scientific Computation 
(PETSc) algorithms \cite{petsc-web-page,petsc-user-ref,petsc-efficient}.
The grid structure used in our code is a multi-level set of properly nested, 
cell-centered uniform grids. We use a standard second-order finite difference stencil for
the Laplacian operator and first order interpolation across the refinement level boundaries.
The non-linearity of Eq.~\eqref{ham2} is addressed by performing Newton-Raphson iterations.
A brief description of our FMR implementation is summarized in Appendix~\ref{appA}, to which we refer the 
interested reader.


\subsubsection{Diagnostics}

To check the consistency of solutions obtained with our FMR code
 we calculate the following diagnostic quantities:

The ADM mass is given by
\labeq{MADM}{
M_{\rm ADM}  = -\frac{1}{2\pi}\oint_\infty \partial^i \Psi dS_i
              = -\frac{1}{2\pi}\int \nabla^2\Psi d^3x,
}
where we have applied Gauss' theorem to convert the surface
 integral into a volume integral. 
The actual expression we use
to calculate the ADM mass volume integral is \eqref{MADM} with $\nabla^2\Psi$
replaced by the right-hand-side of Eq.~\eqref{ham2}.

The Komar mass is given by 
\labeq{MKomar}{
  M_{\rm K} = \frac{1}{4\pi}\oint_\infty \partial^i \alpha dS_i
            = \frac{1}{4\pi}\int \nabla^2\alpha d^3x,
}
where again we have applied Gauss' theorem in the last step. 
By use of Eqs.~\eqref{ham2} and \eqref{lap2} we find
\labeq{maximal_slicing_lapse}{
\nabla^2 \alpha = -\frac{2}{\Psi}\nabla_i\alpha\nabla^i \Psi
                 + 4\pi\alpha\Psi^4(\rho+S).
}
The actual expression we use to calculate the Komar mass volume integral 
is \eqref{MKomar} with $\nabla^2\alpha$ replaced by the 
right-hand-side of Eq.~\eqref{maximal_slicing_lapse}.

Finally, the total baryon mass is given by \cite{Baum98_NSNS}
\labeq{Mo}{
M_{\rm 0} = \int_{{\cal M}} \rho_o \alpha u^t \Psi^6 d^3x,
}
where ${\cal M}$ means that the integration is
 carried over the support of the matter. 

\begin{figure}[t]
\centering
{\includegraphics[width=0.495\textwidth,angle=0]{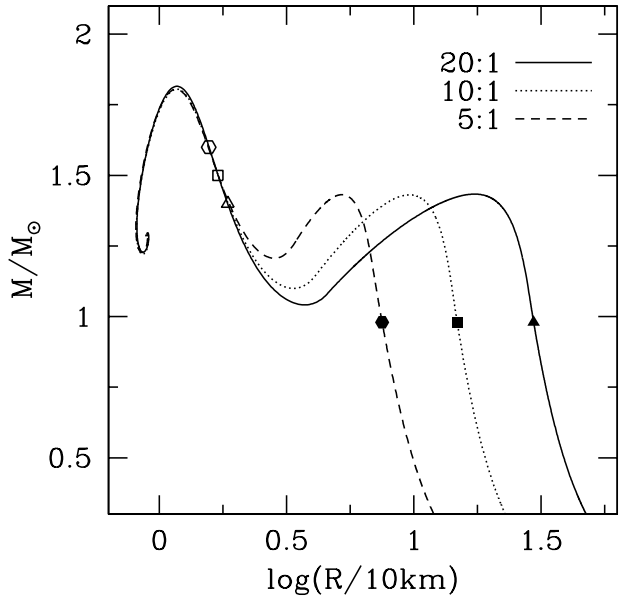}}
\caption{{ Mass -- radius relationship of TOV stars generated by
the piecewise polytropic EOS \eqref{EOS}. Plotted are curves
corresponding to three versions of the EOS where the ratio
 of the isotropic radius of a $0.98M_\odot$ pWD to that of a $1.5M_\odot$ NS 
is  20:1, 10:1 and 5:1. The corresponding EOS parameters 
are given in Table~\ref{tab:eosparams}. The open points correspond to the
 NS models studied in this paper, which have masses $1.4M_\odot$,
 $1.5M_\odot$ and $1.6M_\odot$.  The solid points correspond to the
 pWD models considered in this work, which all have the same mass: $0.98M_\odot$. 
 \label{fig:models}}}
\centering
\end{figure}

\subsubsection{Code Testing}

\begin{figure}[t]
\centering
\includegraphics[width=0.49\textwidth]{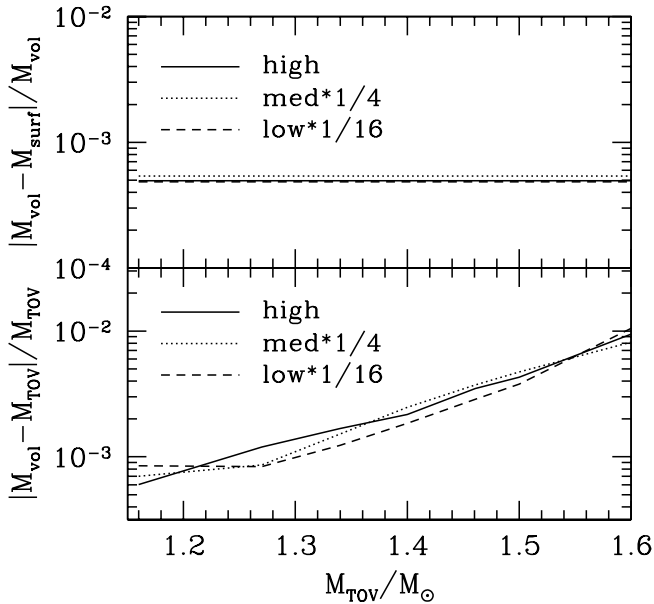}
\caption{{ Convergence test for the FMR elliptic code using TOV stars. 
    Five levels of refinement have been used for this test.  
Upper panel: Fractional difference between the volume integral $M_{\rm vol}$ and 
surface integral $M_{\rm surf}$ for the ADM mass, calculated 
with our FMR elliptic code, versus the gravitational mass 
$M_{\rm TOV}$ calculated with a 1D integrator of the TOV equations. 
The plot demonstrates satisfaction of Gauss' theorem and second-order convergence. 
Lower panel: Fractional difference between $M_{\rm vol}$ and $M_{\rm TOV}$ 
versus $M_{\rm TOV}$. The plot demonstrates
 equality of $M_{\rm vol}$ and $M_{\rm TOV}$ and second-order
 convergence of our FMR elliptic code to the 1D (exact) result. 
In both panels three resolutions are plotted:
 ${\rm low}\equiv 32^3$, ${\rm med}\equiv 64^3$,
${\rm high}\equiv 128^3$. Resolutions $32^3$ and $64^3$
 have been rescaled with a factor of $1/16$ and $1/4$ respectively, 
so that they overlap with resolution $128^3$.
 \label{fig:convergence}}}
\centering
\end{figure}

Gauss' theorem constitutes a strong consistency check for our FMR code. 
To demonstrate that the solutions obtained with our elliptic code satisfy
Gauss' theorem and achieve second-order convergence  we performed the following test. 
Employing the $10:1$ piecewise polytropic EOS
we constructed TOV NS solutions of various masses with a 1D TOV integrator. 
We then used second-order polynomial interpolation
to set up the rest-mass density profiles in our FMR elliptic code and
 solve Eqs. \eqref{ham2} and \eqref{lap2}
for the  conformal factor and lapse function. 
We set up grids with five levels of refinement ($nl = 5$) centered on
 the NS, and three different resolutions 
$nx=ny=nz=32,\ \Delta x_5= 2.2 \rm km$, $nx=ny=nz=64,
\ \Delta x_5= 1.1 \rm km$ and $nx=ny=nz=128,\ \Delta x_5= 0.55 \rm km$. 
In Fig.~\ref{fig:convergence} we 
demonstrate that our numerical solutions satisfy Gauss' theorem, are in agreement 
with the TOV integration, and that our code is 2nd-order convergent for this test.
To generate the plot we used the results of the ADM mass integration
but the convergence properties remain the same when we use the results of the Komar mass integration.

\section{Evolution of WDNS systems}
\label{sec:evolutions}

\subsection{Basic Equations}
\label{sec:basic_eqns}

The formulation and numerical scheme for our simulations are
the same as those already reported 
in~\cite{PhysRevD.72.024028,Etienne08a,Illinois_new_mhd}, to which
the reader may refer for details.  Here we introduce our notation and
summarize our method.

We use the 3+1 formulation of general relativity where
 the metric is decomposed in the form of Eq.~\eqref{ADMmetric}, and where
the fundamental dynamical variables for the metric evolution are the spatial
three-metric $\gamma_{ij}$ and extrinsic curvature $K_{ij}$. We adopt
the Baumgarte-Shapiro-Shibata-Nakamura (BSSN) 
formalism~\cite{ShibNakamBSSN,BaumShapirBSSN,BSBook}, in which 
the evolution variables are the conformal exponent $\phi
\equiv \ln (\gamma)/12$, the conformal 3-metric $\tilde
\gamma_{ij}=e^{-4\phi}\gamma_{ij}$, three auxiliary functions
$\tilde{\Gamma}^i \equiv -\tilde \gamma^{ij}{}_{,j}$, the trace of
the extrinsic curvature $K$, and the trace-free part of the conformal extrinsic
curvature $\tilde A_{ij} \equiv e^{-4\phi}(K_{ij}-\gamma_{ij} K/3)$.
Here, $\gamma={\rm det}(\gamma_{ij})$. The full spacetime metric $g_{\mu \nu}$
is related to the three-metric $\gamma_{\mu \nu}$ by $\gamma_{\mu \nu}
= g_{\mu \nu} + n_{\mu} n_{\nu}$, where the future-directed, timelike
unit vector $n^{\mu}$ normal to the time slice can be written in terms
of the lapse $\alpha$ and shift $\beta^i$ as $n^{\mu} = \alpha^{-1}
(1,-\beta^i)$. The evolution equations of these BSSN variables are 
given by Eqs.~(9)--(13) in~\cite{Etienne08a}.

We adopt standard puncture gauge conditions: an advective
``1+log'' slicing condition for the lapse and a 
``Gamma-freezing'' condition for the shift~\cite{PunctureGauge}. 
Thus, we have 
\beqn
  \partial_0 \alpha &=& -2\alpha K  \ , \label{eq:1+log} \\ 
  \partial_0 \beta^i &=& (3/4) B^i \ , \\ 
  \partial_0 B^i &=& \partial_0 \tilde{\Gamma}^i - \eta B^i \ ,
\label{puncturegauge}
\eeqn
where $\partial_0 \equiv \partial_t - \beta^j \partial_j$. We
set the $\eta$ parameter to $0.01\rm km^{-1}$ 
for all simulations presented in this work.

The fundamental matter variables are the rest-mass density 
$\rho_0$, specific internal energy $\epsilon$, pressure $P$, and 
four-velocity $u^{\mu}$. We write the stress-energy tensor as
\beq
  T_{\mu \nu} = \rho_0 h u_\mu u_\nu + P g_{\mu \nu} \ ,
\eeq
where $h=1+\epsilon+P/\rho_0$ is the specific enthalpy
and $\epsilon$ is the total energy density. 
In our numerical implementation of the hydrodynamics 
equations, we evolve the ``conservative'' variables 
$\rho_*$, $\tilde{S}_i$, and $\tilde{\tau}$. They are 
defined as 
\beqn
&&\rho_* \equiv - \sqrt{\gamma}\, \rho_0 n_{\mu} u^{\mu} \ ,
\label{eq:rhos} \\
&& \tilde{S}_i \equiv -  \sqrt{\gamma}\, T_{\mu \nu}n^{\mu} \gamma^{\nu}_{~i}
\ , \\
&& \tilde{\tau} \equiv  \sqrt{\gamma}\, T_{\mu \nu}n^{\mu} n^{\nu} - \rho_* \ .
\label{eq:S0} 
\eeqn
The evolution equations for these variables are given by Eqs.~(21)--(24) 
in~\cite{Etienne08a}.

The EOS we adopt for the evolution has both a thermal and cold contribution, i.e.,
\labeq{Ptot}{
P = P_{\rm th} + P_{\rm cold},
}
where $P_{\rm cold}$ is given by Eq.~\eqref{EOS} and
 the thermal pressure is given by
\labeq{}{
P_{\rm th} = (\Gamma_{th} - 1)\rho_0(\epsilon-\epsilon_{\rm cold}),
}
where  
\labeq{}{
\epsilon_{\rm cold} = - \int P_{\rm cold}d(1/\rho_0).
}
We set $\Gamma_{\rm th} = 1.66$ ($\simeq 5/3$) in all our simulations, i.e., we set it equal
to the $\Gamma_1$ exponent of the $10:1$ EOS,
appropriate either for nonrelativistic cold degenerate electrons or (shock)
heated, ideal nondegenerate baryons.
Eq.~\eqref{Ptot} reduces to our piecewise polytropic law 
Eq.~\eqref{EOS} for the initial (cold) NS and pWD matter. 

\subsection{Evolution of the metric and hydrodynamics}
\label{sec:num_metric_hydro}

We evolve the BSSN equations using
fourth-order accurate, centered finite-differencing stencils,
except on shift advection terms, where fourth-order accurate
upwind stencils are applied.  We apply Sommerfeld outgoing wave boundary
conditions on all BSSN fields, as in~\cite{Etienne08a}.  Our code is embedded in
the Cactus parallelization framework~\cite{Cactus}, and our
fourth-order Runge-Kutta timestepping is managed by the {\tt MoL}
(Method of Lines) thorn, with the CFL number
set to 0.45 in all pWDNS simulations.  We use the
Carpet~\cite{Carpet} infrastructure to implement the moving-box
adaptive mesh refinement. In all AMR simulations presented here, we
use second-order temporal prolongation, coupled with fifth-order
spatial prolongation, and impose equatorial symmetry to reduce the computational 
cost.

We write the general relativistic hydrodynamics equations in
conservative form. They are evolved via a high-resolution
shock-capturing (HRSC) technique~\cite{PhysRevD.72.024028,Illinois_new_mhd}
 that employs the
piecewise parabolic (PPM) reconstruction scheme~\cite{PPM}, coupled to
the Harten, Lax, and van Leer (HLL) approximate Riemman solver~\cite{HLL}. 
The adopted hydrodynamic scheme is second-order accurate. 
To stabilize our hydrodynamic scheme in regions where there is no
matter, a tenuous atmosphere is maintained on our grid, with a density
floor $\rho_{\rm atm}$ set to $10^{-10}$ times the initial
maximum density on our grid. The initial atmospheric pressure
$P_{\rm atm}$ is set by using the cold EOS~\eqref{EOS}.
Throughout the evolution, we impose limits on the pressure to prevent
spurious heating and negative values of the internal energy
$\epsilon$. Specifically, we require $P_{\rm min}\leq P \leq P_{\rm max}$, 
where $P_{\rm max}=10 P_{\rm cold}$ and $P_{\rm min}=0.8P_{\rm cold}$,
 where $P_{\rm cold}$ is 
the pressure calculated using the cold EOS~\eqref{EOS}.
Whenever $P$ exceeds $P_{\rm max}$ or drops below $P_{\rm min}$, we 
reset $P$ to $P_{\rm max}$ or $P_{\rm min}$, respectively.  
Following~\cite{Etienne09} we impose the upper pressure limits only 
in regions where the rest-mass density remains very low ($\rho_0 < 100
\rho_{\rm atm}$), but we impose the lower limit everywhere on our grid.

At each timestep, 
the ``primitive variables'' 
$\rho_0$, $P$, and $v^i$ must be recovered from the ``conservative'' variables 
$\rho_*$, $\tilde{\tau}$, and $\tilde{S}_i$. We perform the 
inversion numerically as specified in~\cite{Illinois_new_mhd}.  We use 
the same technique as
in ~\cite{Etienne08a} to ensure that the values of $\tilde{S}_i$ and
$\tilde{\tau}$ yield physically valid primitive variables,
except we reset $\tilde{\tau}$ to
$10^{-10}\tilde{\tau}_{0,{\rm max}}$ (where
$\tilde{\tau}_{0,{\rm max}}$ is the maximum value of $\tilde{\tau}$
initially) when either $\tilde{S}_i$ or $\tilde{\tau}$ is 
unphysical [i.e., violate one of the inequalities~(34)~or~(35)
in ~\cite{Etienne08a}]. The restrictions are usually imposed only
in the low-density atmosphere.

It is instructive to discuss
the mathematical structure of the system of hydrodynamic 
equations when a piecewise polytropic EOS~\eqref{EOS} is used. 
According to \cite{Leveque,VossAlexander}, when the fluxes of the conservation laws
are non-smooth, split waves and composite structures may be present in the solutions.
In these cases a numerical solution may not converge to the correct solution.
In our case the fluxes are not smooth everywhere because EOS \eqref{EOS} is non-smooth 
(it is continuous but not differntiable) at the 
turning points $\rho_i\ \ i=1,2$. Away from the turning points the fluxes are smooth, 
therefore there may be some concern when non-linear waves cross the transition densities.
However, according to \cite{VossAlexander} our adopted numerical scheme
should be able to handle such composite structures, if they ever arise, 
and our solutions should converge to the correct continuum solution. 

To study this effect we constructed an EOS which is similar
to Eq.~\eqref{EOS}, but where the pressure discontinuities are ``smoothed out'' at the turning 
points $\rho_i, (i=1,2)$, such that the EOS becomes a smooth, once (or twice) differentiable function of the rest-mass density.
We perform such a smoothing operation using a cubic (or quintic) spline over a density
interval $[\rho_i(1-\epsilon), \rho_i(1+\epsilon)]$, where $\epsilon>0$. We chose
$\epsilon$ to be sufficiently small so that the smoothed EOS mimics as closely as possible EOS \eqref{EOS}, 
but large enough to avoid roundoff errors due to very large gradients. Setting up several generalized Riemman 
problems, we found that the numerical solutions obtained using EOS \eqref{EOS} converge to those obtained
when using its smooth counterpart and the two can hardly be distinguished for the resolutions considered. 
Therefore, our numerical schemes in conjunction with
EOS \eqref{EOS} are able to capture the correct solution, in that they are almost identical to the solutions
obtained with the smooth counterpart of \eqref{EOS}. For the details of this analysis, we refer the interested reader to Appendix \ref{appB}.

\begin{center}
\begin{table*}[t]
\caption{Summary of initial configurations. $M_{\rm NS}$ ($M_{\rm WD}$)
 stands for the ADM mass of an isolated NS (pWD)$^{(a)}$, $R_{\rm NS}$ ($R_{\rm WD}$) 
is the isotropic radius of an isolated NS (pWD), $C_{\rm NS}$ ($C_{\rm WD}$)  
is the compaction of an isolated NS (pWD), where the compaction is 
defined as the ratio of the ADM mass of the isolated star to its areal radius. 
$M_{\rm ADM}$ is the ADM mass of the system and $A$ the initial binary separation 
in isotropic coordinates. 
All cases have exactly the same coordinate separation of $586.9\rm km$
 to allow for comparison. Cases A1, A2, A3 have been
 produced with the 10:1 EOS, while cases B and C have
 been produced with the 5:1 and 20:1 EOSs, respectively.}
\begin{tabular}{ccccccccc}\hline\hline
\multicolumn{1}{p{1.8cm}}{\hspace{0.45 cm} Case } & 
\multicolumn{1}{p{1.8cm}}{\hspace{0.15 cm} $M_{\rm NS}/M_\odot$ \quad}  &
\multicolumn{1}{p{1.8cm}}{\hspace{0.1 cm} $M_{\rm WD}/M_\odot$ \quad}  &
\multicolumn{1}{p{1.8cm}}{\hspace{0.4 cm} $C_{\rm NS}$ \quad}  &
\multicolumn{1}{p{1.8cm}}{\hspace{0.35 cm} $C_{\rm WD}$ \quad}  &
\multicolumn{1}{p{1.8cm}}{\hspace{0.05 cm} $R_{\rm WD}/R_{\rm NS}$ \quad}  &
\multicolumn{1}{p{2.1cm}}{\hspace{0.1 cm} $R_{\rm WD}/M_{\rm ADM}$}  &
\multicolumn{1}{p{1.95cm}}{\hspace{0.05 cm} $M_{\rm ADM}/M_\odot$ }  &
\multicolumn{1}{p{1.6cm}}{\hspace{0.2 cm} $A/R_{\rm WD}$ \quad}  \\  \hline 
A1    	               &    1.4    &  0.98  & 0.111    & 0.010      &  8.88    & 41.18    & 2.413  &   4.000        \\  \hline
A2    	               &    1.5    &  0.98  & 0.130    & 0.010      &  9.96    & 39.36    & 2.524  &   4.000        \\  \hline
A3    	               &    1.6    &  0.98  & 0.151    & 0.010      &  11.15   & 37.46    & 2.652  &   4.000        \\  \hline
B    	               &    1.5    &  0.98  & 0.130    & 0.019      &  4.99    & 19.76    & 2.524  &   7.967        \\  \hline
C    	               &    1.5    &  0.98  & 0.130    & 0.005      &  20.01   & 79.08    & 2.524  &   1.991       \\  \hline\hline 
\end{tabular}
\begin{flushleft}
$^{(a)}$ Here we list the ADM masses, isotropic radii and compactions of the isolated (TOV) stars whose rest-mass density profiles we
used to generate initial data for $\Psi$ and $\alpha$ for a given case. 
\end{flushleft}
\label{tab:cases}
\end{table*}
\end{center}

\subsection{Evolution Diagnostics}
\label{sec:diagnostics}

During the evolution, we monitor the normalized Hamiltonian and momentum
constraints as defined in Eqs.~(40)--(43) of ~\cite{Etienne08a}. 

We also monitor the ADM mass and angular momentum of the system, which
can be calculated during the evolution by surface integrals at a large distance (Eqs.~(37)
and (39) of ~\cite{Etienne08a}). The equations used to
calculate the ADM mass and angular momentum with minimal numerical noise
are as follows \cite{BSBook}
\beqn
  M&=& \int_V d^3x \left(\psi^5\rho + {1\over16\pi}\psi^5 \tilde{A}_{ij}
\tilde{A}^{ij} - {1\over16\pi}\tilde{\Gamma}^{ijk}\tilde{\Gamma}_{jik} \right.
\ \ \label{eq:M_sur_vol} \nonumber \\
&& \left. + {1-\psi\over16\pi}\tilde{R} - {1\over24\pi}\psi^5K^2\right), 
\eeqn

\beqn
  J_i &=& {1\over8\pi} \epsilon_{ij}{}^n\int_V d^3x
           \bigl[\psi^6(\tilde{A}^j{}_n + {2\over3}x^j\partial_nK 
\label{eq:J_sur_vol}  \nonumber \\
         && - {1\over2} x^j\tilde{A}_{km}\partial_n\tilde{\gamma}^{km}) 
         + 8\pi x^j S_n\bigr].  
\eeqn
Here
 $V$ is the volume within a distant surface, $\psi = e^\phi$, 
$\rho=n_\mu n_\nu T^{\mu \nu}$, $S_i = -n_\mu \gamma_{i \nu} T^{\mu \nu}$, 
$\tilde{R}$ is the Ricci scalar associated with $\tilde{\gamma}_{ij}$, 
and $\tilde{\Gamma}_{ijk}$ are Christoffel symbols associated with 
$\tilde{\gamma}_{ij}$.

In this work we only focus on head-on collisions, so there
 is no angular momentum involved. 
However, our simulations are three-dimensional, so there is no guarantee
 that $J_i$ will remain 0.
In order to quantify violations of $J_i=0$ we normalize the angular momentum, 
computed via \eqref{eq:J_sur_vol}, with the angular momentum
 a pWDNS system would have, 
if the binary components were Newtonian point masses 
in circular orbit at the initial separation $A$, i.e.,
\labeq{Newt_J}{
J_{\rm z,c} = M_{\rm T}^{3/2}A^{1/2}\frac{q}{(1+q)^2},
}
where the total mass $M_{\rm T}$ is taken to be the
 sum of ADM masses of the isolated stars.

When hydrodynamic matter is evolved on a fixed uniform grid, our
hydrodynamic scheme guarantees that the rest mass $M_0$ is conserved
to machine roundoff error.  This strict conservation is no longer maintained
in an AMR grid, where spatial and temporal prolongation is performed
at the refinement boundaries.  Hence, we also monitor the
rest mass
\beq
  M_0 = \int \rho_* d^3x
\label{eq:m0}
\eeq
during the evolution. Rest-mass conservation is also violated whenever 
$\rho_0$ is reset to the atmosphere value. This usually happens only in the 
very low-density atmosphere.  The low-density regions do not affect rest-mass 
conservation significantly. 

Shocks occur when stars collide. We measure the
 entropy generated by shocks via the quantity
$K\equiv P/P_{\rm cold}\geq 1$, where $P_{\rm cold}$ is
the pressure associated with the cold EOS that characterizes the
initial matter (see Eq.~\eqref{EOS}). 

\begin{table*}[t]
\caption{Grid configurations used in our simulations. Here $M$ is the sum of the
  ADM masses of the isolated stars, Max.~res. is the grid spacing in the innermost refinement 
  box surrounding the NS, $N_{\rm NS}$ denotes the number of
 grid points covering the diameter of  the NS initially, and $N_{\rm WD}$
  denotes the number of grid points covering the diameter
  of the pWD initially. The smallest outer
 boundary distance corresponds to case A3 and is $1028M$.}
\begin{tabular}{cccccc}
  \hline \hline
  Case & $M/M_{\odot}$ & Grid Hierarchy (in units of $M$)$^{(a)}$
  & Max.~res. & $N_{\rm NS}$ & $N_{\rm WD}$ \\
  \hline \hline
  A1  &  2.38   & (534.33, 267.16, 133.58, 66.79, 35.78[N/A], 19.08[N/A], 10.44[N/A],
  7.156[N/A]) & $M/6.71$ & 63 & 35 \\
  A2a &  2.48     & (510.62, 255.31, 127.65 , 63.83, 34.81[N/A], 18.86[N/A], 10.15[N/A],
  6.890[N/A]) & $M/5.52$ & 44 & 28 \\
  A2b &  2.48      & (534.33, 267.16, 133.58, 66.79, 35.78[N/A], 19.08[N/A], 10.44[N/A],
  7.156[N/A]) & $M/6.71$ & 56 & 35 \\
  A3  &  2.58 & (467.27, 233.64,116.82, 58.41, 29.20 [N/A], 15.58[N/A], 8.518[N/A],
  5.841[N/A]) & $M/8.22$ & 56 & 38 \\
  B   &  2.48   & (534.33, 267.16, 133.58, 66.79, 35.78[31.93], 19.08[N/A], 10.44[N/A],
  7.156[N/A]) & $M/6.71$ & 56 & 35 \\
  C   &  2.48   & (534.33, 267.16, 133.58, 66.79[N/A], 35.78[N/A], 19.08[N/A], 10.44[N/A],
  7.156[N/A]) & $M/6.71$ & 56 & 35 \\
  \hline \hline
\end{tabular}

\begin{flushleft}
$^{(a)}$ There are two sets of nested refinement boxes: one centered
  on the NS and one on the pWD.  This column specifies the half side length
  of the refinement boxes centered on both the NS and pWD. When
  the side length around the pWD is different, we specify the pWD half side 
  length in square brackets.  When there is no corresponding pWD
  refinement box (as the pWD is much larger than the NS), 
  we write [N/A] for that box.
\end{flushleft}
\label{table:GridStructure}
\end{table*}

\begin{figure*}
\centering
%
\subfigure{\includegraphics[width=0.325\textwidth]{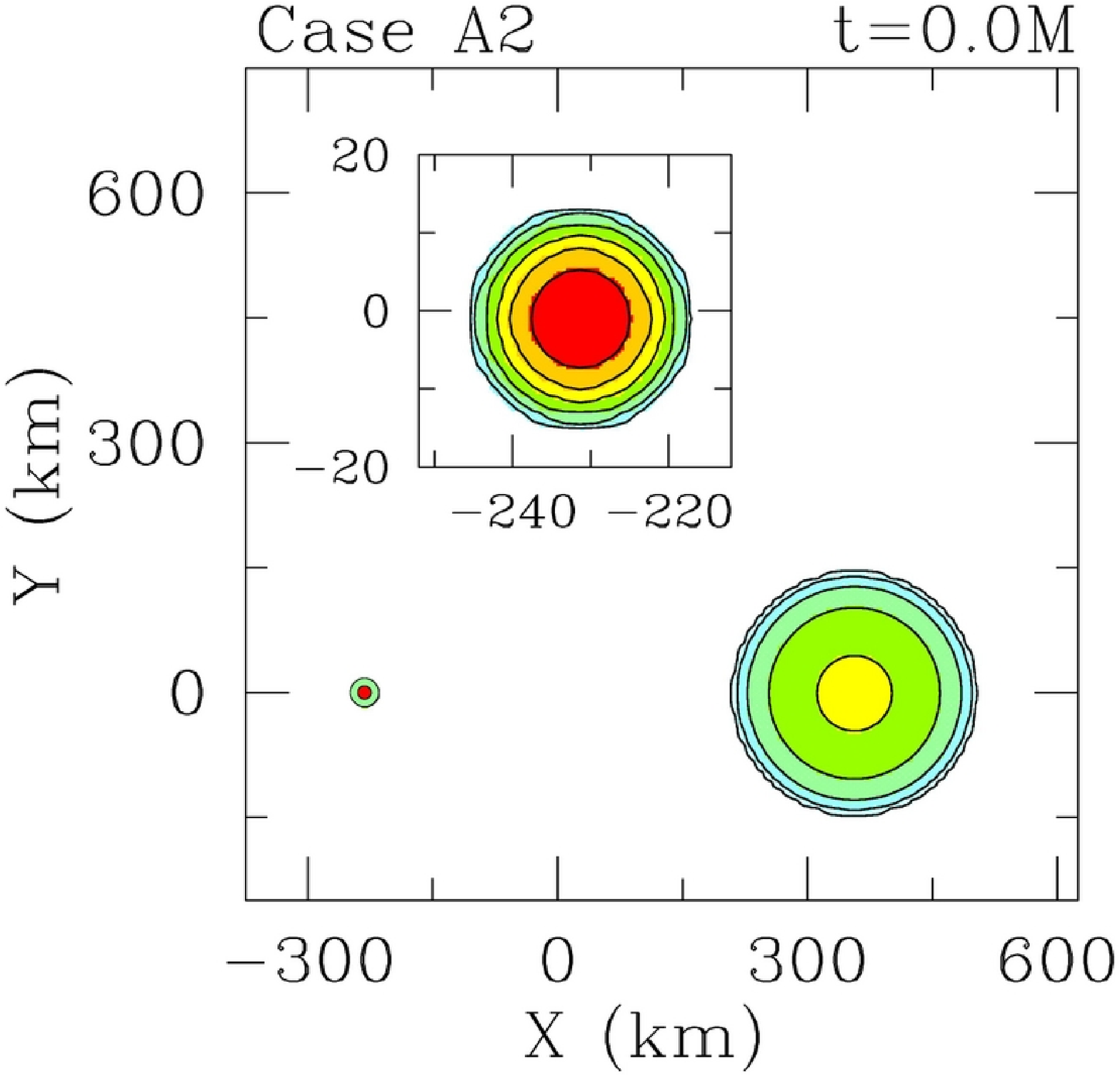}}
\subfigure{\includegraphics[width=0.325\textwidth]{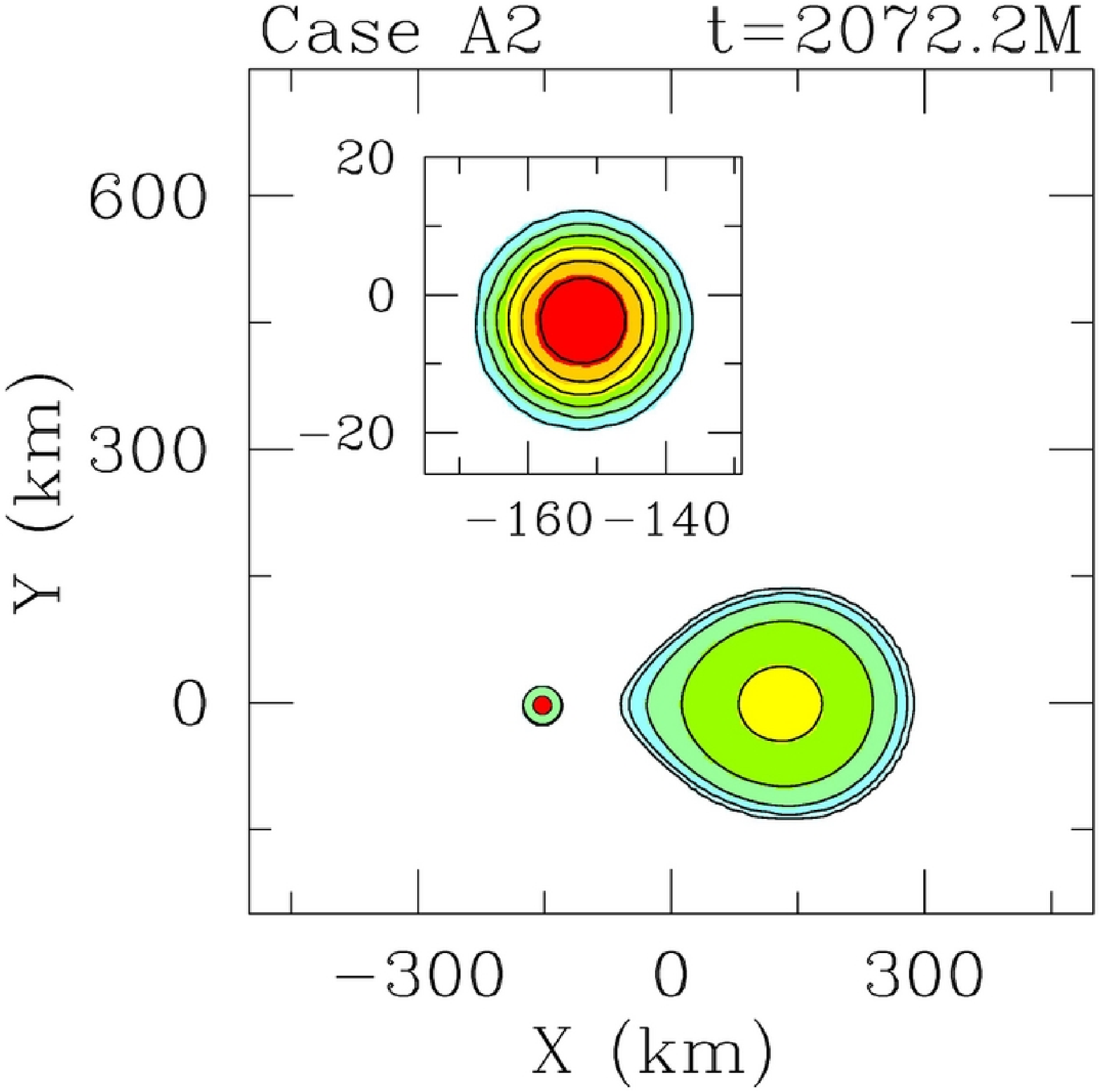}}
\subfigure{\includegraphics[width=0.325\textwidth]{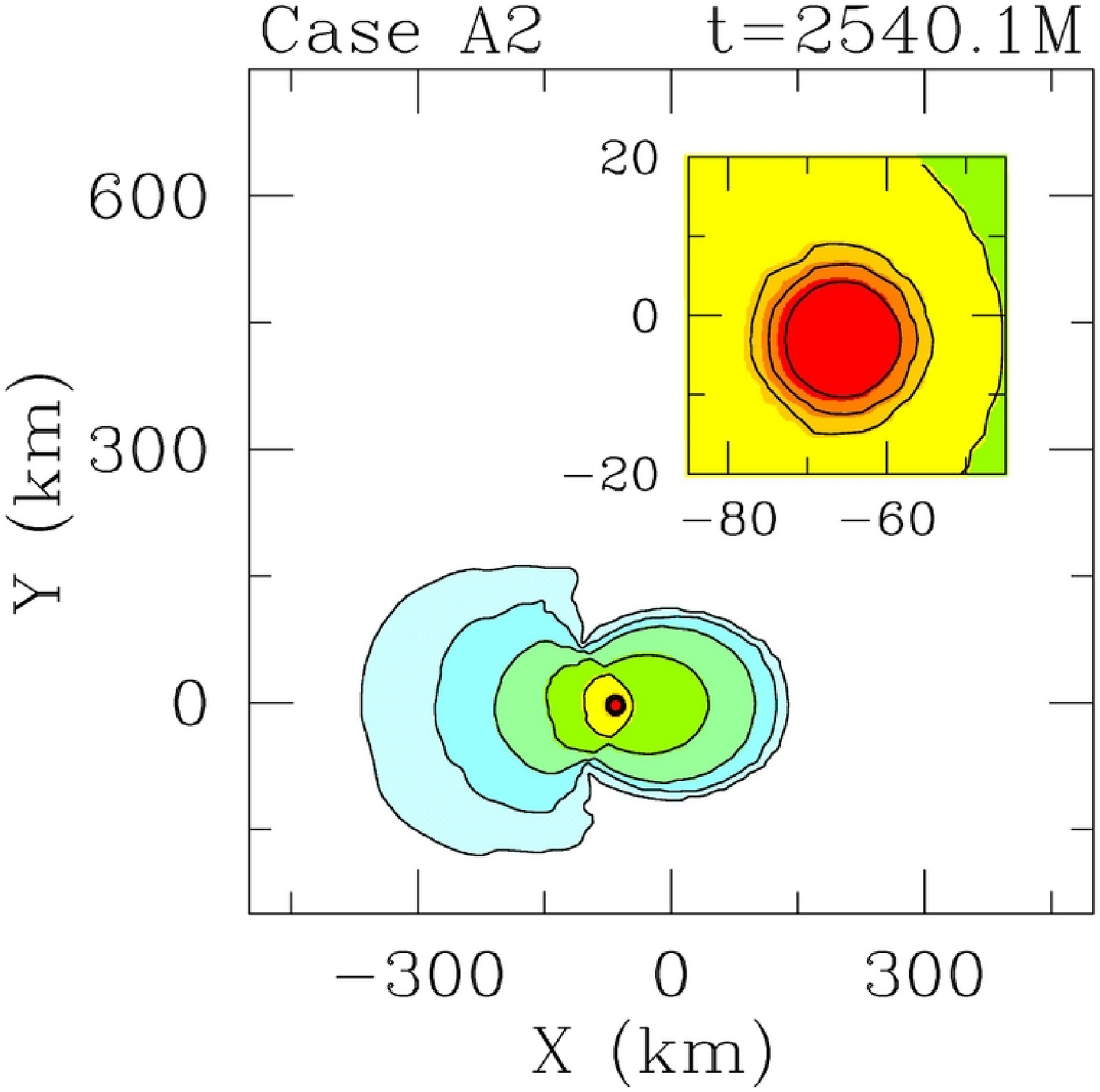}}
\subfigure{\includegraphics[width=0.325\textwidth]{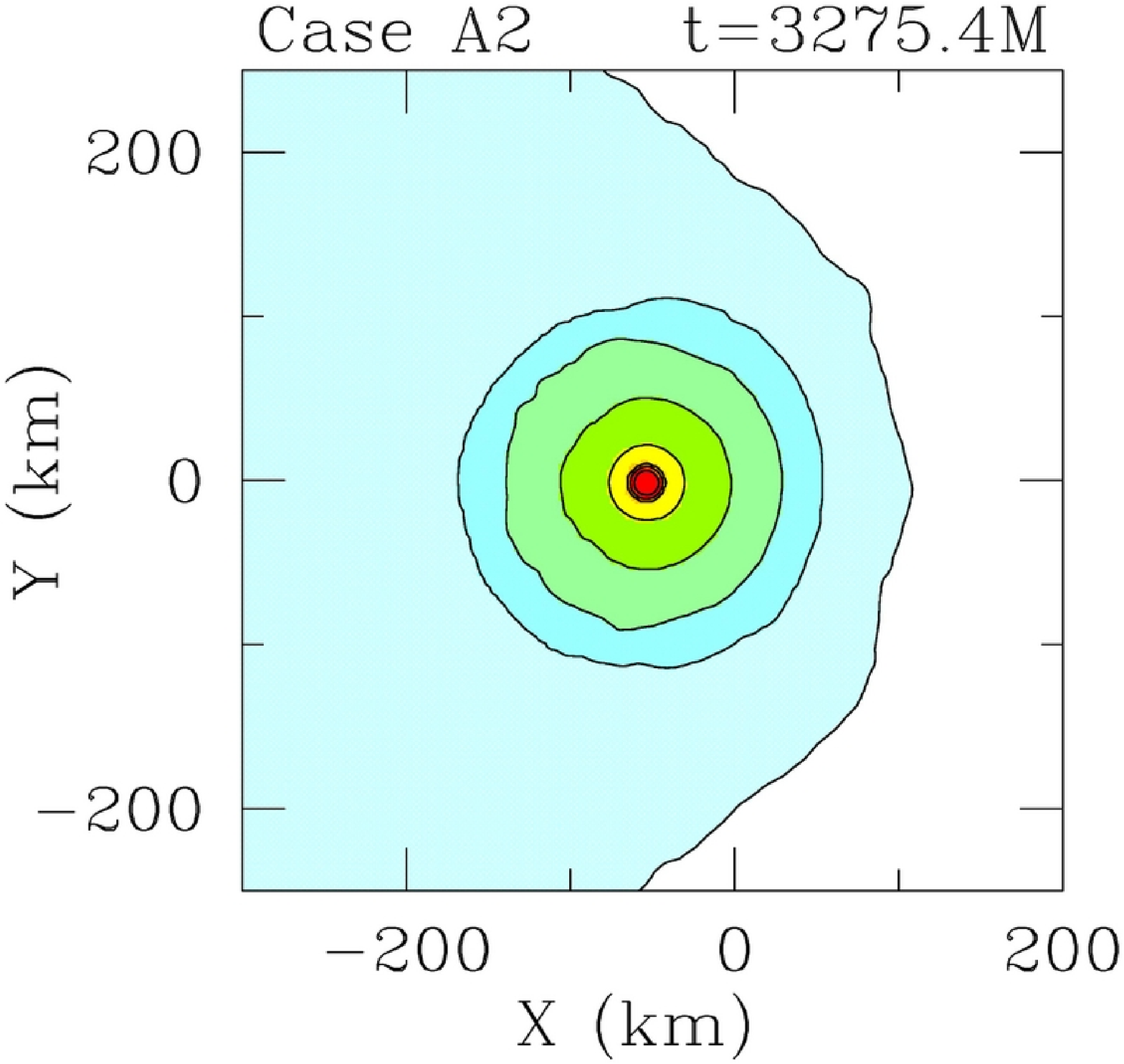}}
\subfigure{\includegraphics[width=0.325\textwidth]{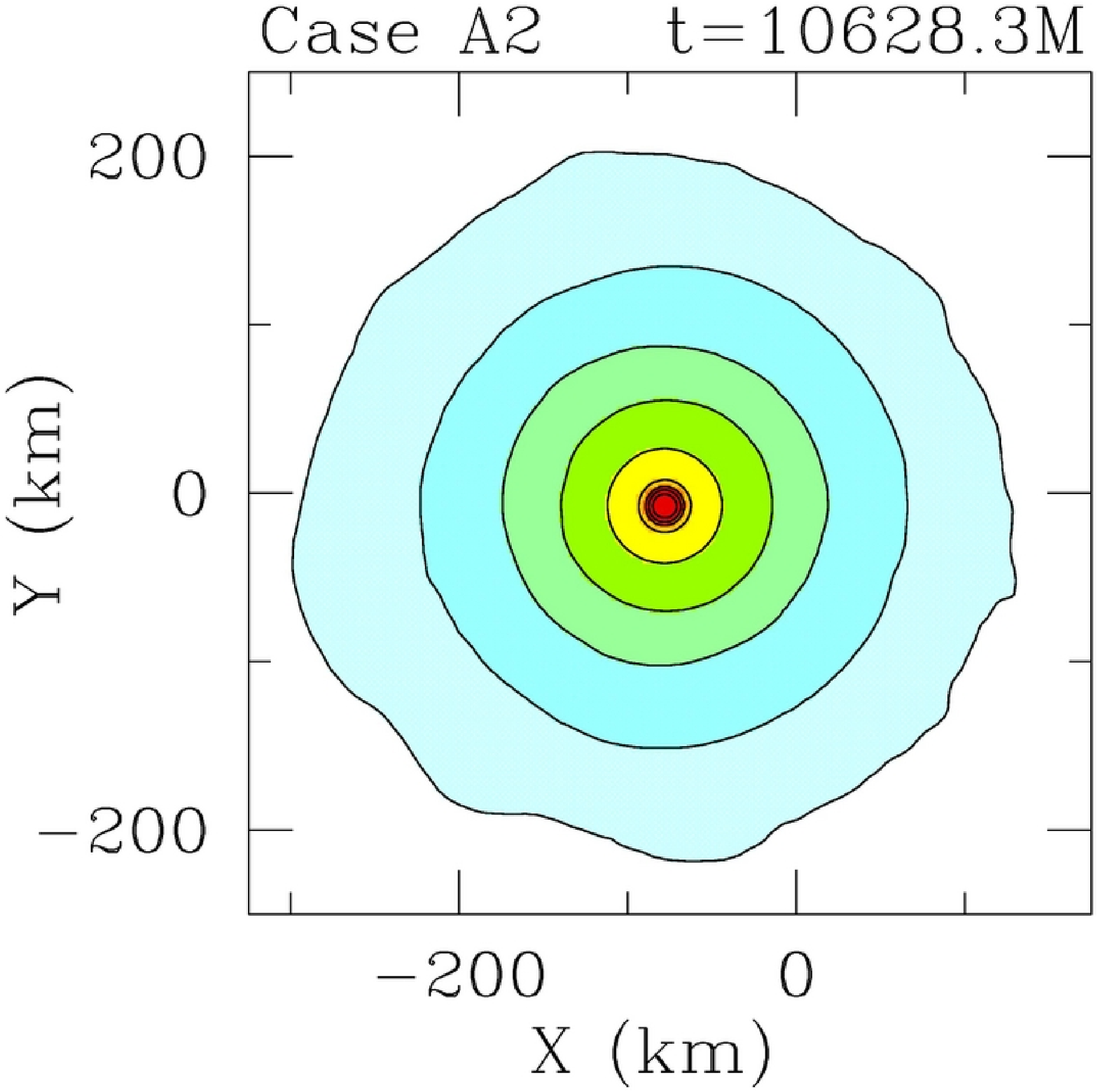}}
\subfigure{\includegraphics[width=0.325\textwidth]{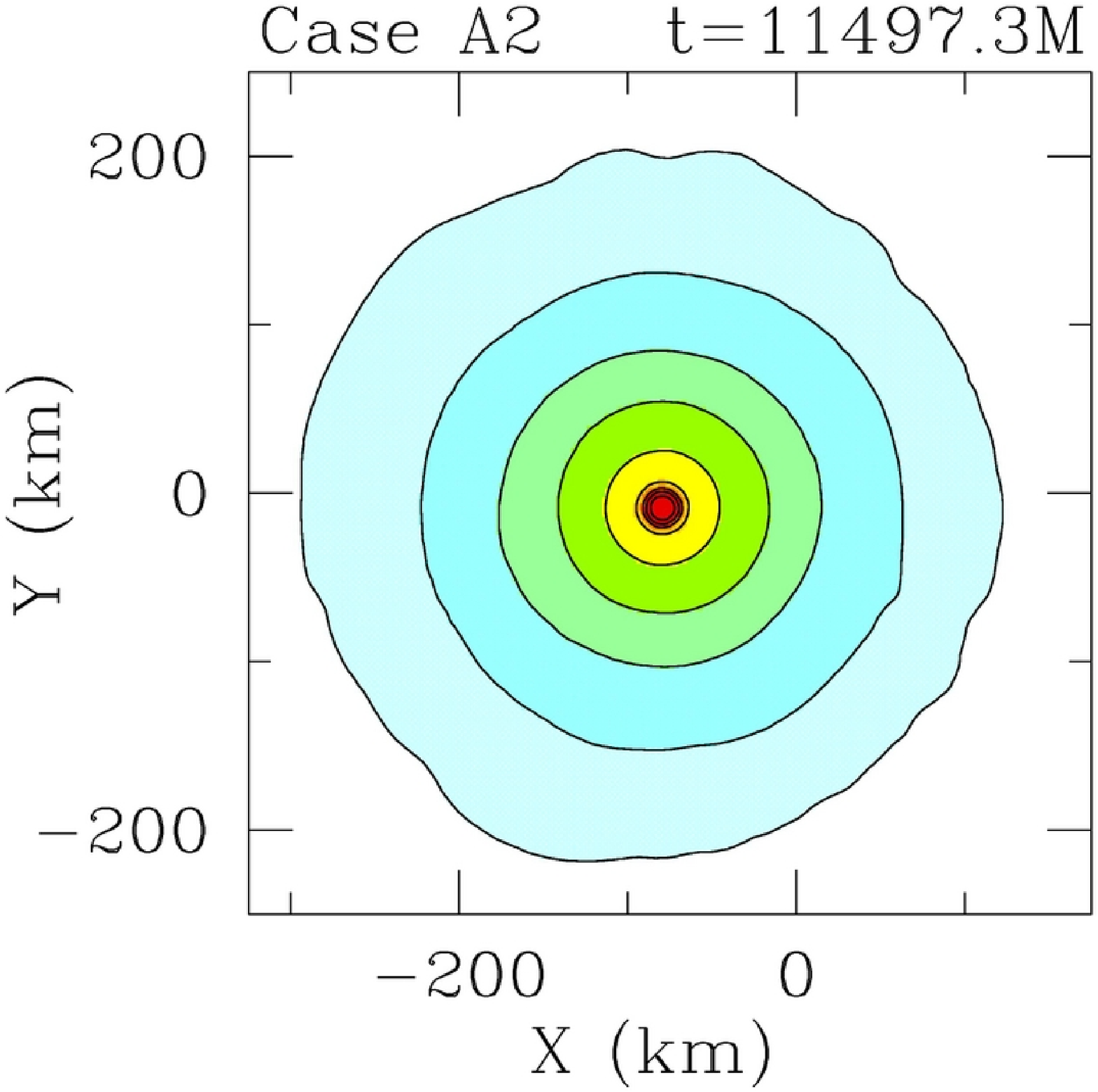}}
\caption{
Snapshots of rest-mass density profiles at selected times for case A2. 
 The contours represent the rest-mass density in the orbital plane, plotted 
according to $\rho_0 = \rho_{0,\rm max} 10^{-0.66j-0.16}\ (j=0,1,\ldots, 9)$. The color sequence
dark red, red, orange, yellow, green, light green, blue and light blue 
implies a sequence from higher to lower values. This roughly corresponds to darker greyscaling for higher values.
The maximum initial NS density is $\rho_{0,\rm max} = 4.6454\rho_{\rm nuc}$.
The last two snapshots are near the end of the simulation, and they demonstrate
that the density contours within a radius of about $150\rm km$ remain unchanged. 
Here $M = 2.48M_\odot = 3.662\rm km = 1.222\times 10^{-5} s $ 
is the sum of the ADM masses of the isolated stars. 
\label{fig:A2xy}
}
\centering
\end{figure*}

\section{Cases and Results}

\label{sec:results}


\subsection{Initial configurations}

We perform a number of pWDNS head-on collision simulations varying the 
initial configurations, so that we can study
 the effect of the pWD compaction and NS mass on the final outcome
 separately. Table~\ref{tab:cases} outlines the physical parameters for 
the cases considered in this work, and Table~\ref{table:GridStructure} 
presents the AMR grid structure used in each case.

To generate initial data for our cases we first choose the ADM mass $M_{\rm NS}$ of the NS 
and the ADM mass $M_{\rm WD}$ of the pWD (see Table~\ref{tab:cases}) and solve the TOV
equations for each star in isotropic coordinates to prepare the rest-mass density distribution 
for the NS and the pWD separately. We then use second-order polynomial
 interpolation to interpolate the rest-mass
 density profiles onto the nested grids of our FMR elliptic initial value code and solve
 Eqs.~\eqref{ham2} and~\eqref{lap2} for $\Psi$ and $\alpha$.  
The two stars are placed at coordinate separation $A$ and such
 that the Newtonian center of mass of the system is identified
 with the origin of the coordinate system. Once a solution
 is achieved by the FMR code for the initial metric, we map
$\rho_0$, $\Psi$ and $\alpha$ from the elliptic code
 grids onto the evolution grids using second-order polynomial interpolation.
We always make sure that the resolution of the initial data grids is higher 
than the resolution of the evolution grids.  
The surfaces of the stars are a locus of rapidly decreasing density gradients.
As a result, small oscillations due to interpolation may arise and lead to negative
rest-mass density. To cure this, we set the density $\rho_0$ equal to the 
tenuous atmosphere density that we maintain on the grid 
whenever $|\rho_0/\rho_{i,c}|<10^{-10}$, where $\rho_0$ is the value of the density 
after the interpolation and $\rho_{i,c},\ i={\rm WD,NS,}$ 
is the central density of the WD or NS. 
We do not find such oscillations when interpolating the gravitational fields,
which is most likely due to the fact that these are sufficiently smooth.


\begin{figure}
\centering
{\includegraphics[width=0.495\textwidth,angle=0]{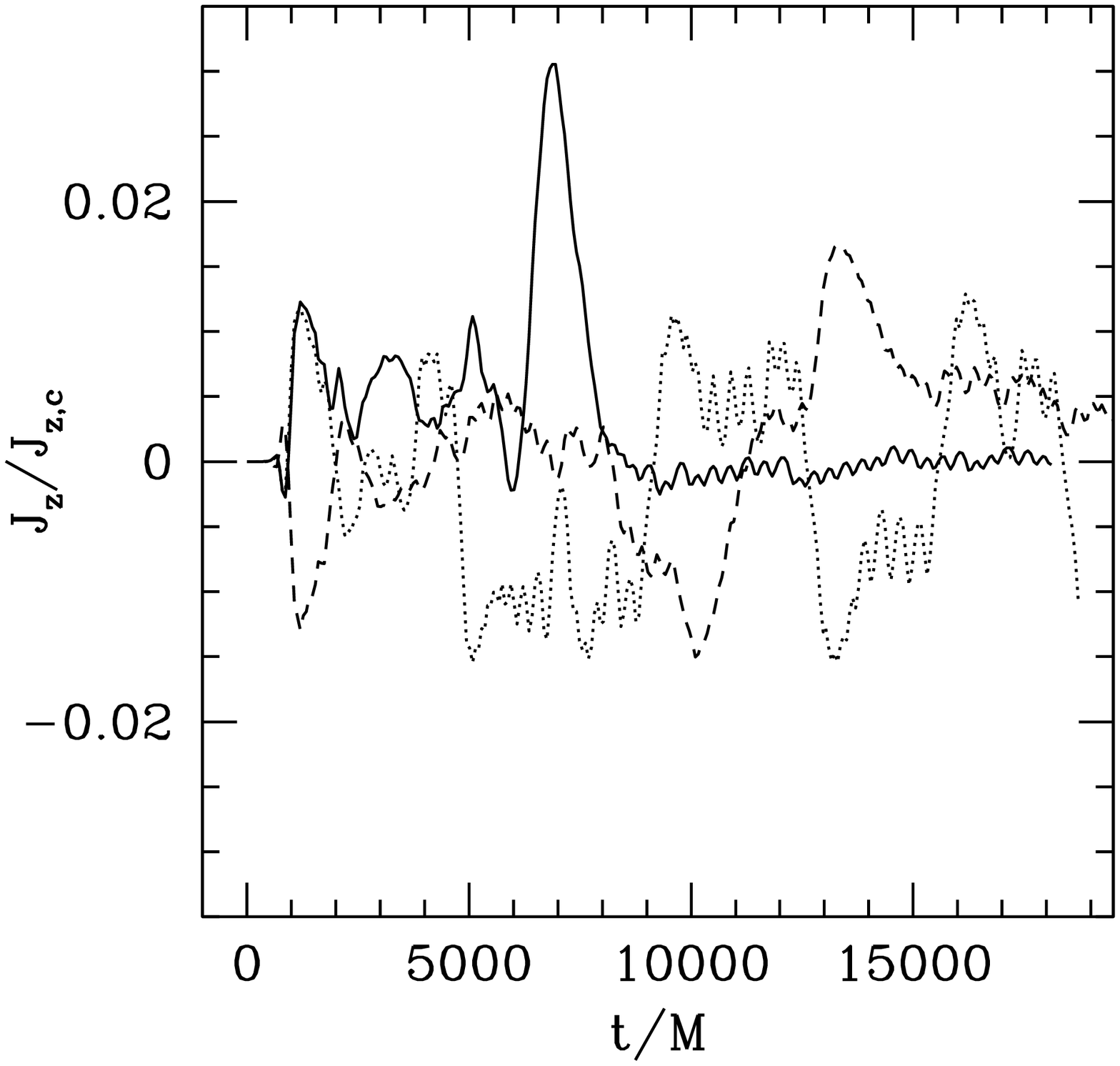}}
\caption{{Normalized angular momentum vs time. $J_{\rm z}$ and $J_{\rm
      z,c}$  are given by Eqs.~\eqref{eq:J_sur_vol} and
      \eqref{Newt_J}, respectively. The solid, dashed and dotted
      lines correspond to cases A2, B and C, respectively. 
Here $M = 2.48M_\odot = 3.662\rm km = 1.222\times 10^{-5} s $. 
 \label{fig:ang_mom}}}
\centering
\end{figure}

\begin{figure*}
\centering
%
\subfigure{\includegraphics[width=0.45\textwidth]{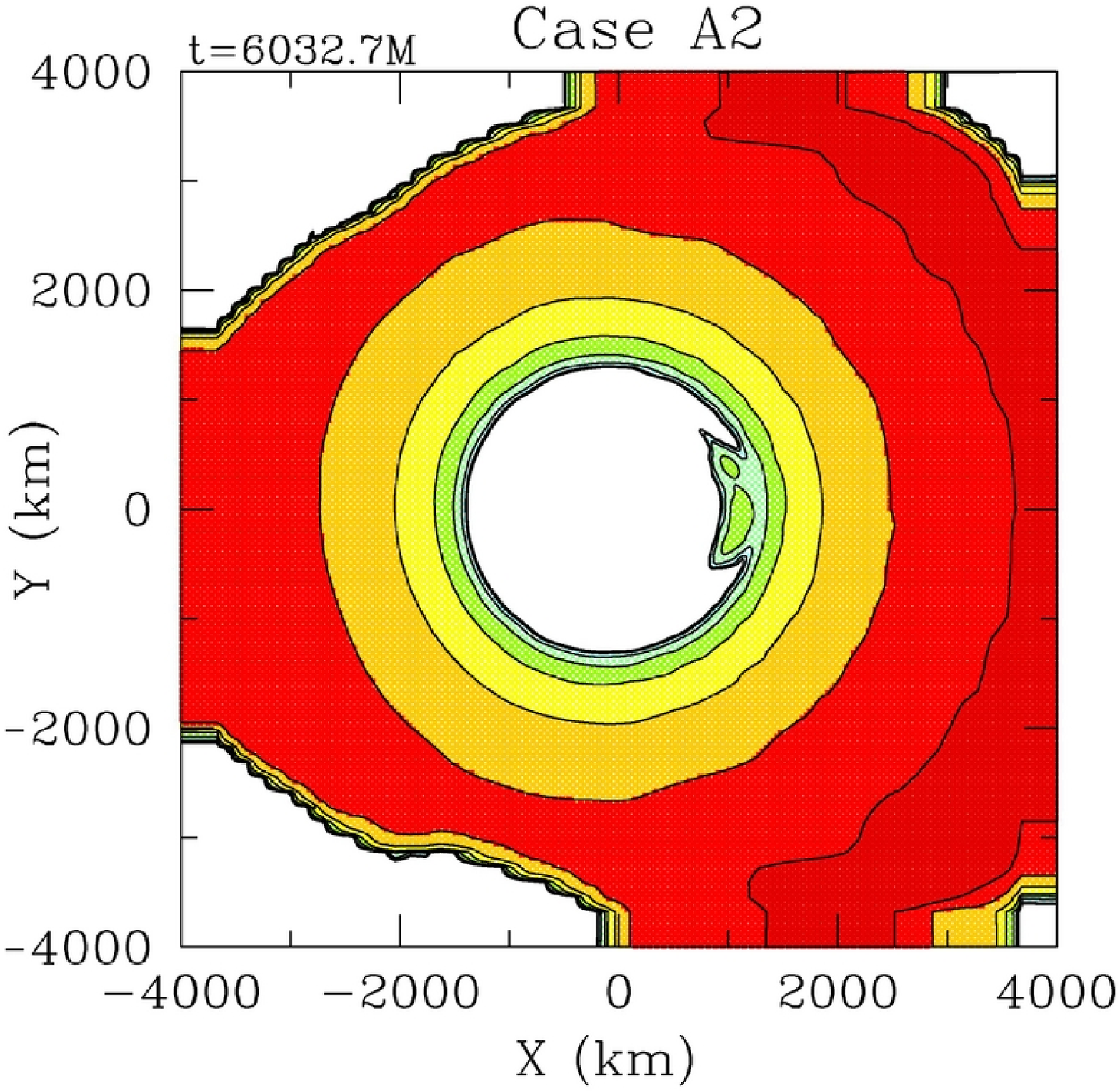}}
\subfigure{\includegraphics[width=0.45\textwidth]{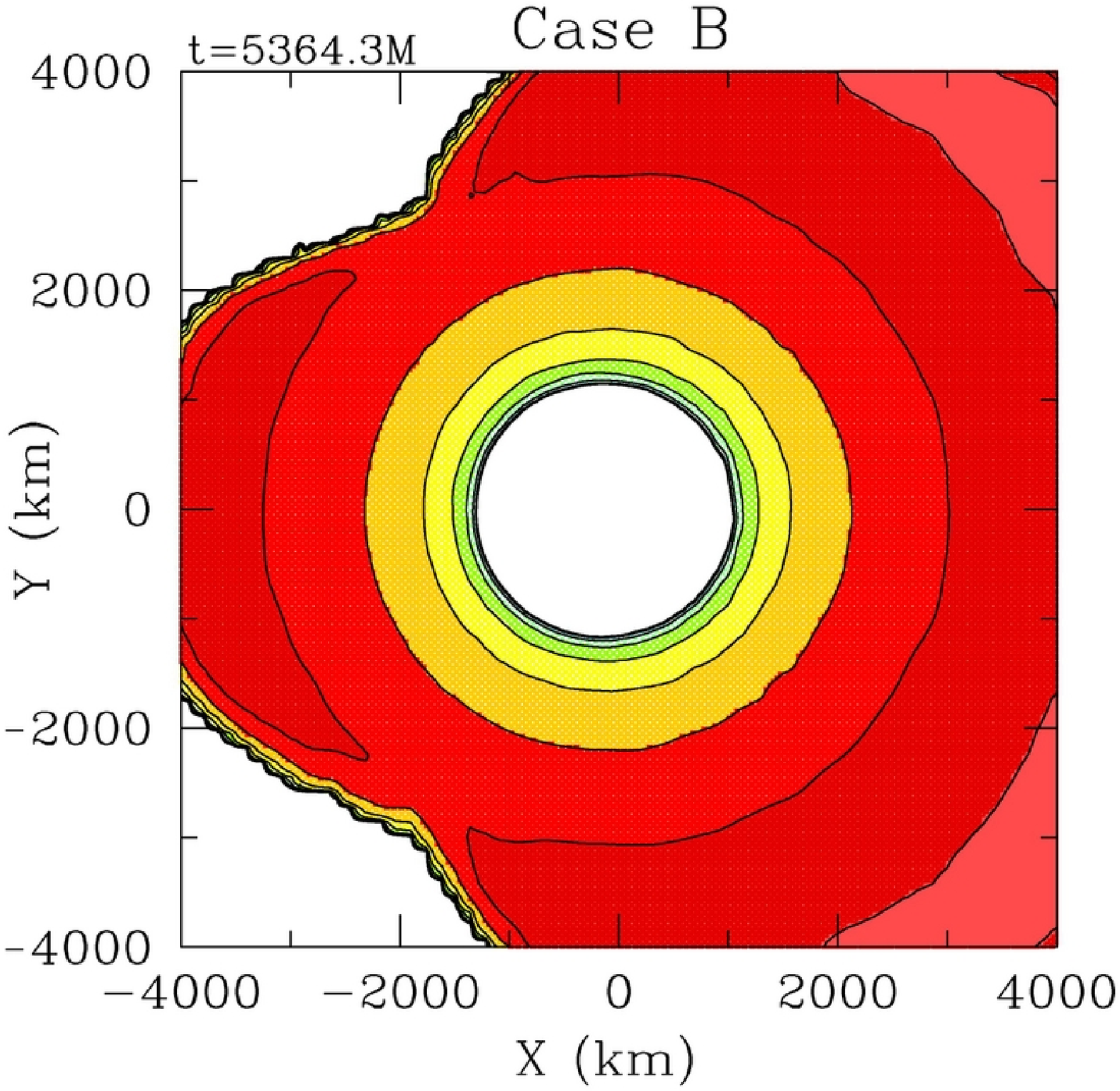}}
\caption{
      Snapshot of total energy per unit mass $U=-u_0-1$ (subtracting the rest-mass energy)
      for cases A2 and B,
      when matter has already started crossing 
      the outer boundary of the computational domain. Matter with $U < 0$
      is bound, while $U > 0$ implies that matter is unbound. 
      The contours represent the total energy per unit mass in the 
      equatorial plane, plotted according to $U = U_{\rm min} 10^{0.37j}\ 
      (j=0,1,\ldots, 9)$. The color code here is the same as that defined in Fig.~\ref{fig:A2xy}.
      The white spaces in the centers of the plots correspond
      to bound matter ($U<0$). 
      We chose the cutoff value $U_{\rm min} = 10^{-4}$. The size of the
      bound matter area is insensitive to the choice of 
  $U_{\rm min}$. Here $M = 2.48M_\odot = 3.662\rm km = 1.222\times 10^{-5} s $.
      \label{fig:u_0casesA2B}
}
\centering
\end{figure*}

\begin{figure}
\centering
{\includegraphics[width=0.495\textwidth,angle=0]{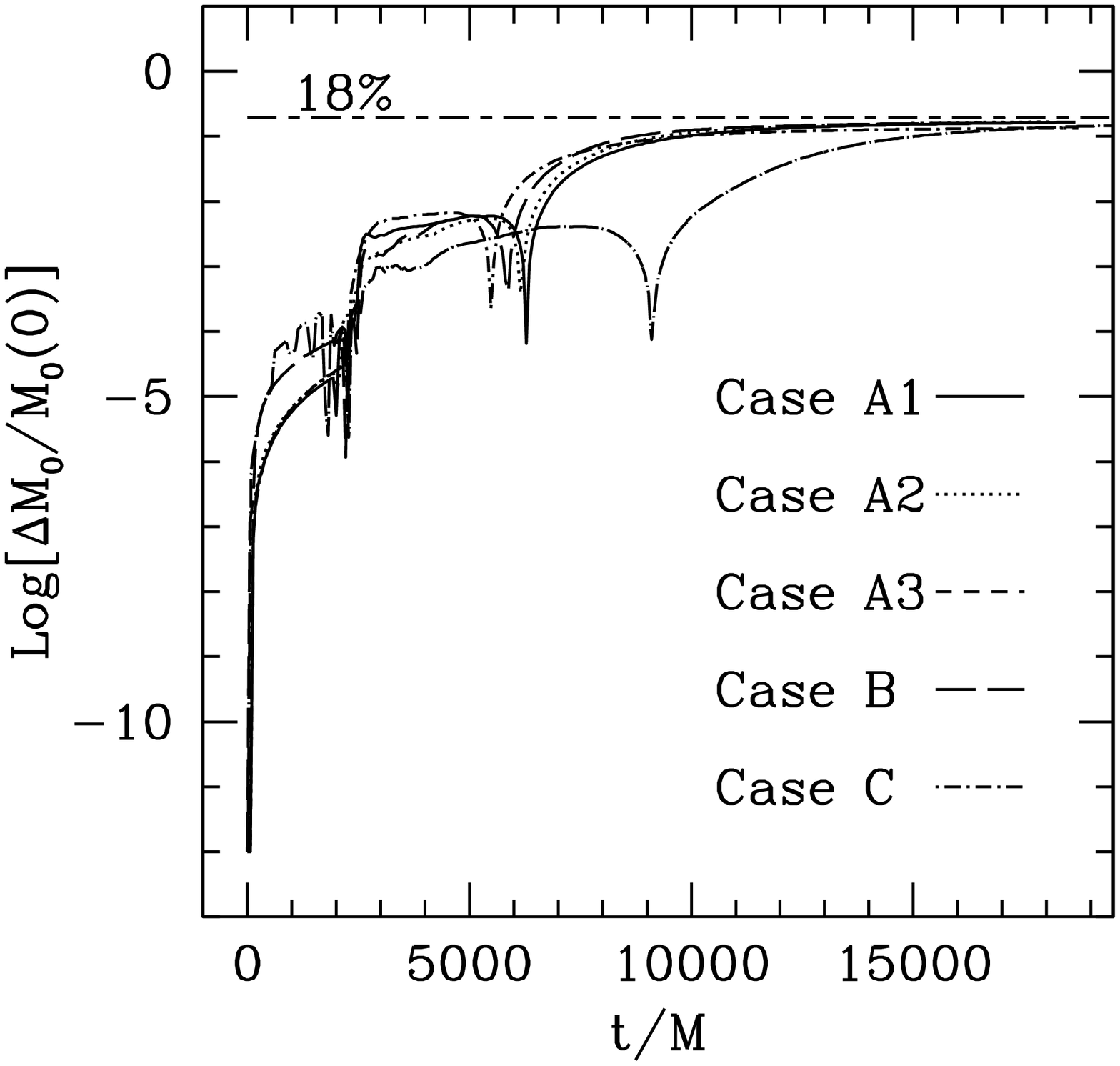}}
\caption{{Fraction of rest mass lost {\it vs} time. Here $\Delta M_0 = |M_0 - M_0(0)|$,
    where $M_0(0)$ is the initial total rest mass.
    Small changes in the rest mass until approximately $5000M$ for cases A1, 
    A2, A3, B and $10000M$ for case C 
    are due to interpolations when matter crosses refinement levels and
    inaccuracies in evolving the very low-density atmosphere.
    At the end of the simulations the amount of mass ejected is 13.4\% of
    the initial rest mass in case B and ranges
    from 16.1\%-16.7\% of the initial rest mass for the other cases. 
    Extrapolating the results to late $t$ we find that
    in case B $\Delta M_0/M_0(0) \sim 14\%$,
    $\Delta M_0/M_0(0) \sim  18\%$ in all other cases. 
    Here $M = 2.48M_\odot = 3.662\rm km = 1.222\times 10^{-5} s $. 
 \label{fig:mass_loss}}}
\centering
\end{figure}

In all cases we require that the sum of the ADM masses of the isolated stars be larger than
the maximum gravitational mass of $1.8M_\odot$ that our
cold EOS can support (see Fig.~\ref{fig:models}). 
There exist at least 18 observed WDNS systems that satisfy this requirement.
Since typical NS and WD masses in massive WDNS binaries lie in the range $1.3M_\odot-1.6M_\odot$ and
$0.5M_\odot-1.1M_\odot$ respectively (see Sec.~\ref{sec:introduction}), we choose 
pWD rest-mass density profiles that correspond to an ADM mass of $0.98M_\odot$ in isolation 
and keep it fixed in all cases we study. This almost fixes the pWD rest mass, because
 of the small compaction ($<0.02$) of the pWDs we consider.
The pWD rest-mass variation from case B to case C, due to fixing the ADM mass, is $0.7\%$.
We vary only the pWD compaction, i.e., the EOS, and the NS mass. 
The reason why we chose the pWD mass to be $0.98M_\odot$ is that
 the ratio of the isotropic radius of a pWD of such mass to the 
isotropic radius of a $1.5M_\odot$ NS is $\approx 10$ for the 10:1 EOS.

Another quantity that we fix in all our simulations is
the initial coordinate separation $A$ of the two components. 
This almost fixes the
kinetic energy of the stars when they collide. 
In particular, we set $A= 4R_{\rm WD,A}$. Here $R_{\rm WD,A}$ denotes 
the isotropic radius of the spherical $0.98M_\odot$ pWD used in cases A1, A2, A3.
We choose the initial separation  this way because we want the stars to be 
sufficiently far apart so that spherical TOV initial models
remain in near equilibrium, and at the same time, simulate the collision within reasonable time scales,
as the head-on collision time scale varies as $\sim A^{3/2}$ (see Eq.~\eqref{tcoll}).

If $A = 4R_{\rm WD,A}$, the NS tidal field in the vicinity
of the pWD is small, validating our assumption that an equilibrium pWD is
nearly spherical.
To see this,  let us calculate the ratio of the tidal force of
 the NS on the surface of the pWD to the surface gravity of the pWD 
\labeq{Tidalograv}{
\frac{F^{\rm t}_{\rm NS}}{F_{\rm WD}} = \frac{M_{\rm NS}}{M_{\rm WD}}\bigg(\frac{R_{\rm WD,A}}{A}\bigg)^3 \simeq 5\%,
}
where we used $M_{\rm NS}= 1.5 M_\odot$, $M_{\rm WD}= 1 M_\odot$, 
$A =3R_{\rm WD,A}$. Hence, any deviations from sphericity should be small. 
The assumption of sphericity for case B is even better because
 in this case $A \approx 8 R_{\rm WD,B}$, 
but worse for case C, where $A \approx 2 R_{\rm WD,C}$. 
In principle, we could increase the separation so that the 
sphericity approximation becomes better for all cases, but if 
the final remnant does not
collapse promptly to form a BH starting at close separations, 
it is unlikely that it will collapse if the initial separation is larger.
This is due to the fact that for larger separations the kinetic energy
at collision will be larger, generating more shock heating that will
work to prevent prompt collapse.

To summarize, the set of cases A1, A2 and A3 probe the
 effect of the NS mass on the final outcome, 
whereas the set of cases A2, B and C probe the effect of
 the pWD compaction on the final outcome. 
In the following sections we summarize the results of our simulations.

\begin{figure*}
\centering
\subfigure{\includegraphics[width=0.325\textwidth]{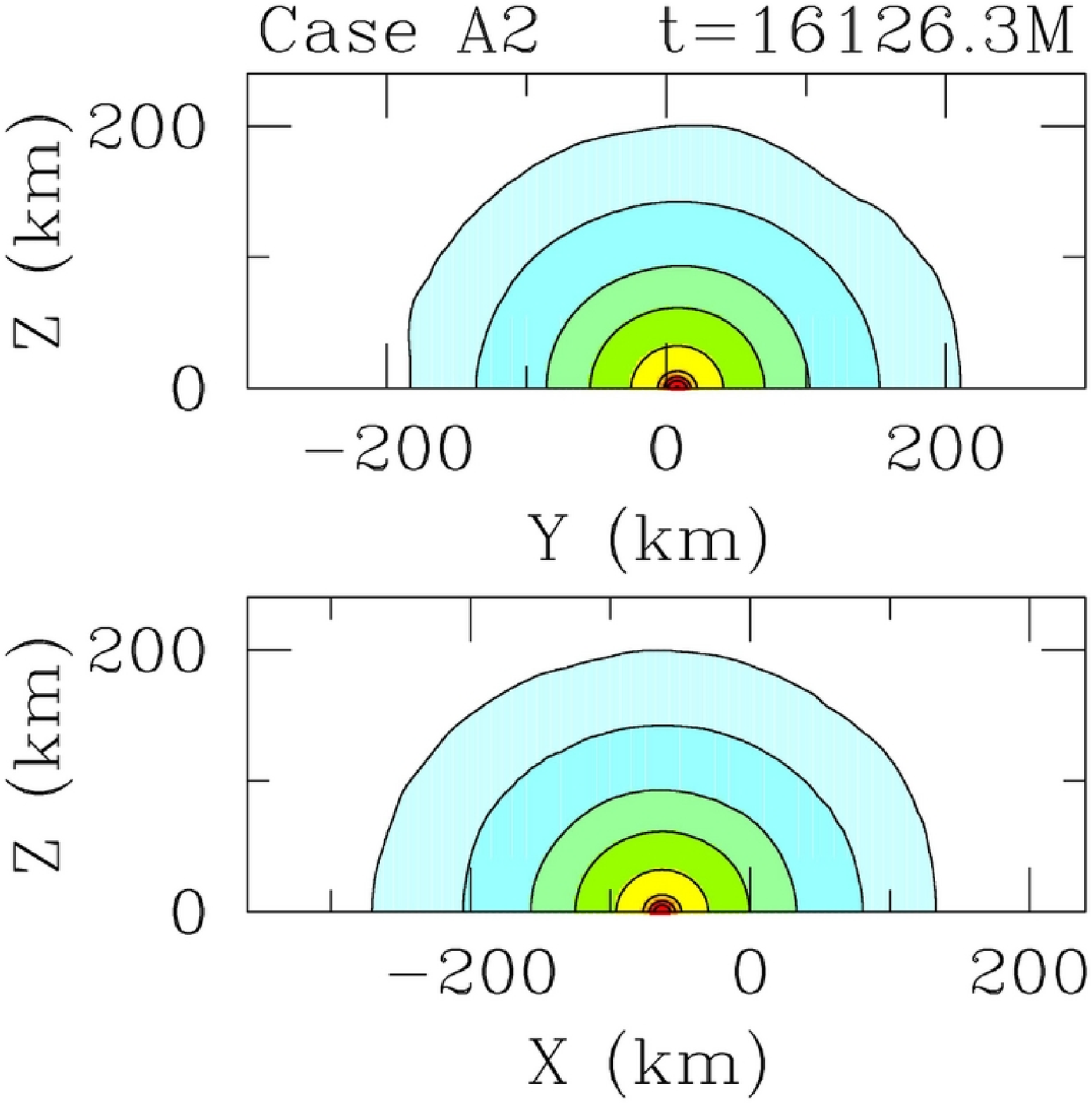}}
\subfigure{\includegraphics[width=0.325\textwidth]{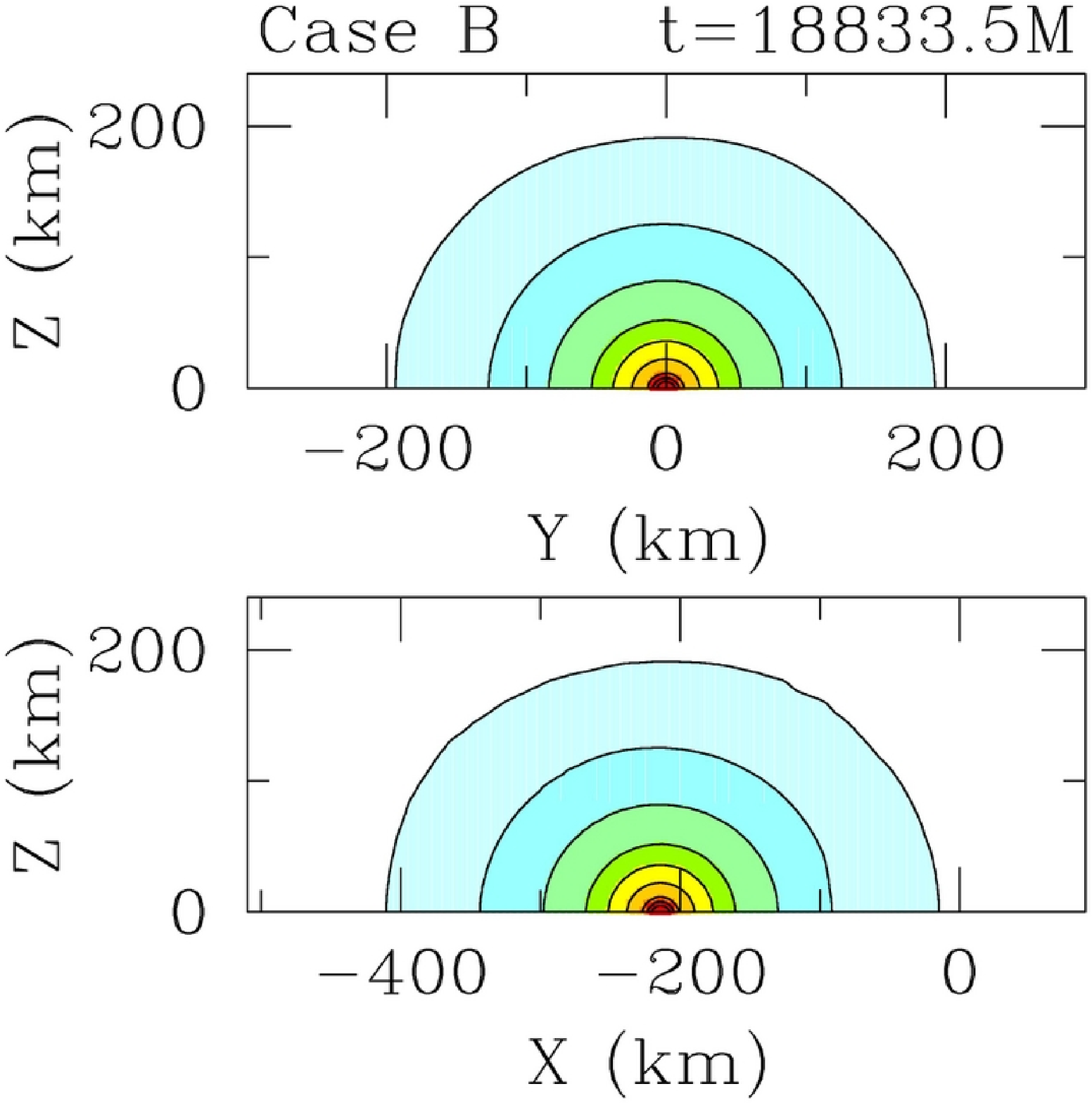}}
\subfigure{\includegraphics[width=0.325\textwidth]{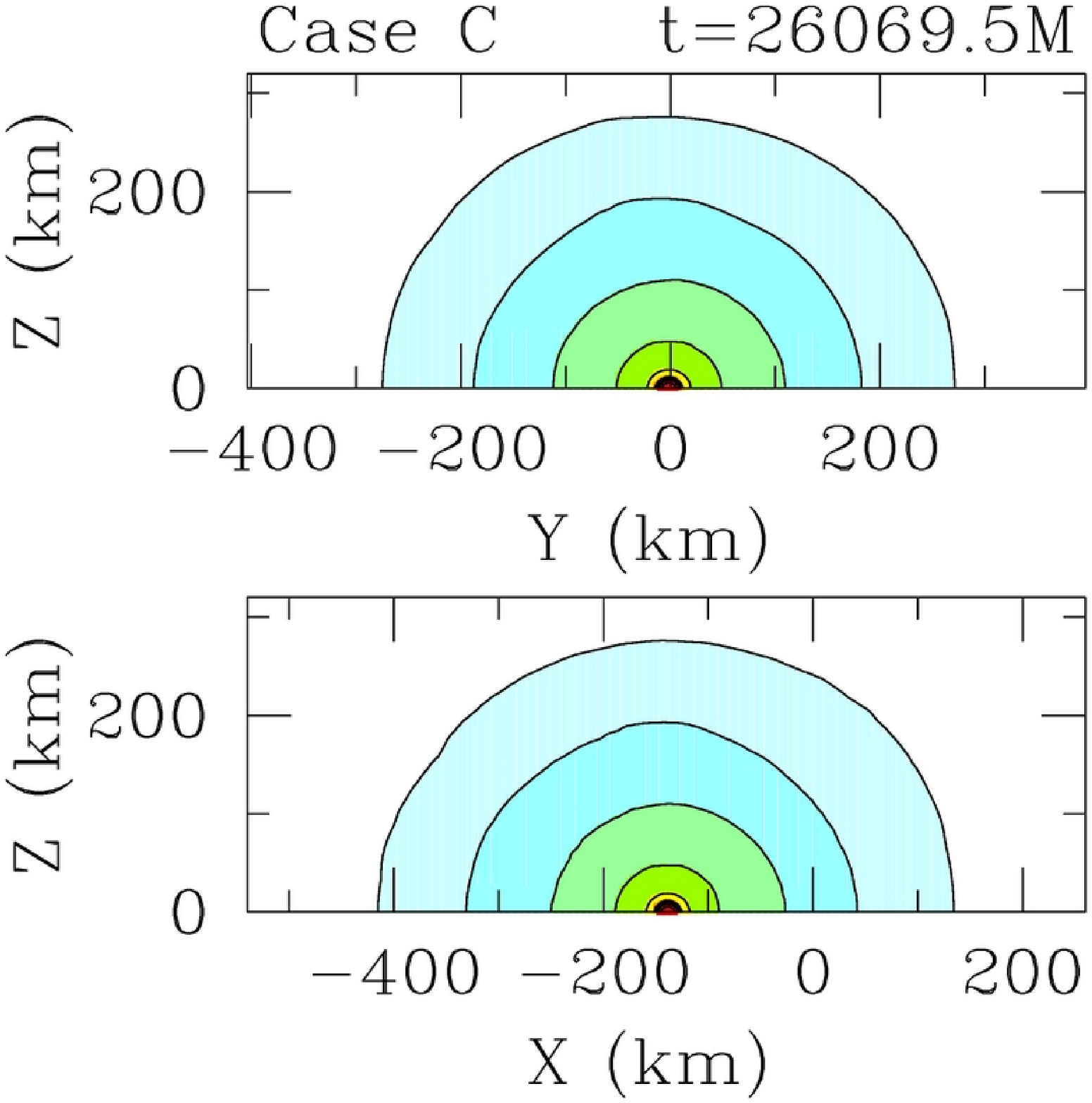}}
\subfigure{\includegraphics[width=0.325\textwidth]{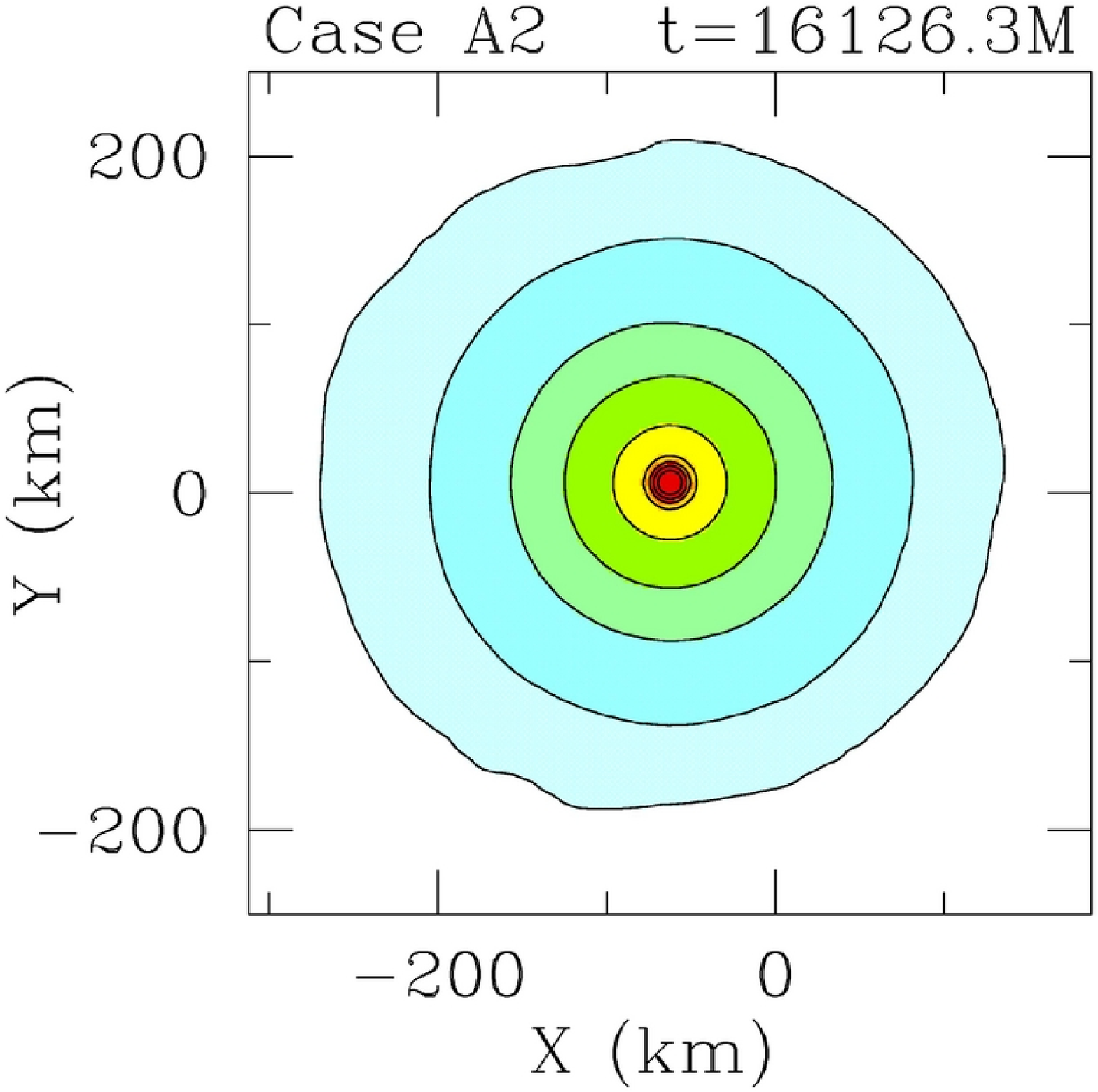}}
\subfigure{\includegraphics[width=0.325\textwidth]{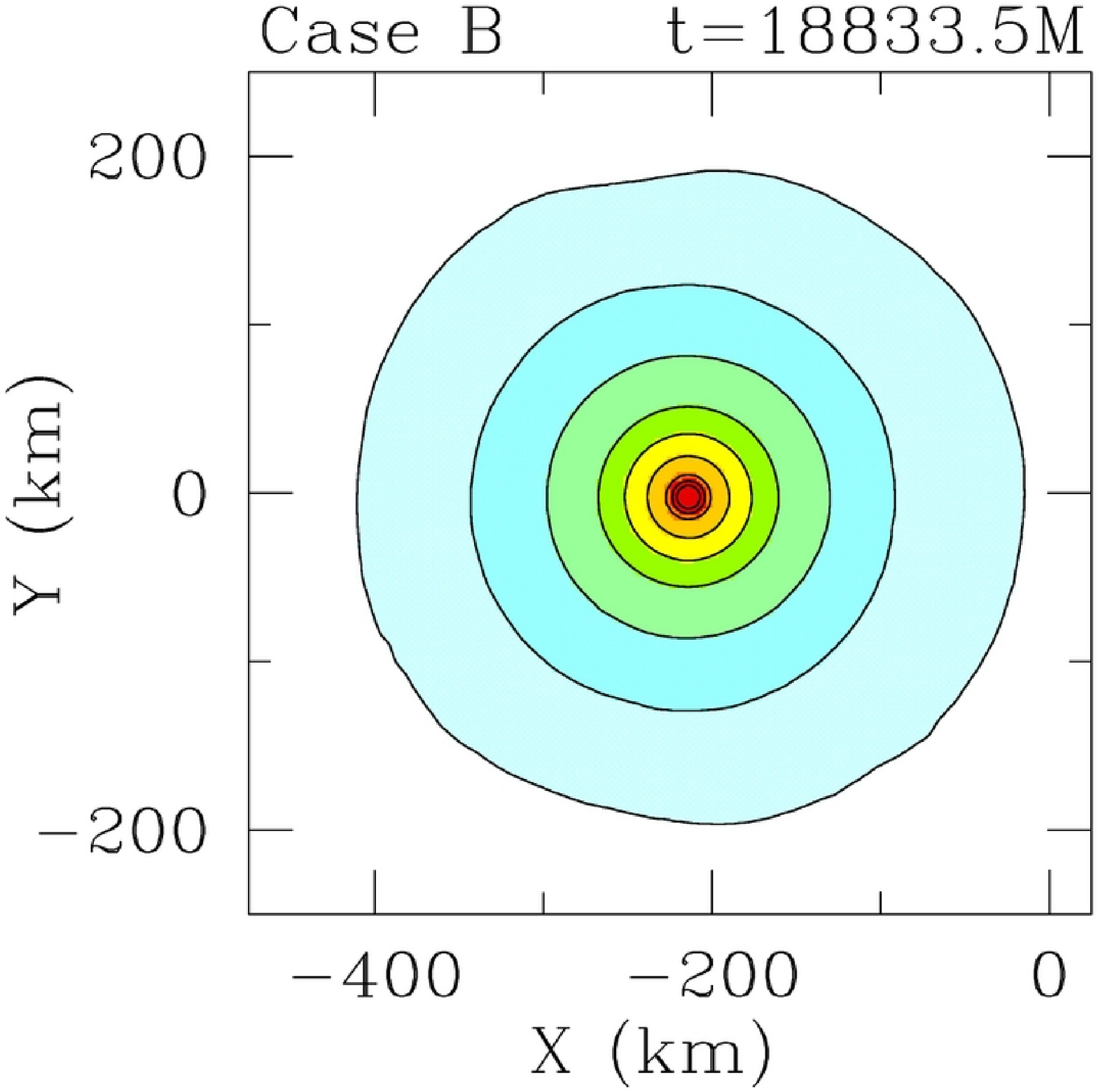}}
\subfigure{\includegraphics[width=0.325\textwidth]{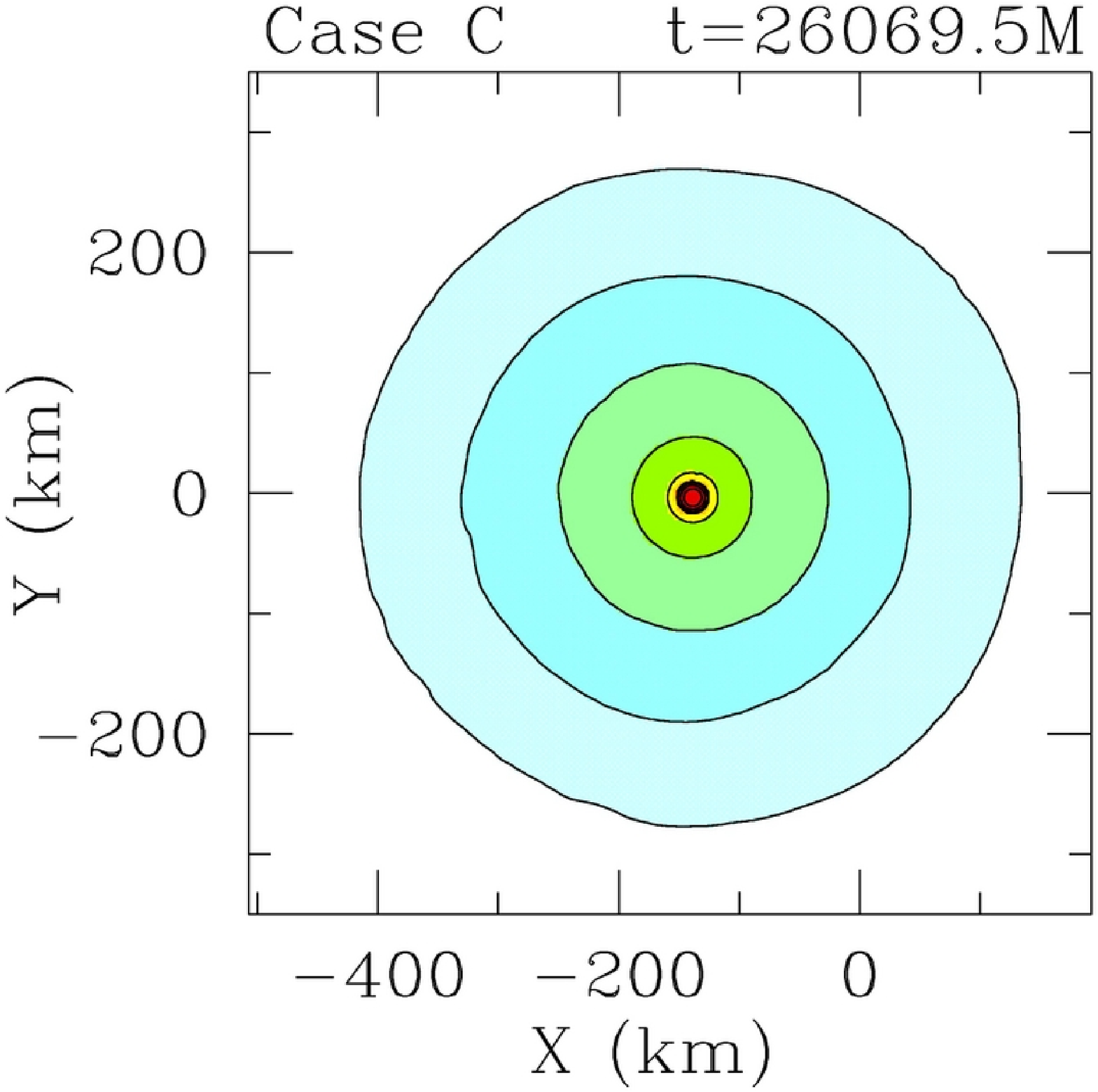}}
\subfigure{\includegraphics[width=0.325\textwidth]{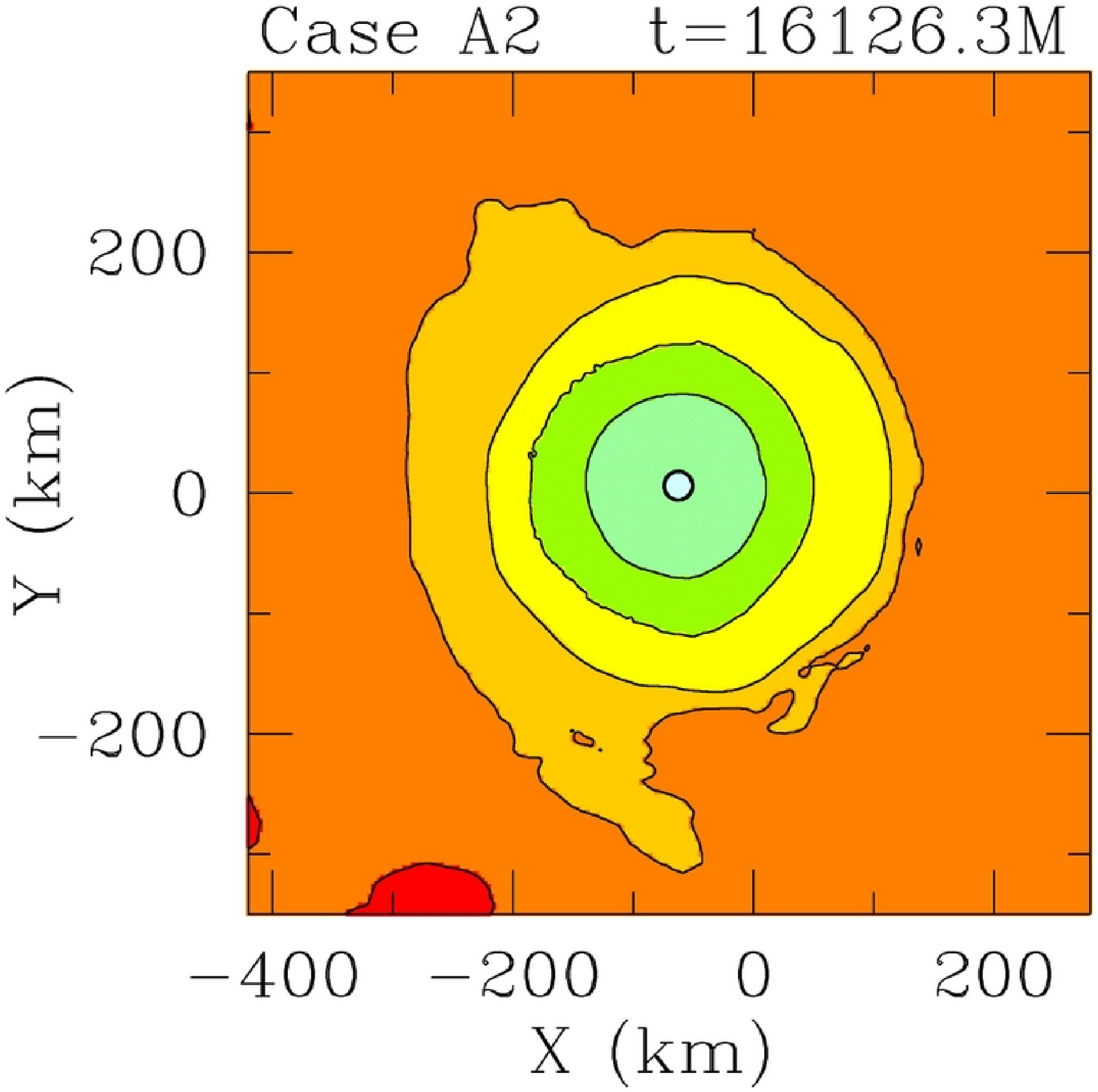}}
\subfigure{\includegraphics[width=0.325\textwidth]{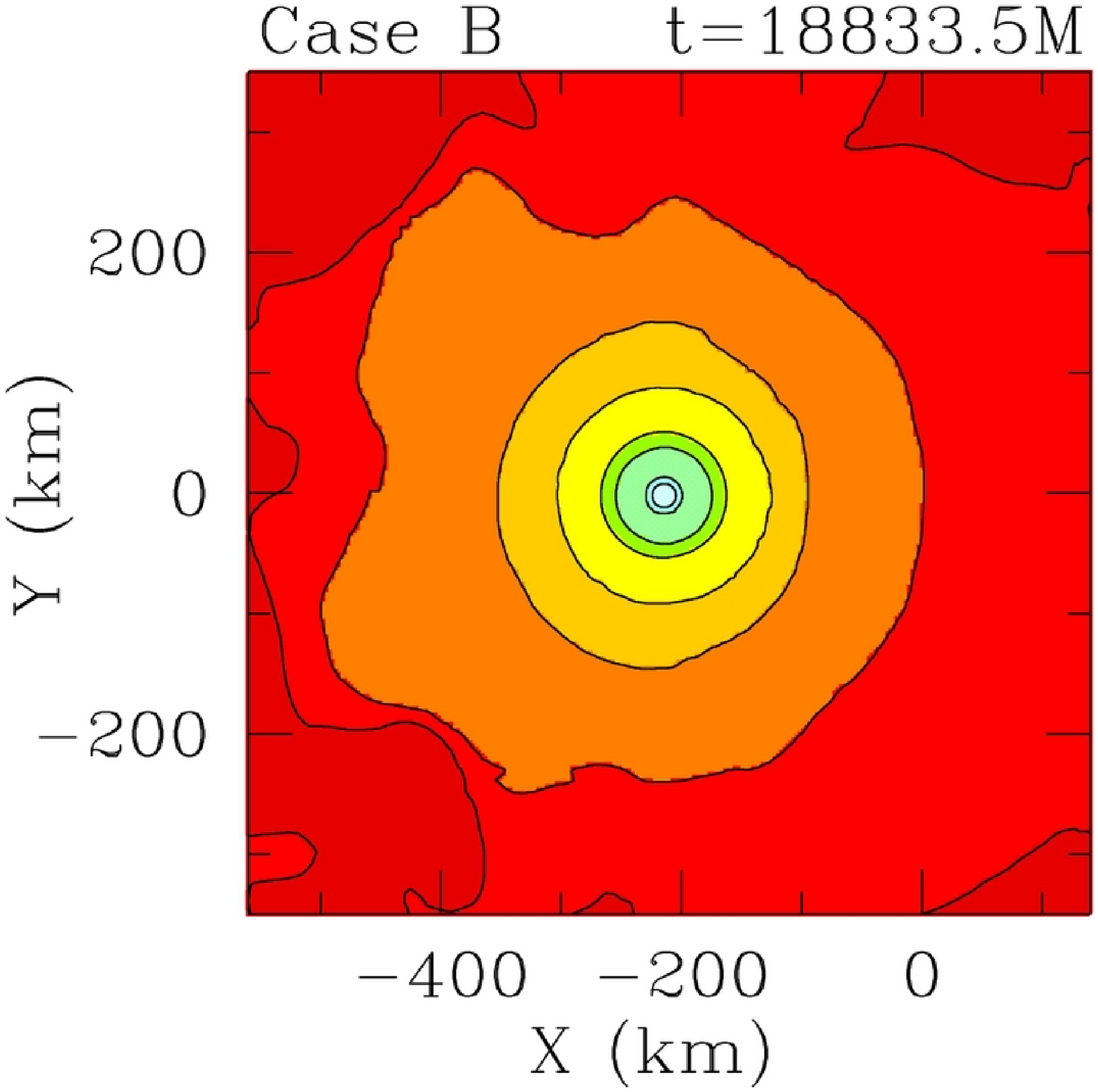}}
\subfigure{\includegraphics[width=0.325\textwidth]{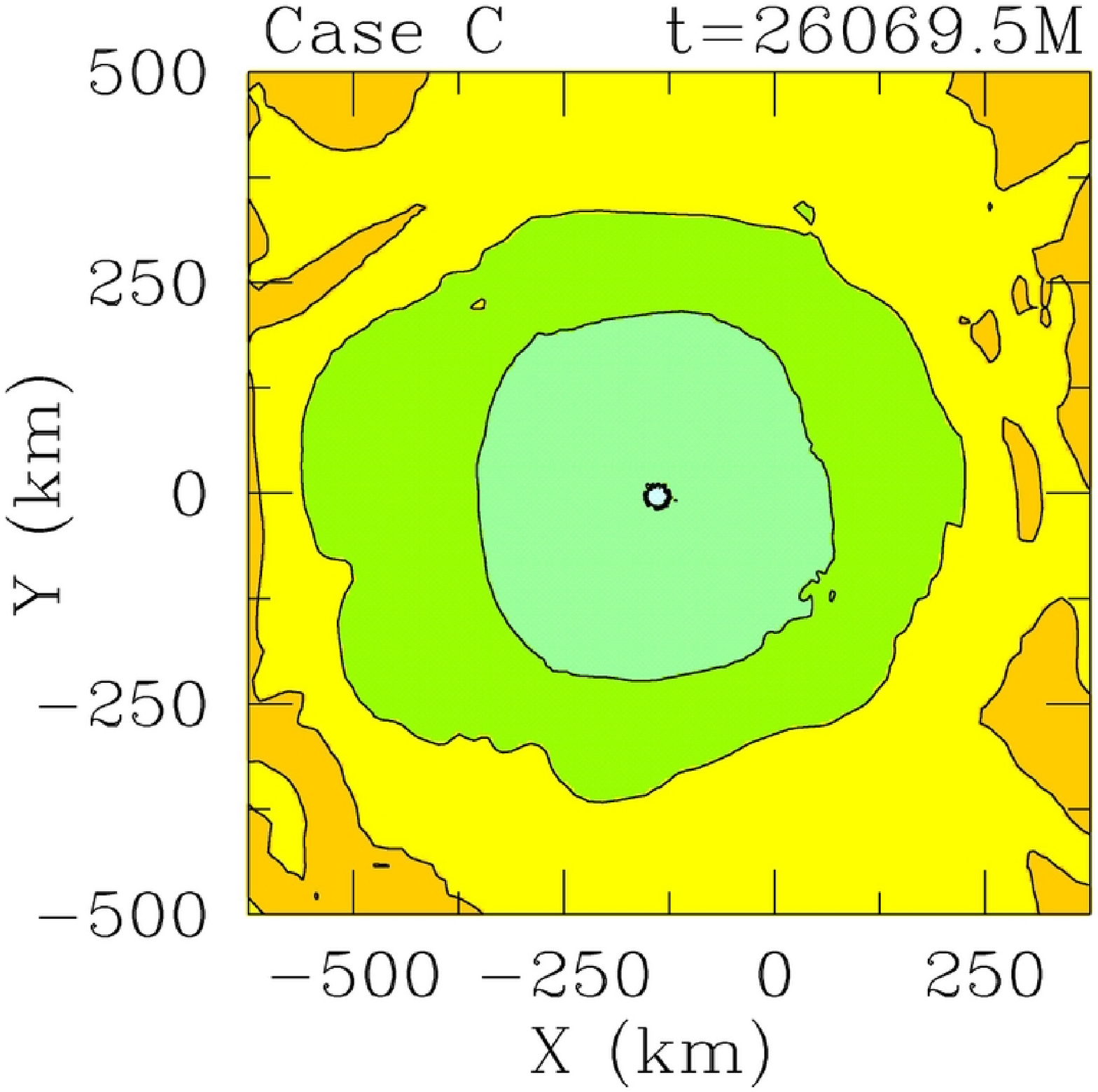}}
\caption{
First three rows: Snapshots of rest-mass density profiles at the
 end of the simulation for cases A2, B, C.
 The contours represent the rest-mass density in the YZ 
plane (first row), in the XZ plane (second row) and in the XY plane (third row) 
plotted according to $\rho_0 = \rho_{0,\rm max} 10^{-0.72j-0.16}\ 
(j=0,1,\ldots, 9)$. 
The maximum initial NS density is $\rho_{0,\rm max} = 4.6454\rho_{\rm nuc}$.
These snapshots demonstrate that in the adopted gauge,
 the final object is roughly spherical.
Last row: Snapshots of $K=P/P_{\rm cold}$ profiles at
 the end of the simulation for cases A2, B, C.
 The contours represent $K$ in the  XY plane plotted according to 
$K = 10^{0.288j}\ (j=0,1,\ldots, 9)$.
It is evident that the core of the
 remnant remains cold ($K\simeq 1$). $K$ becomes larger than unity as we move outwards from the 
center of the objects, and shock heating is more
 intense in case B and less intense in case C. 
The color code used is the same as that defined in Fig.~\ref{fig:A2xy}. 
Here $M = 2.48M_\odot = 3.662\rm\ km = 1.222\times 10^{-5}\ s $.
\label{fig:xyxzCasesA}
}
\centering
\end{figure*}

\subsection{Dynamics of collision and effects of the NS mass}

Here we describe the effects of the NS mass on the dynamics
 of pWDNS head-on collisions. We find that about 18\% of 
the initial total mass escapes to infinity for all cases A1, A2 and A3. Nevertheless, the 
initial total mass  in these cases is large enough to 
guarantee that the final total mass of the pWDNS remnant still
 exceeds the maximum mass that our cold EOS can support.
However, prompt collapse to a black hole does not take place
 in any of the cases studied because strong shock heating
 gives rise to a hot remnant. 
Ultimate collapse to a BH is almost certain after the remnant has cooled.
The outcome of the three cases A1, A2, A3 is a TZlO.

 Overall, cases A1, A2 and A3 are qualitatively similar 
and for this reason we mainly 
describe case A2 as representative of this class of our
simulations. Furthermore, our study of case A2 with two different resolutions
(see Table~\ref{table:GridStructure}) shows the results to be qualitatively 
insensitive to resolution indicating that the resolutions used in our simulations 
are sufficiently high. In what follows all case A2 plots correspond to 
 the high resolution run of case A2, i.e., case A2b. 


In general, the head-on collision of pWDNS systems can
 be decomposed into three phases: the acceleration, 
the plunge, and the quasiequilibrium phase. 

During the acceleration phase, the two stars accelerate toward
one another starting from rest. The separation decreases as a function of time and 
this phase ends when the two stars first make contact.

As the separation decreases, the increasing NS tidal field strongly distorts
the pWD. This can be seen in the equatorial rest-mass
 density contours of Fig.~\ref{fig:A2xy}. In the insets of Fig.~\ref{fig:A2xy},
the NS interior is almost unchanged during this phase. In reality, it
oscillates but is not tidally distorted by the pWD. 
Nevertheless, the NS atmosphere does expand. 
The insets also show that due to numerical errors the NS veers slightly off the x-axis,
 which is the collision axis in our simulations. 
In general, in all our simulations both the NS and the pWD
 wiggle around the x-axis. The amplitude of the NS wiggling motion is 
at most 1\% of the pWD radius, while the amplitude
of the pWD off-axis motion is less than 0.1\% of its radius. 
Hence, the collision is practically head-on.

It is likely that this lack of symmetry is due to small asymmetries
introduced when mapping the initial 
data onto the evolution grids via 2nd order
interpolation. This is because 2nd order interpolation requires the 
use of 3 grid points (per direction) that 
surround the point to which one interpolates. However, 
the effect is small and our results cannot change qualitatively due to this
small asymmetry.

Along with this small off-axis motion, the pWDNS system acquires a
small amount of spurious angular momentum. Fig.~\ref{fig:ang_mom}
shows the normalized z component of the angular momentum of the
system and demonstrates that it is always less than
3\% (see Eq.~\eqref{Newt_J}).  In addition to conserving 
the angular momentum to within 3\%, the normalized Hamiltonian
constraint violations remain smaller than 1\% and the normalized
momentum constraint violations smaller than 3\%.  These results hold
for all cases studied in this work.

During the plunge phase the NS penetrates the pWD, 
launching strong shocks that sweep
through and heat the interior of the pWD.  The NS outermost
 layers are stripped when they encounter the dense
 central parts of the pWD, and the NS is compressed. 
We find that at maximum compression in case A2, the NS
 central density only increases by about 8\% of the initial central density.
 
Eventually, strong shocks sweep through the entire pWD
 interior and then transfer linear momentum to the pWD outer layers, 
a large fraction of which receives 
sufficient momentum to escape to infinity. 
This can be seen in Fig.~\ref{fig:u_0casesA2B}, where a snapshot is
shown of the total energy per unit mass $U=-u_0-1$ (subtracting the rest-mass energy)
on the equator long
after the collision, when ejected material has already started
crossing the outer boundary of the computational domain.  
Unbound ($U>0$) matter covers most of the computational domain, as shown in
Fig.~\ref{fig:u_0casesA2B}. The rest-mass density of 
the ejected material is of order $10^{-9}\rho_{\rm nuc}$, 
but the total mass that escapes to infinity is large.
This is demonstrated in Fig.~\ref{fig:mass_loss}, which shows the
fractional change in the rest mass
as a function of time.  We find the amount of matter that escapes in cases
A1, A2 and A3 is $\simeq 18\%$ of the initial rest mass when we
extrapolate our results to late times.

The thermal energy deposited into the 
ejected material is significant, with $K=P/P_{\rm cold}> 40$. 
As the ejected matter
comes from the WD outermost layers, its density is very low.
This implies that its initial pre-shocked sound speed is small. As a result,
the Mach number of the ejected material can be very large prior to shock heating,
and so shock heating is very strong, i.e., K increases from 1 
initially to greater than 40 (see also discussion in Appendix
B of \cite{Etienne08}).

The time scale for shock heating to equilibrate
must be of order few times the dynamical (free-fall) time scale 
of the WD (see Eq.~\eqref{tdynWD}), as this is the only relevant timescale. 
For cases A, the WD dynamical timescale is 
roughly 400M. Our computations show that it actually takes about 800M-1000M for 
the star to equilibrate, which is consistent with the 
estimate above.

\begin{figure*}[t]
\centering
%
\subfigure{\includegraphics[width=0.325\textwidth]{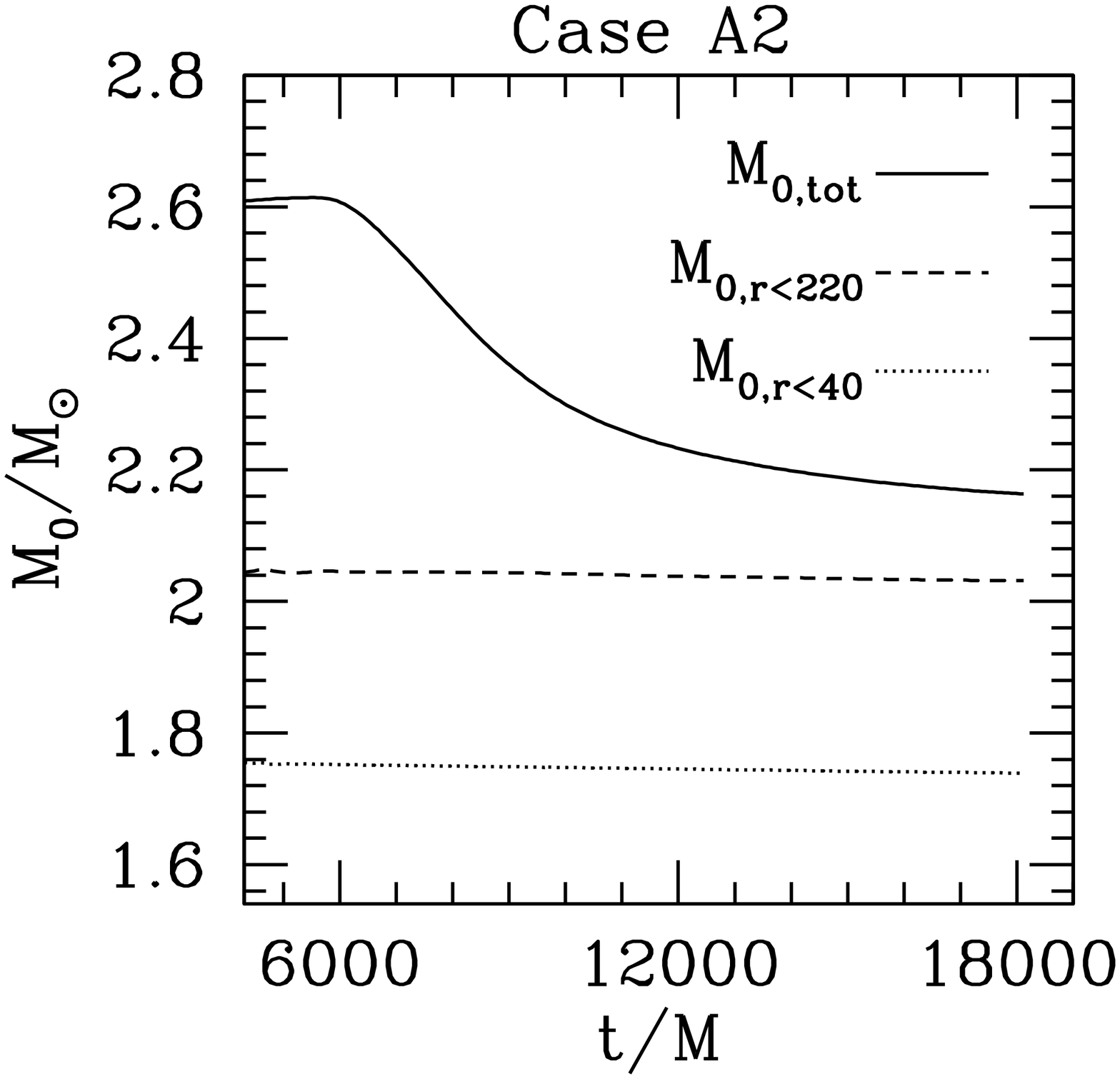}}
\subfigure{\includegraphics[width=0.325\textwidth]{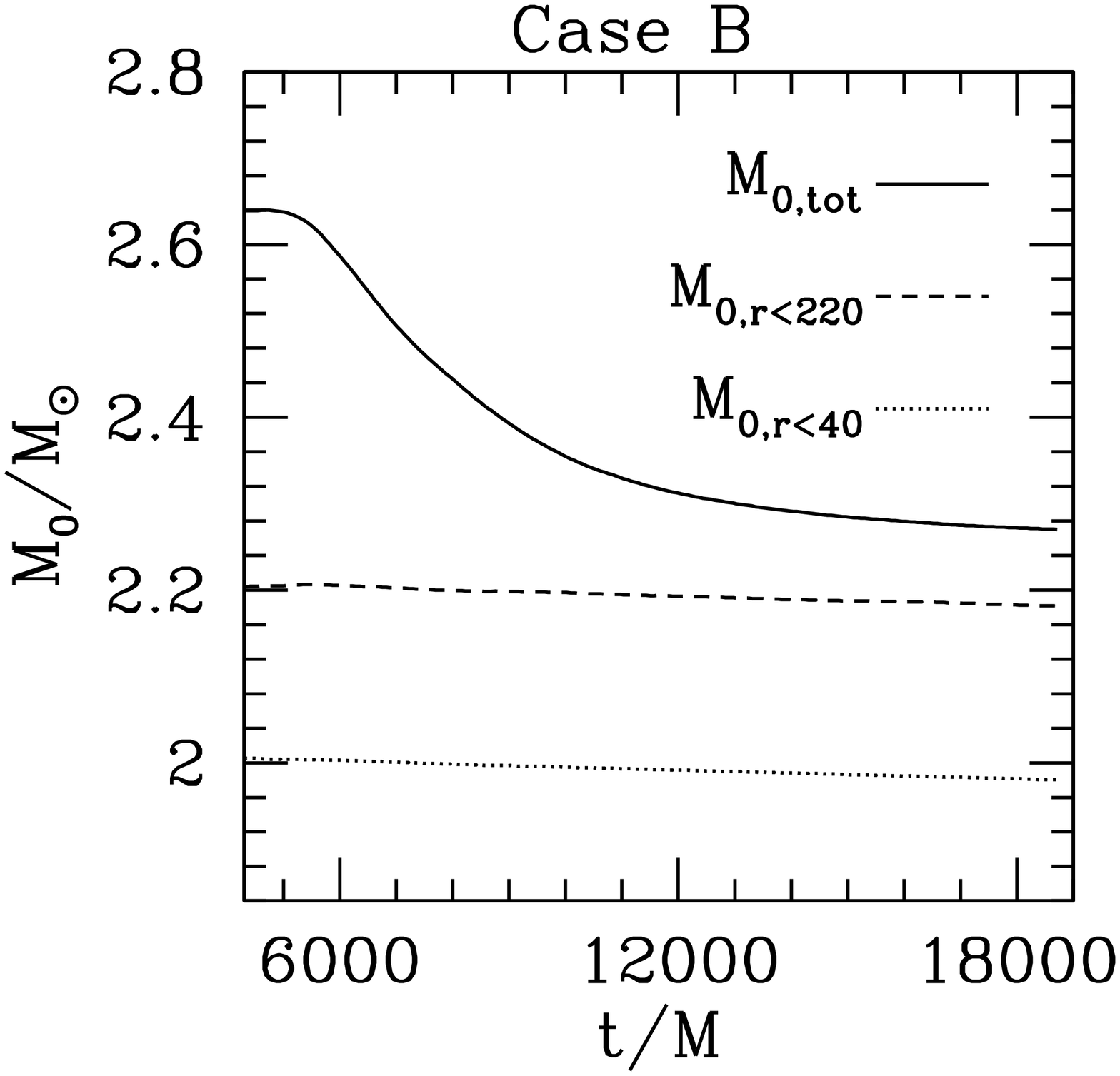}}
\subfigure{\includegraphics[width=0.325\textwidth]{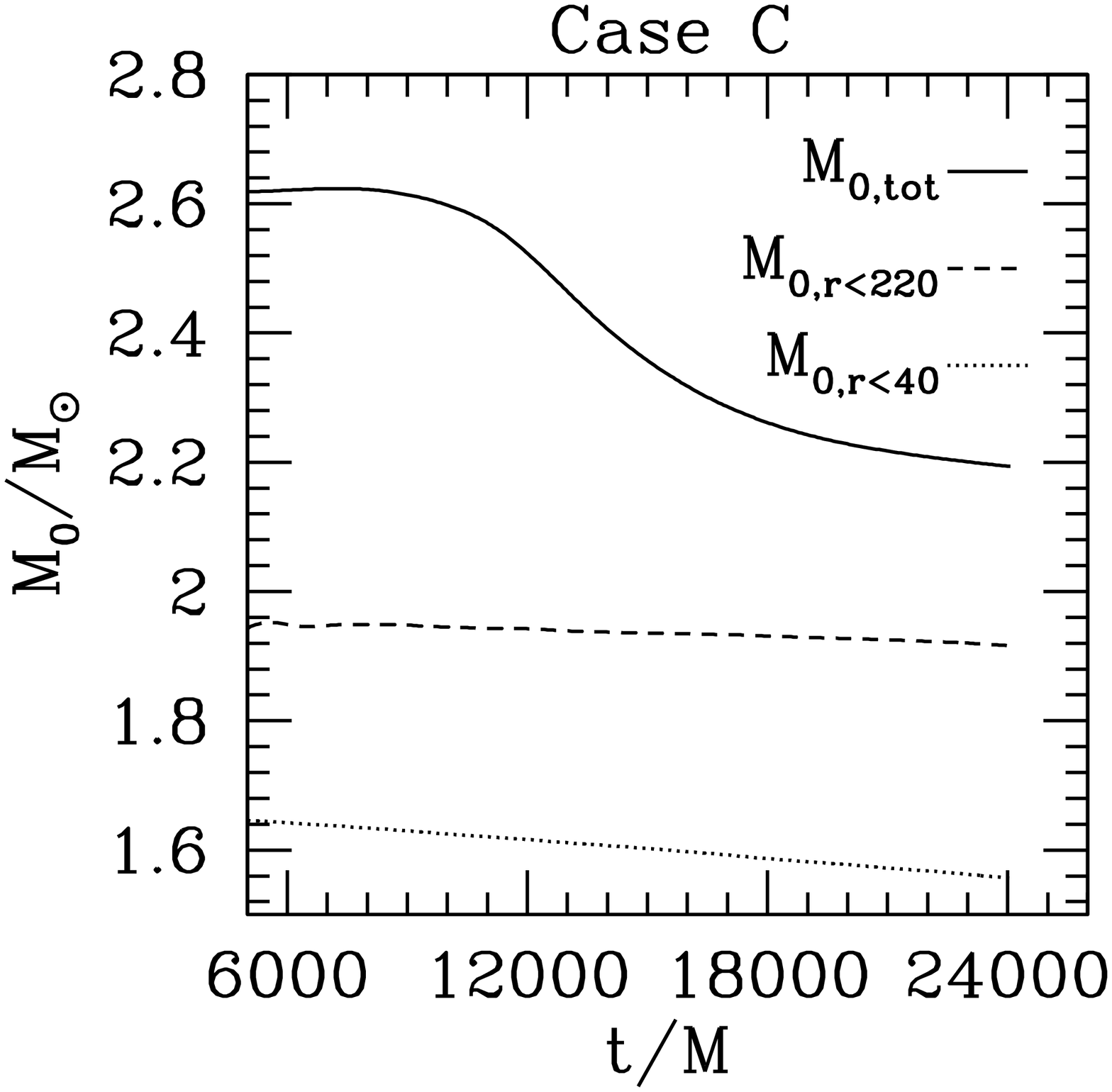}}
\caption{
Post-merger rest mass as a function of time. 
Here $M_{0,\rm tot}$ is the total rest mass in
 the entire computational domain and $M_{0,\rm r < r_0}$ 
stands for the rest mass contained within a coordinate sphere of
radius $r_0$ in units of km, centered on the remnant's center of
mass. In all cases $M_{0,\rm r < 220}$ accounts for more 
than $90\%$ of the final total rest mass and it is always
 greater than $1.92M_\odot$ - the maximum rest mass that our cold EOS can support.
Here $M = 2.48M_\odot = 3.662\rm km = 1.222\times 10^{-5} s $.
\label{fig:rest_mass_evol}
}
\centering
\end{figure*}

Material that did not receive sufficient thrust to escape to infinity starts to rain
down onto the NS and pWD matter and the plunge phase ends when this process is over.

During the quasiequilibrium phase the remnant settles into a spherical quasiequilibrium 
object whose outer layers oscillate. This can be seen in the two final snapshots 
of Fig.~\ref{fig:A2xy}, where we show that the inner equatorial
 rest-mass density contours do not change with time, while
 the outer layers change only a little. 
Fig.~\ref{fig:xyxzCasesA} plots xy, xz and yz rest-mass density
contours.  Notice that in the adopted gauge, the remnant
appears to be spherical.

The pWDNS final remnant consists of a cold NS core
 with a hot mantle on top. This is demonstrated by the
 plots in the last row of Fig.~\ref{fig:xyxzCasesA}, where 
we quantify the results of shock heating by showing
 contours of $K=P/P_{\rm cold}$.  Within a radius of $100$ km from the center of mass
of the remnant,  $K$ ranges
 from unity to about 15 for case A2.
In all cases $K \simeq 1$ at the center of the remnant, 
while it becomes larger than unity as we move outwards from the center. 
We refer to this spherical configuration as a Thorne-Zytkow-like object.

Even though a large fraction of the initial mass escapes,
 the final total rest mass well exceeds the maximum rest mass of $1.92M_\odot$ our
cold EOS can support. In Fig.~\ref{fig:rest_mass_evol} we 
show the rest mass of the remnant as a function of time and
for various spatial domains. This figure demonstrates that 
the rest mass within a radius of $220$km accounts for more
 than 90\% of the final total rest mass
and is greater than  $1.92M_\odot$. However, the pWDNS remnant 
does not collapse promptly to form a black hole, because of extra support provided
 thermal pressure. Delayed collapse 
to a black hole is almost certain after the pWDNS remnant has cooled.

Finally, we note that it has been suggested in 
 \cite{0004-637X-707-2-1173,PhysRevD.80.064001} that GWs may
arise from shocks. Even though the discussion in these studies focused on
core collapse supernovae, the appearance of strong shocks in our case
can also generate GWs. However, here we do not calculate the GW signature
because what is really interesting from an observational and astrophysical point of view 
is the GW signature in the circular binary WDNS case, not the head-on case we consider. 
In \cite{WDNS_PAPERI} we did calculate GWs from the inspiral phase of binary WDNS systems.
General relativistic computations of the merger of circular binary WDNSs will
be the subject of a forthcoming paper.


\begin{center}
\begin{table*}[t]
\caption{Summary of pWD compaction study.
Here $C_{\rm WD}$ is the compaction of an isolated pWD (see Table~\ref{tab:cases}), $K=P/P_{\rm cold}$ at the end of simulations$^{(a)}$,
$T_p$ is the peak temperature at the end of simulations, $M_0(0)$ is the initial total rest mass, 
$\Delta M_0 = |M_{0,\rm f}-M_0(0)|$, where $M_{0,\rm f}$ is the final total rest mass, $M_{0,\rm r<220}$ 
is the mass enclosed within $220$ km from the center of mass of the remnant at the end of the simulations, $\rho_{{0,\rm c}}$ is the 
final value of the central rest-mass density$^{(b)}$, and $\alpha_{\rm min}$ the final value of the minimum of the lapse function.
}
\begin{tabular}{cccccccc}\hline\hline
\multicolumn{1}{p{1.8cm}}{\hspace{0.45 cm} Case } & 
\multicolumn{1}{p{1.8cm}}{\hspace{0.35 cm} $C_{\rm WD}$ \quad}  &
\multicolumn{1}{p{1.2cm}}{\hspace{0.35 cm}  $K$ \quad}  &
\multicolumn{1}{p{2.35cm}}{\hspace{0.1 cm}  $T_p (10^{11}{}\ ^o$K)$^{(c)}$ \quad}  &
\multicolumn{1}{p{2.1cm}}{\hspace{0.1 cm}  $\Delta M_0/M_0(0)$ \quad}  &
\multicolumn{1}{p{2.2cm}}{\hspace{0.05 cm} $M_{0,\rm r<220}/M_{\odot}$ \quad}  &
\multicolumn{1}{p{2.2cm}}{\hspace{0.43 cm} $\rho_{0,\rm c}/\rho_{\rm nuc}$ }  &
\multicolumn{1}{p{1.6cm}}{\hspace{0.35 cm} $\alpha_{\rm min}$ \quad}  \\  \hline 
B    	               & 0.0190    & [1,35]  & 3.7  &  14\%    & 2.180    & 4.91  &   0.570        \\  \hline
A2    	               & 0.0100    & [1,15]  & 3.2  &  18\%    & 2.035    & 4.49  &   0.595        \\  \hline
C    	               & 0.0049    & [1,10]  & 3.0  &  18\%    & 1.900    & 4.10  &   0.609       \\  \hline\hline 
\end{tabular}
\begin{flushleft}
$^{(a)}$ The $K$ column lists the range of values which $K$ obtains within a radius of $100$ km from the centers of 
mass of the remnants.\\
$^{(b)}$ $\rho_{\rm nuc} = 2\times 10^{14} g/cm^3$.\\
$^{(c)}$ For realistic WDNS collisions we expect $T_p \sim 10^{9}{}\ ^o$K (see discussion following Eq.~\eqref{Temperature}).
\end{flushleft}
\label{tab:results}
\end{table*}
\end{center}

\begin{figure*}
\centering
%
\subfigure{\includegraphics[width=0.325\textwidth]{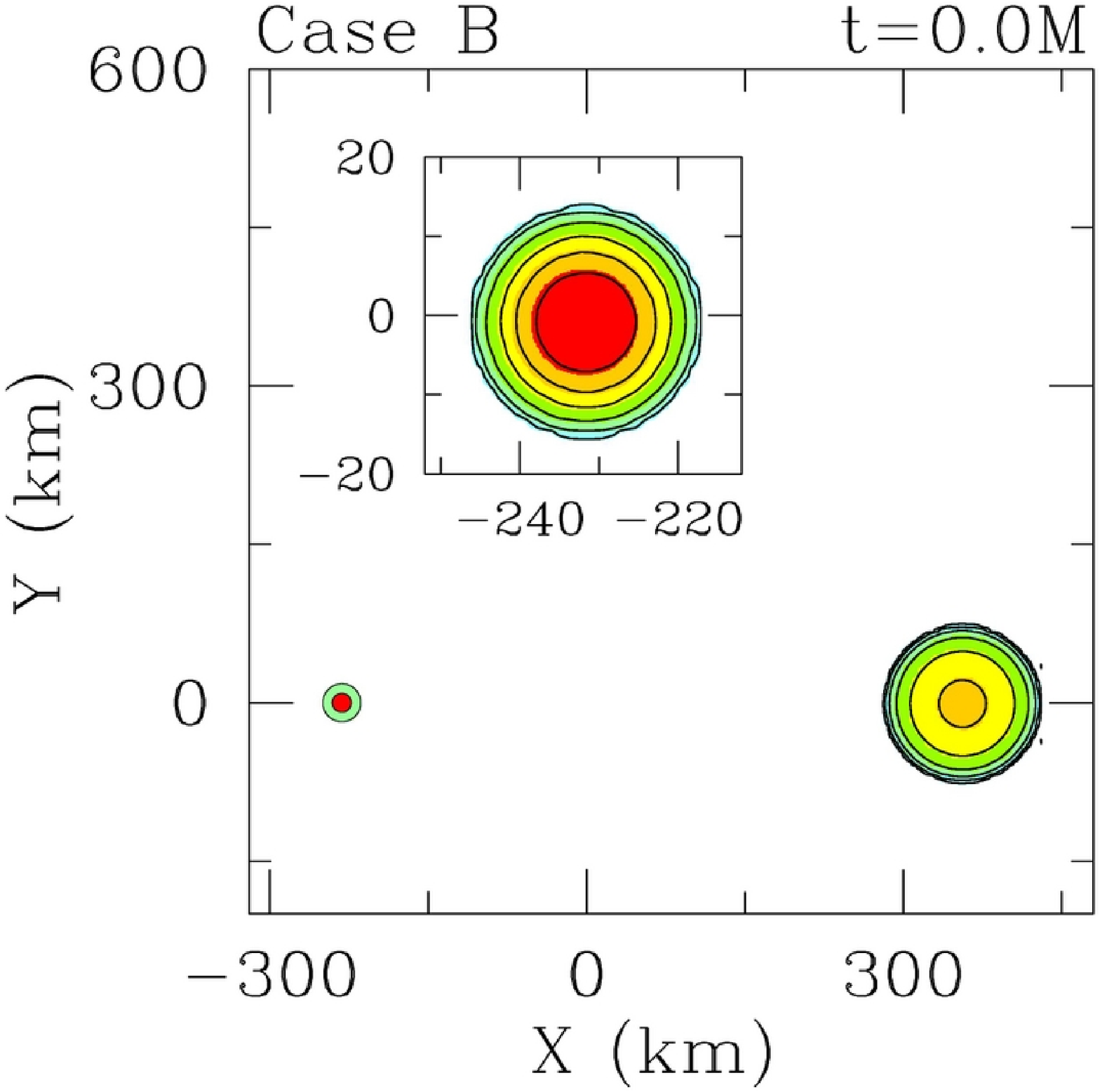}}
\subfigure{\includegraphics[width=0.325\textwidth]{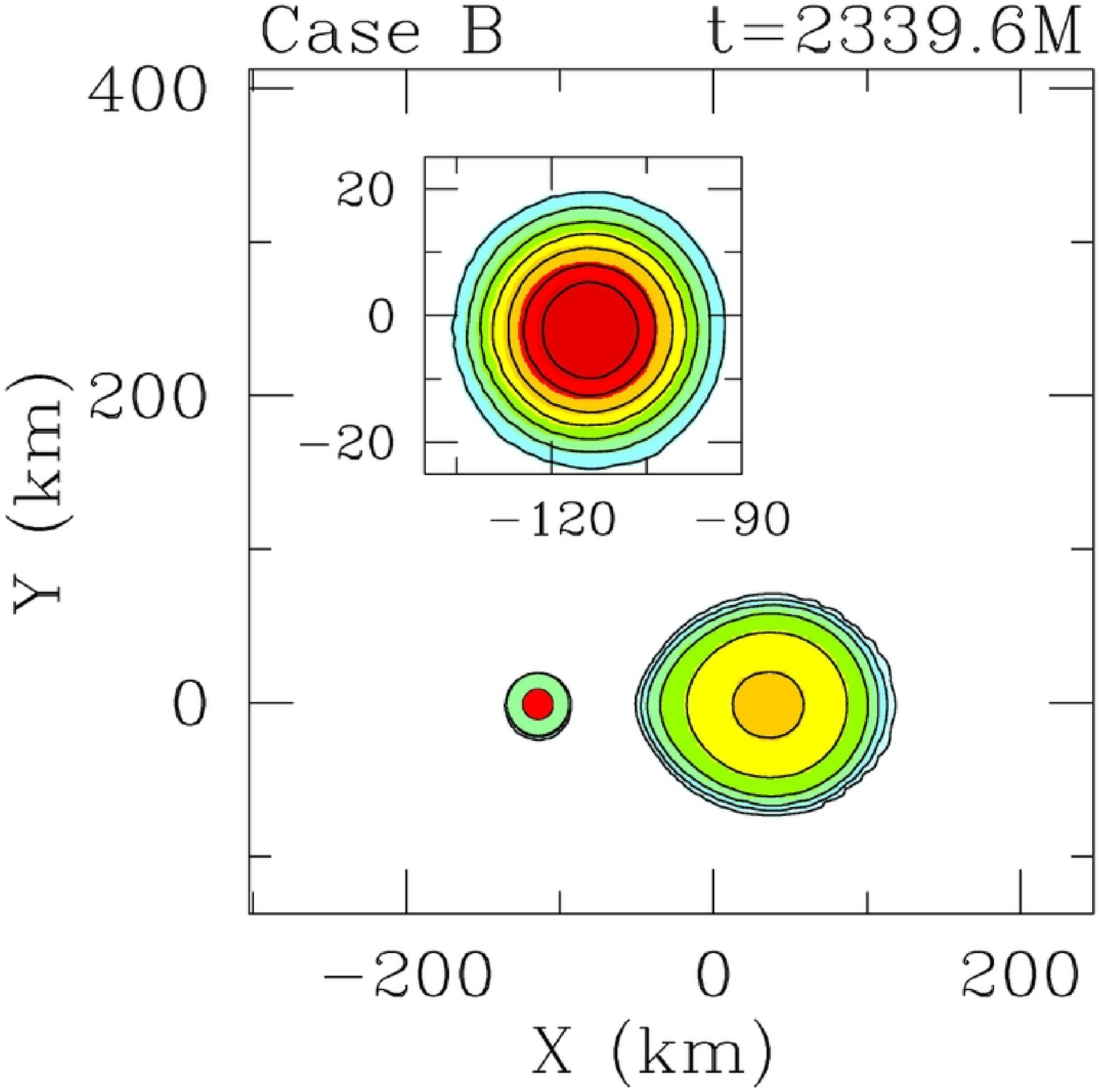}}
\subfigure{\includegraphics[width=0.325\textwidth]{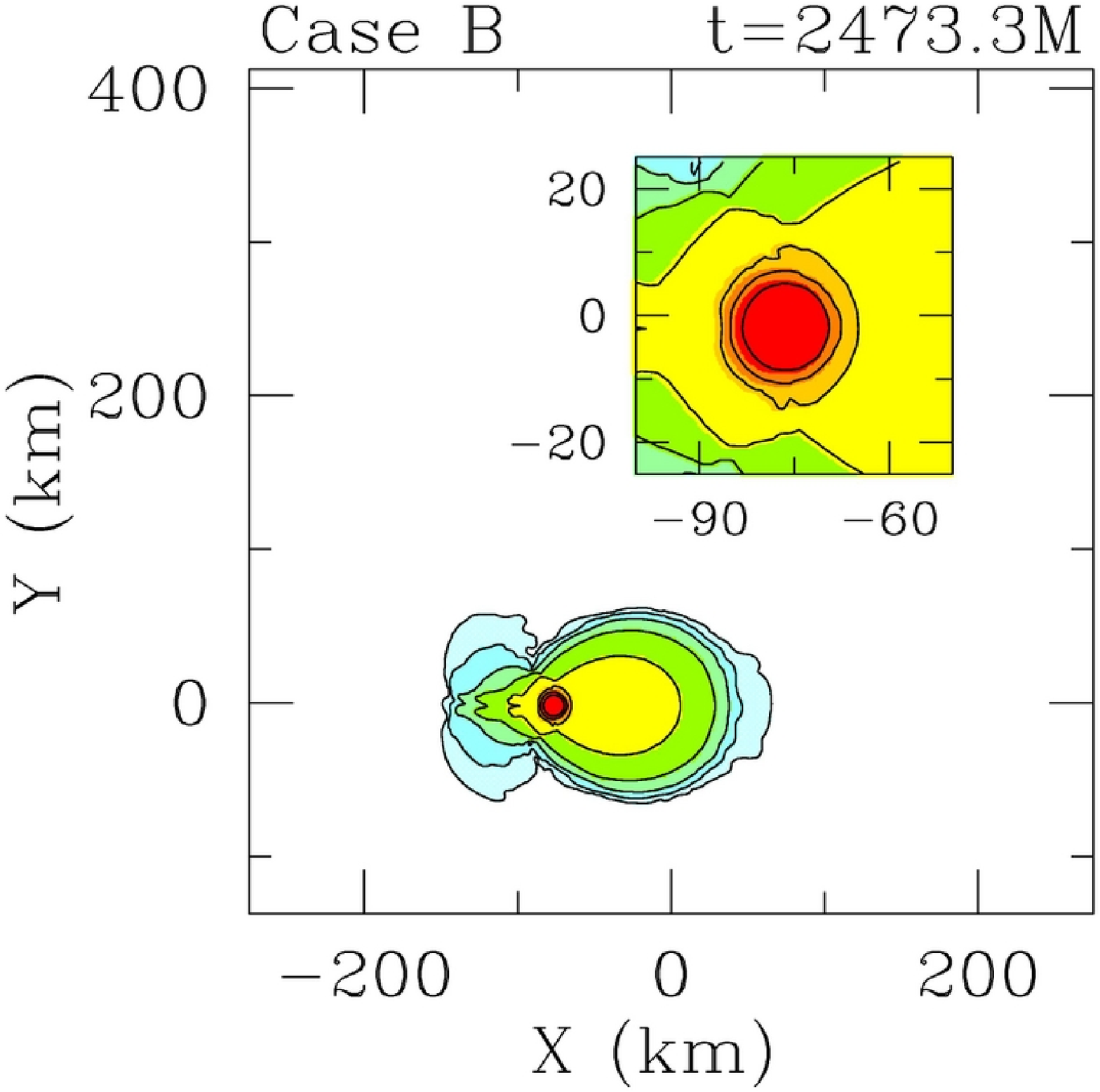}}
\subfigure{\includegraphics[width=0.325\textwidth]{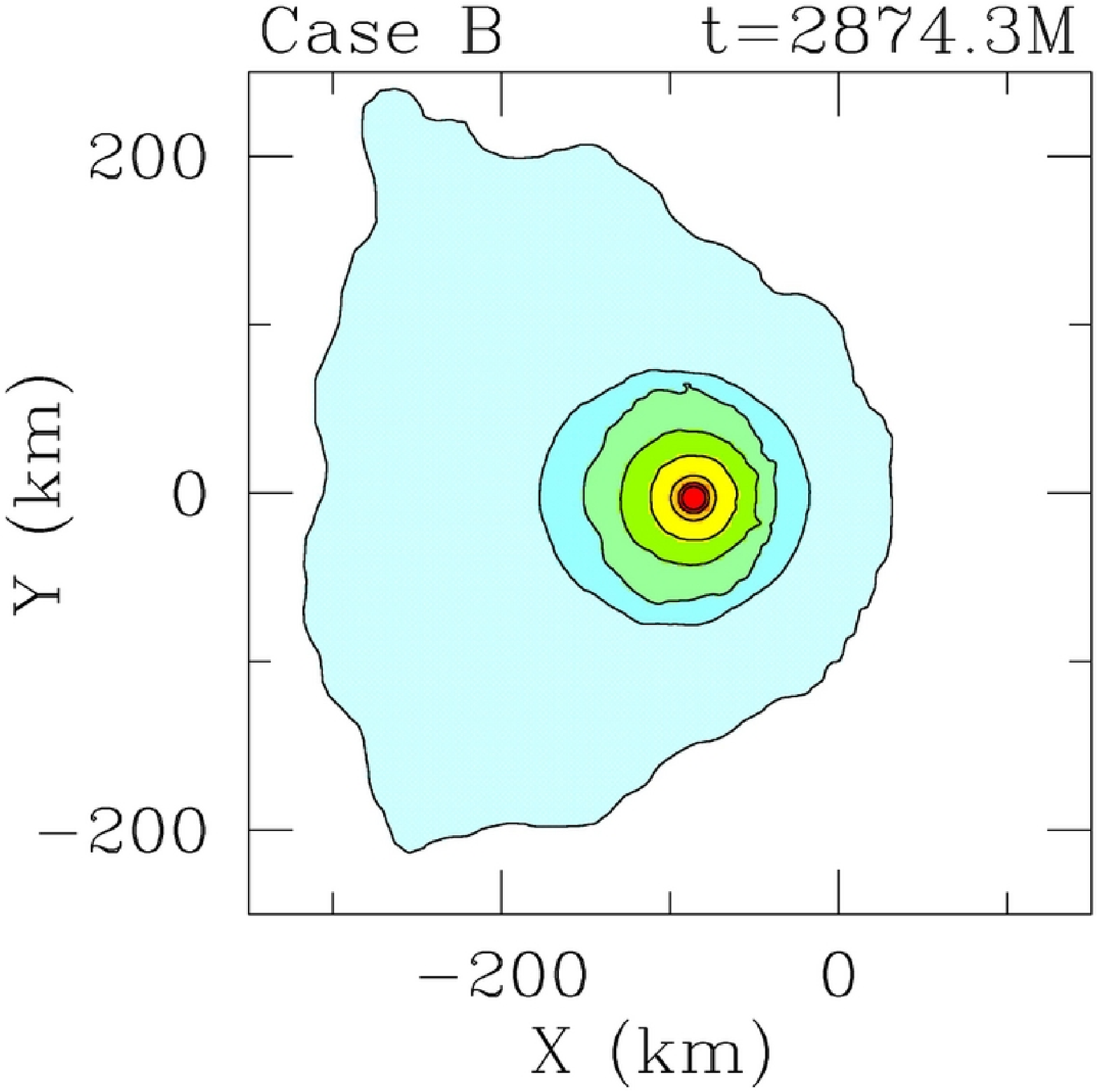}}
\subfigure{\includegraphics[width=0.325\textwidth]{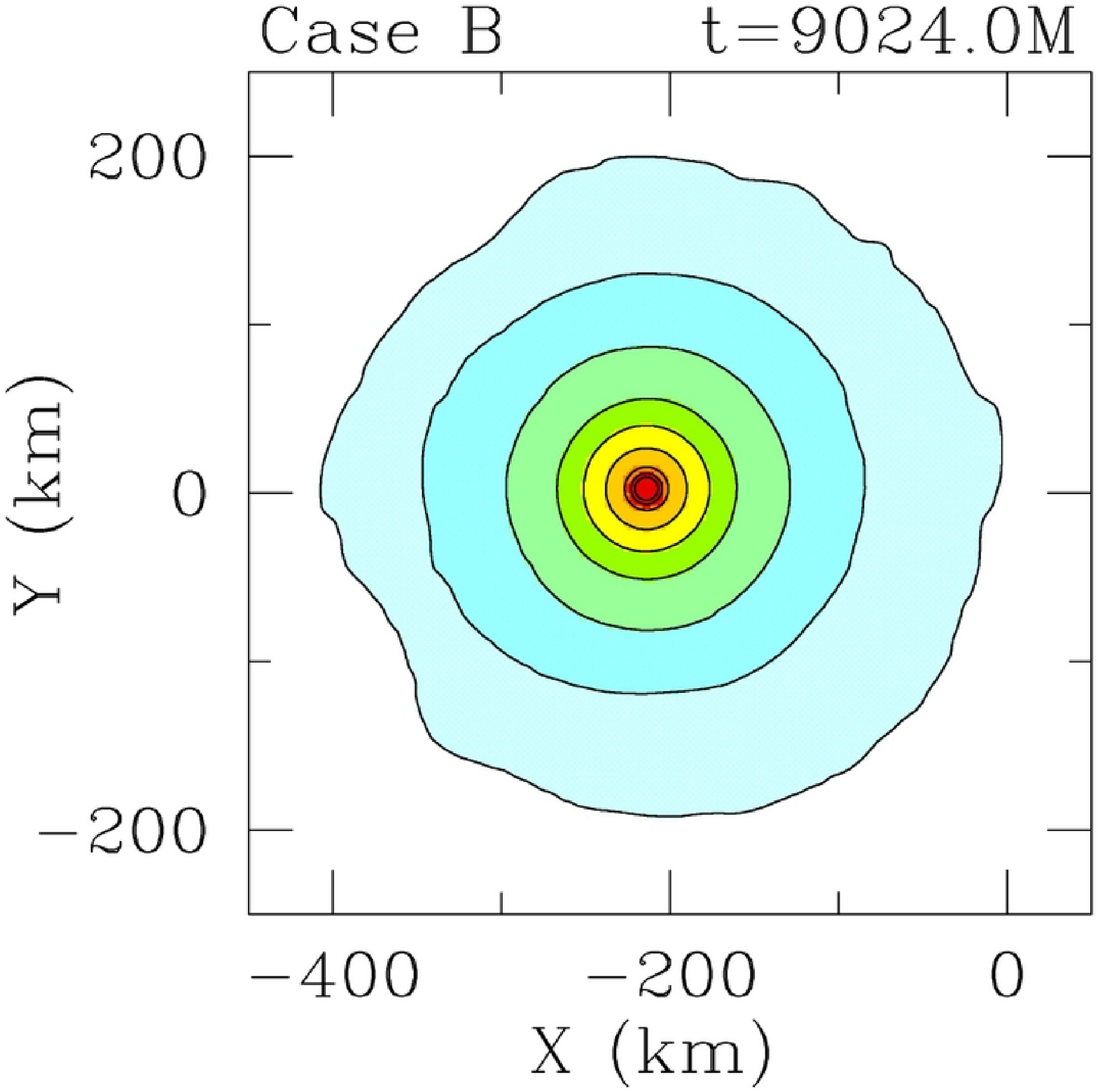}}
\subfigure{\includegraphics[width=0.325\textwidth]{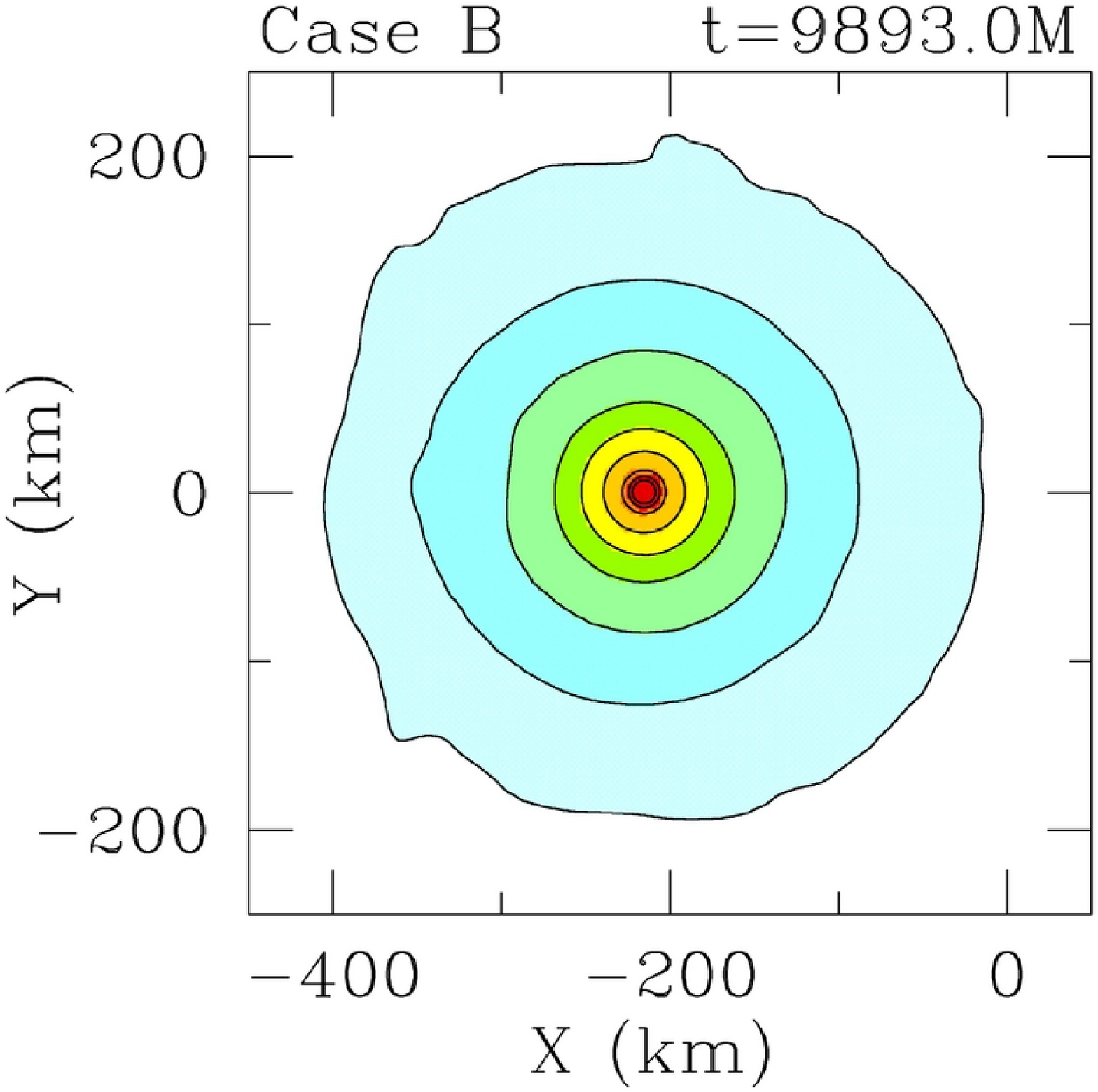}}
\caption{
Snapshots of rest-mass density contours at selected times for case B. 
The contours are plotted in the orbital plane, according to $\rho_0 =
\rho_{0,\rm max} 10^{-0.61j-0.16}\ (j=0,1,\ldots, 9)$.
The color code used is the same as that defined in Fig.~\ref{fig:A2xy}.
The maximum initial NS density is $\rho_{0,\rm max} = 4.6454\rho_{\rm nuc}$.
The last two snapshots, which take place near the end of
 the simulation, demonstrate
that the density contours within a radius of about $150\rm km$ remain unchanged. 
Here $M = 2.48M_\odot = 3.662\rm km = 1.222\times 10^{-5} s $ 
is the sum of the isolated stars' ADM masses.
\label{fig:Bxy}
}
\centering
\end{figure*}

\begin{figure*}
\centering
%
\subfigure{\includegraphics[width=0.325\textwidth]{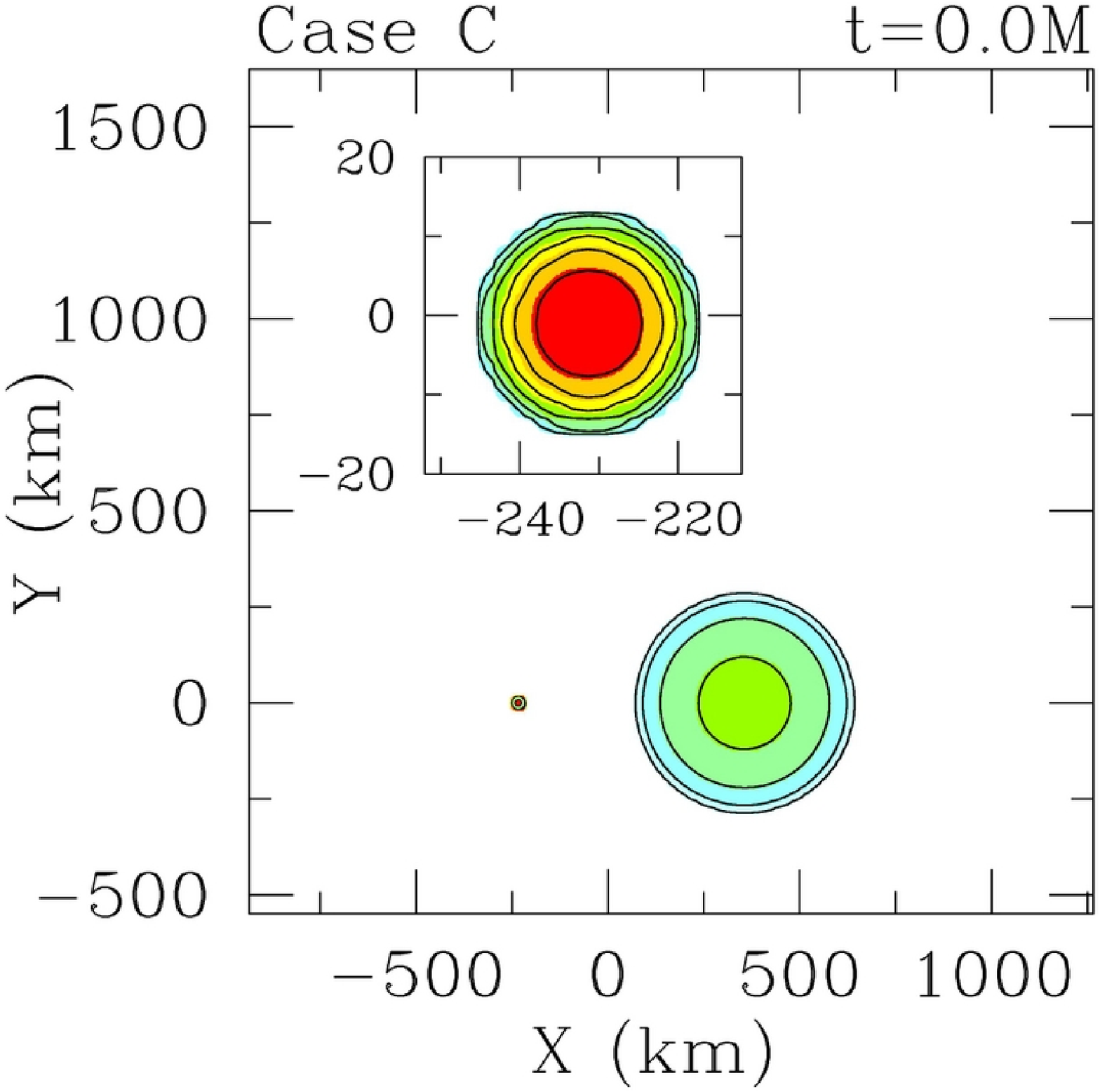}}
\subfigure{\includegraphics[width=0.325\textwidth]{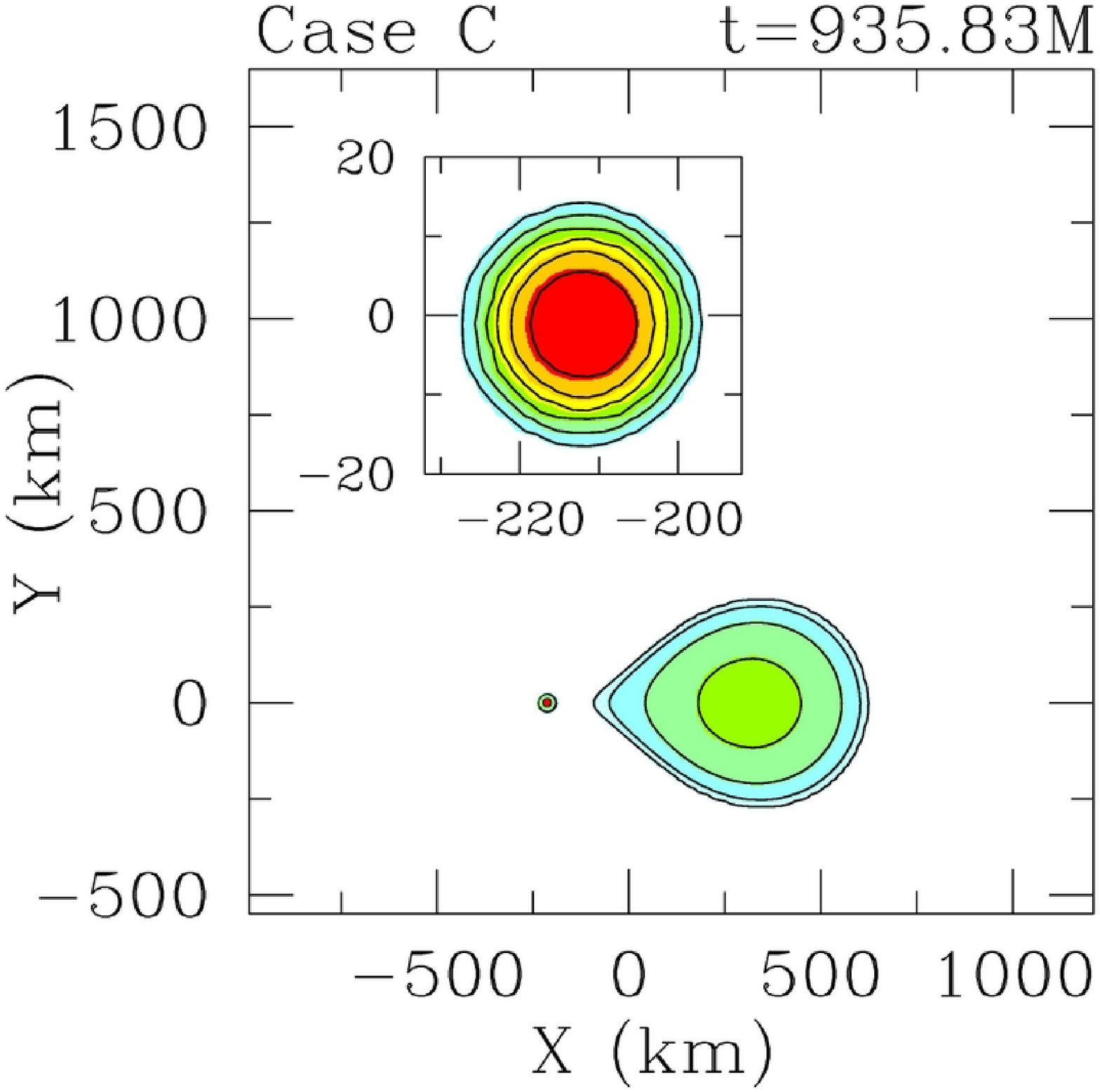}}
\subfigure{\includegraphics[width=0.325\textwidth]{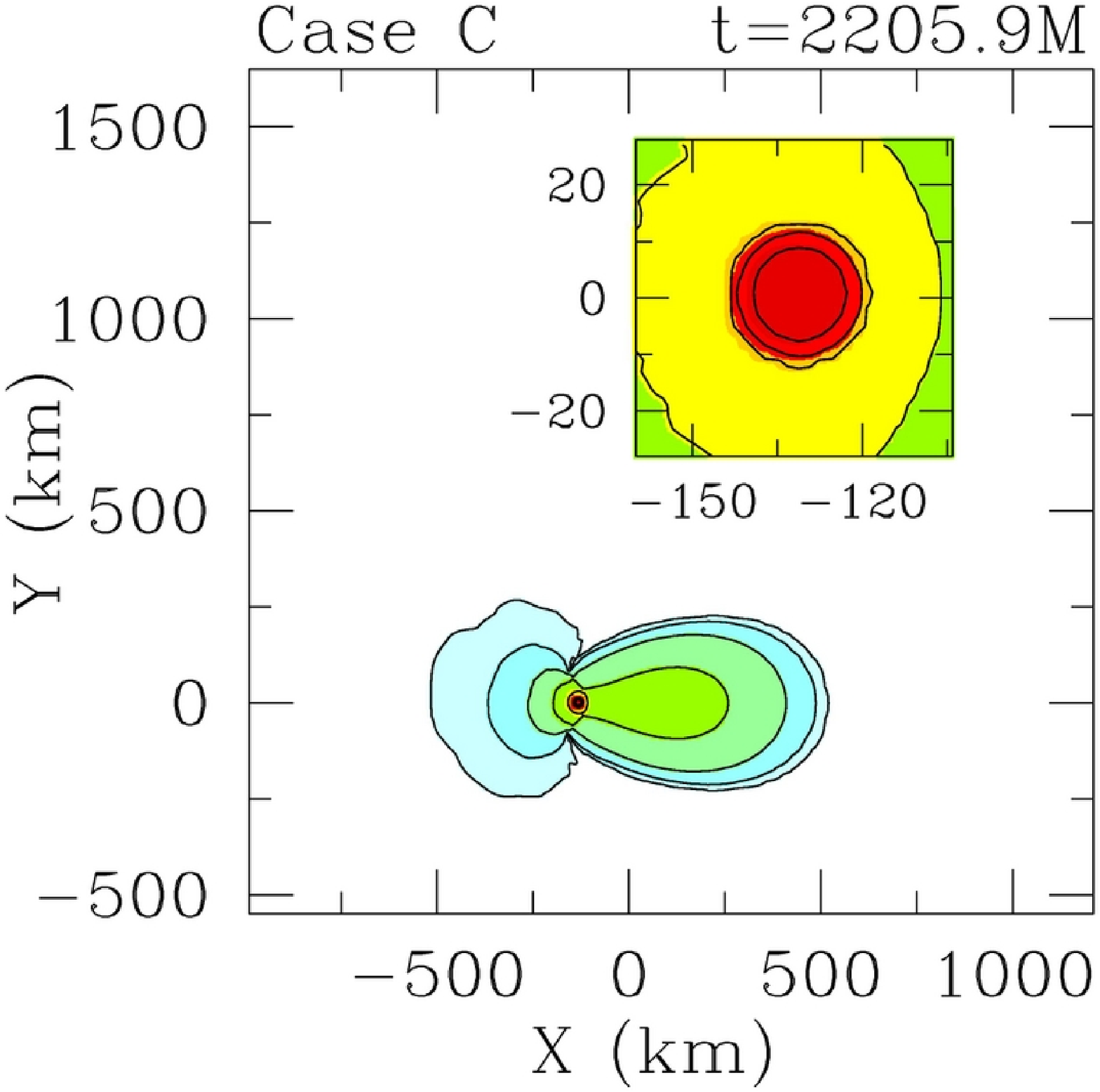}}
\subfigure{\includegraphics[width=0.325\textwidth]{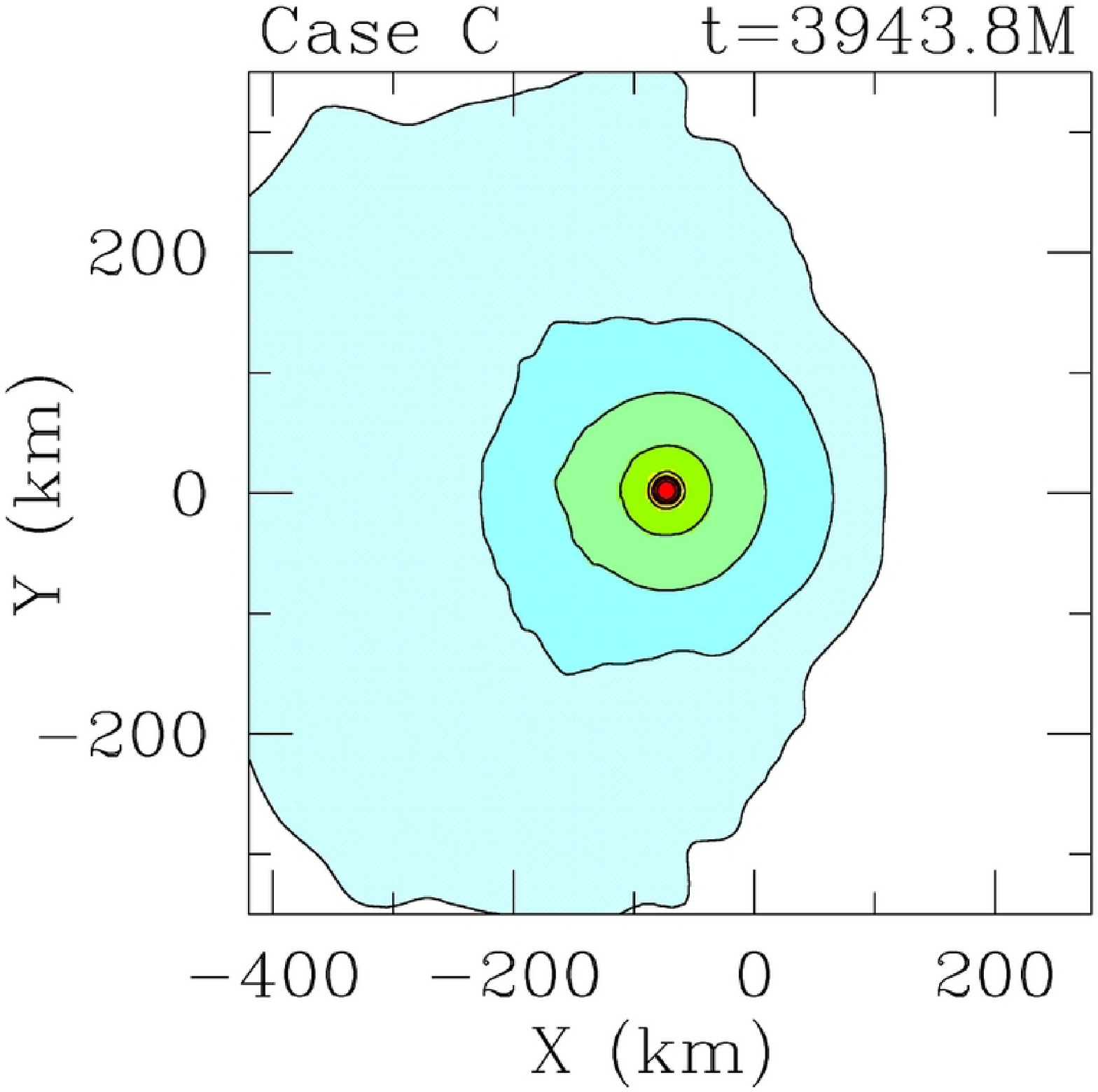}}
\subfigure{\includegraphics[width=0.325\textwidth]{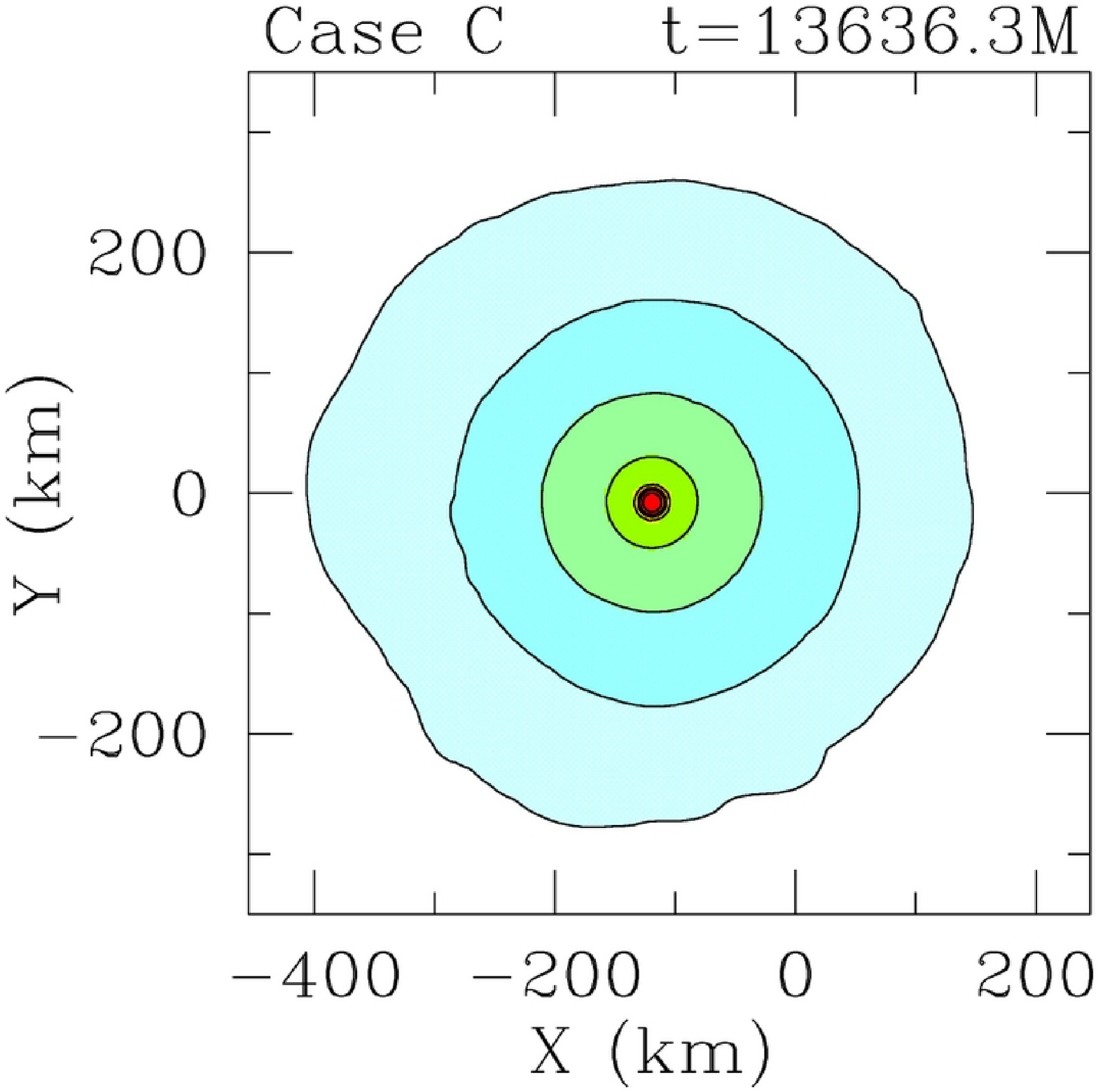}}
\subfigure{\includegraphics[width=0.325\textwidth]{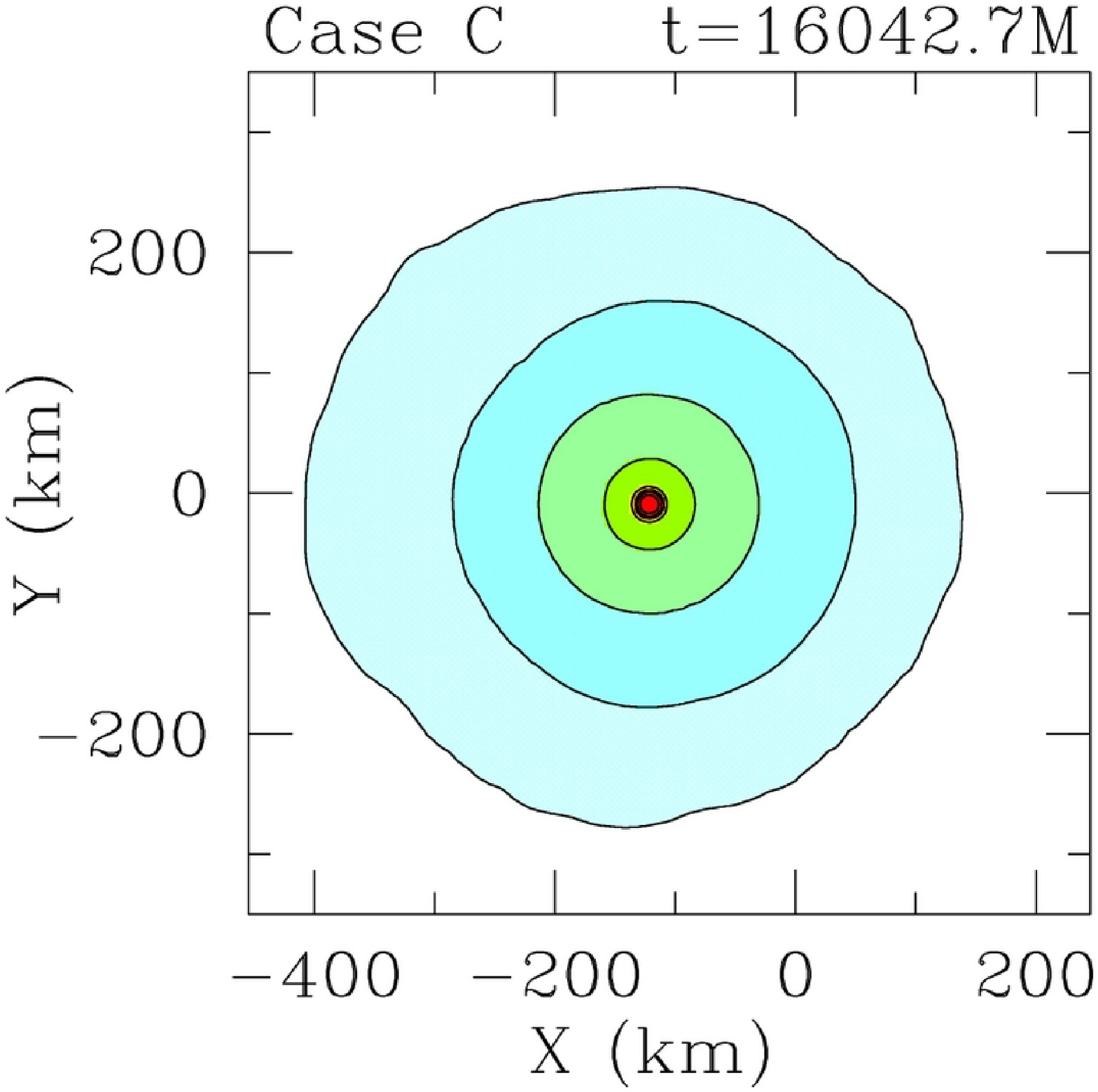}}
 \caption{
Rest-mass density contours in the orbital plane at selected times for case C. 
The contours are plotted 
according to $\rho_0 = \rho_{0,\rm max} 10^{-0.72j-0.16}\ (j=0,1,\ldots, 9)$.
The color code used is the same as that defined in Fig.~\ref{fig:A2xy}.
The maximum initial NS density is $\rho_{0,\rm max} = 4.6454\rho_{\rm nuc}$.
The last two snapshots, which take place near the end of the simulation,
 demonstrate that the density contours within a radius of
 about $150\rm km$ remain unchanged. 
Here $M = 2.48M_\odot = 3.662\rm km = 1.222\times 10^{-5} s $ 
is the sum of the isolated stars' ADM masses. 
\label{fig:Cxy}
}
\centering
\end{figure*}

\subsection{Effects of the pWD compaction}
\label{sec:pWD_compaction_study}

Here we describe the effects of the pWD compaction on the dynamics of pWDNS
head-on collisions. Overall, our findings are qualitatively
similar to those of case A2 described in the previous section.
An appreciable fraction of 
the initial total mass escapes to infinity, but the final total
 mass of the pWDNS remnant still exceeds the maximum mass that our cold EOS can support.
Prompt collapse to a black hole does not take place
 either in case B or in case C, because strong shock heating
 gives rise to a hot remnant. 
The outcome of cases B and C is again a TZlO.

The three phases of the head-on collision we described
 in the previous section apply here, too. For this reason
 we now focus our discussion on describing the
differences between cases B, A2 and C, i.e., in order
 of decreasing pWD compaction. 

The tidal acceleration, which the pWD experiences due to the
 NS tidal field, increases as the pWD compaction decreases. 
This is because the initial coordinate separation is the same
 for cases B, A2 and C. As a result, the acceleration phase
is shorter for larger pWD compaction. This can be seen
in Figs.~\ref{fig:Bxy} and~\ref{fig:Cxy}, where equatorial rest-mass
density contours are plotted.

Shock heating far from the core of the remnants, as quantified by $K$,  is somewhat less intense as the pWD
 compaction decreases. The shorter acceleration phase implies that the
 relative speed of the two components at the beginning of the plunge 
phase is a little smaller, which in turn leads to weaker shocks. 

During the plunge phase, the NS interior is less
 affected by decreasing pWD compaction. This can be seen 
(a) in the insets in Figs.~\ref{fig:Bxy} and~\ref{fig:Cxy},
where in case C the post-plunge structure of the NS core
 is almost the same as that showed in the pre-plunge snapshots, 
while this is not true for 
case B, and (b) by the variation in the NS central density ($\rho_{0,\rm c}$); 
in particular, we find that at maximum compression the NS 
central density increases by 
about 42\% in case B, 8\% in case A2 and 5\% in case C.
These results can be easily interpreted because
in a sequence of pWDNS head-on collisions where the NS is
 fixed and the size of the pWD increases with fixed mass, the
 NS gradually encounters
thinner and thinner material, and hence changes to the NS structure
become less and less important. 

Were the system mass loss to vary appreciably with pWD size, we might expect a 
corresponding variation in $\rho_{0,\rm c}$. However, such a mass loss variation is not 
observed, as we discuss next.

As in case A2, in both cases B and C, a large fraction
 of the initial mass eventually escapes to infinity 
(see Fig.~\ref{fig:u_0casesA2B} for case B), but we
do not find strong variations in the mass lost among cases B, A2 and C.
 We find that the amount of matter that escapes in case B 
is $14\%$ of the initial rest mass, while case C
 loses $18\%$, (case A2 loses $18\%$) (see Fig.~\ref{fig:mass_loss}) of the initial rest mass. 
Given our earlier discussion that shocks  in case B
 are stronger than those in case A2, 
and in turn shocks in case A2 are stronger than those
 in case C, this last result may sound contradictory, 
because one might expect that stronger shocks would eject more
matter to infinity. The apparent contradiction can be resolved, if one considers
that as the pWD compaction decreases, the pWD outer layers become less
and less bound, and hence, it requires less energy to eject them to infinity.

As in case A2, the remnants in cases B and C
 eventually settle into spherical quasiequilibrium objects 
with oscillating outer layers 
(see Figs.~\ref{fig:Bxy},~\ref{fig:Cxy}).  The sphericity of the
remnants (in the adopted gauge) in cases B and C is demonstrated by
 the xy, xz and yz rest-mass density contours shown in Fig.~\ref{fig:xyxzCasesA}.

The pWDNS remnants in both case B and case C consist of
 a cold NS core with a hot mantle on top. Thus, all 
cases lead to the formation of a TZlO.
This is again demonstrated in the last row of Fig.~\ref{fig:xyxzCasesA}, where 
contours of $K=P/P_{\rm cold}$ are shown. 
Within a radius of about $100$ km from the center of mass of the final remnants, 
$K \in[1,35]$ in case B, and $K \in[1,10]$ in case C ($K \in [1,15]$ in case A2). 
Thus, shock heating is overall strongest in case B, weaker in
 case A2 and even weaker in case C. 

\begin{figure}[t]
\centering
\includegraphics[width=0.495\textwidth,angle=0]{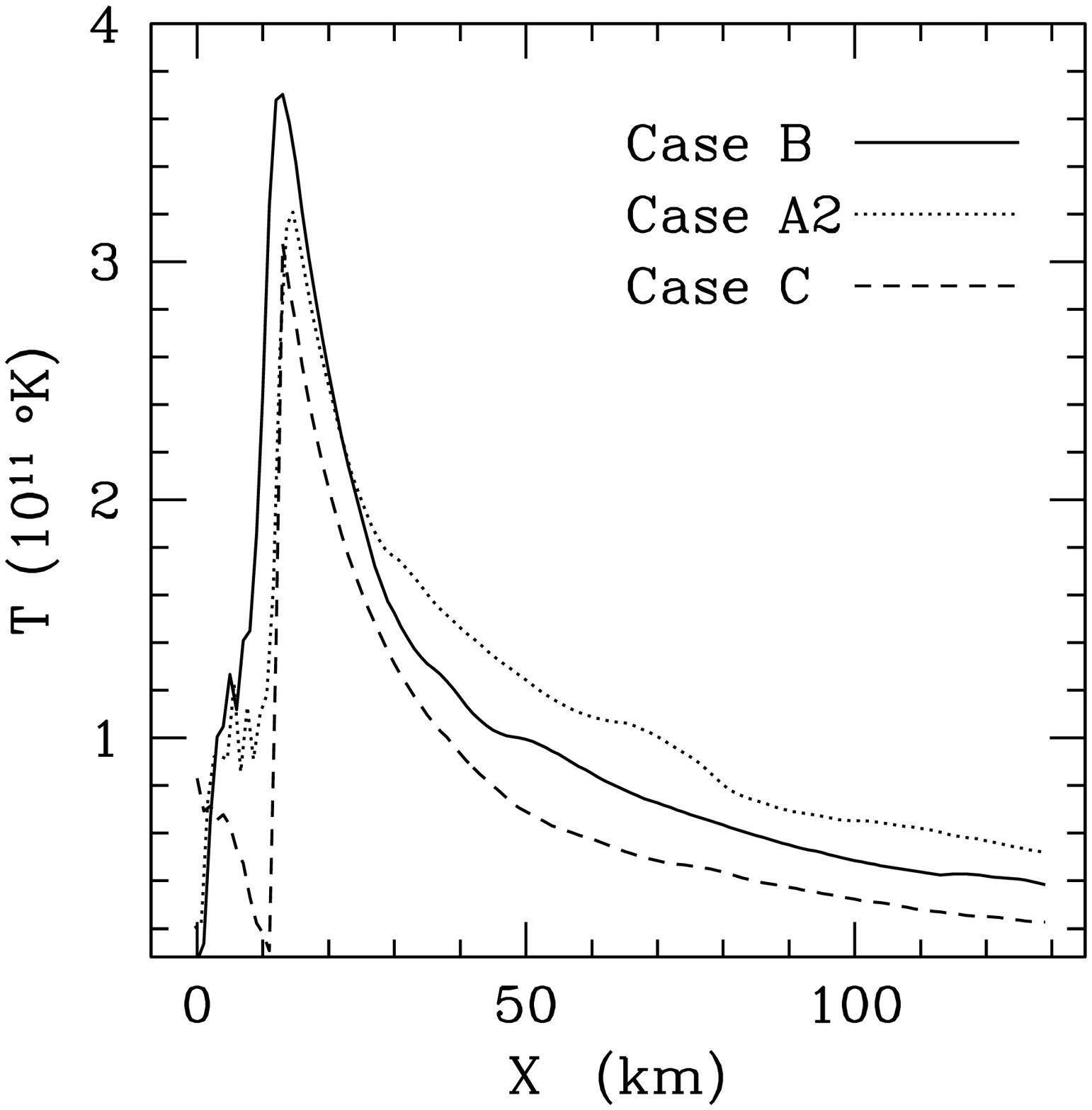}
\caption{Temperature ($T$) profiles for cases B, A2 and C. The temperature 
is in units of $10^{11}\ {}^o\rm K$.  The profiles correspond to
 the values of $T$ at the end of the simulations and along the
 x-axis (for $x > x_{\rm c}$, where $x_{\rm c}$ is the x
 position of the center of mass of the remnant in each case). It 
is clear that typical temperatures are of order 
$10^{11}\ {}^o\rm K$. For realistic WDNS collisions we 
expect $T \sim 10^{9}{}\ ^o$K (see discussion 
following Eq.~\eqref{Temperature}).
\label{fig:temp}
}
\centering
\end{figure}

In Fig.~\ref{fig:rest_mass_evol} the remnant rest masses in cases B and
C are plotted as functions of time. The figure demonstrates that in both cases, 
the rest masses within a radius of $220$km account
 for more than 90\% of the final total rest masses
 in both case B and case C, and are greater than
  $1.92M_\odot$, i.e., the maximum rest mass that our 
cold EOS can support. The pWDNS remnant does not collapse promptly to
form a black hole, because of the extra thermal pressure support. 
However, delayed collapse to a black hole is almost certain after the pWDNS remnant has cooled.

Another important feature of Fig.~\ref{fig:rest_mass_evol} is that the amount of mass contained within 
a given radius from the center of mass of the remnant is larger for smaller initial pWDs. 
For example, within a radius of $220$km the remnant mass is
$2.18M_\odot$ in case B, $2.035M_\odot$ in case A2, and $1.90M_\odot$ in case C.
This in turn indicates that the higher the initial pWD compaction the higher 
the core densities of the pWDNS remnant. This is supported by the rest-mass density contours 
shown in Fig.~\ref{fig:xyxzCasesA} and by the values of the final central rest-mass density. 
In particular, we find that the final central rest-mass density is $4.10\rho_{\rm nuc}$
 in case C, $4.49\rho_{\rm nuc}$ in case A2, and $4.91\rho_{\rm nuc}$ in case B. 
Thus, there is a variation in the final central density
 of $9.2\%$ from case B to case A2, and $9.5\%$ from case A2 to case C. 
Finally, it is also worth noting that the final minimum value
of the lapse function, which is a good indicator of collapse, increases 
with increasing initial pWD size, too. Specifically, we find this value
to be $0.57$ in case B, $0.595$ 
in case A2, and $0.609$ in case C. All these facts
seem to indicate that as the pWD size increases towards realistic 
WD sizes the less likely it is for the pWDNS remnant to collapse.  
To demonstrate this trend more clearly we compile all aforementioned
 physical parameters of the final configurations in cases B, A2, and C in Table~\ref{tab:results}.

Hence, given the consistency in the results of cases B,
 A2 and C, i.e., the sequence of increasing pWD size with fixed pWD mass,
we expect that as the parameters of our EOS vary, 
so as to describe realistic WDs, the 
result of the head-on collision of a
massive WDNS system will most likely lead to formation of a
quasiequilibrium TZlO. If the initial total mass of the
system exceeds the maximum mass a cold EOS can support, then
 the TZlO may have mass
that exceeds the maximum mass a cold EOS can support, so it will
eventually undergo collapse to a black hole, but only after the remnant has cooled.

To identify the dominant cooling mechanisms and/or relevant nuclear reaction networks, one would need to estimate
the temperatures of these TZlOs. We can do this as follows.

Using Eq. \eqref{Ptot} and the definition of $K$ we can calculate the specific thermal energy as
\labeq{}{
\epsilon_{\rm th} = \frac{(K -1 )P_{\rm cold}}{(\Gamma_{th}-1)\rho_0}.
}

To estimate the temperature $T$ of matter, we assume that we can model the temperature dependence of $\epsilon_{\rm th}$ as
\labeq{}{
\epsilon_{\rm th} = \frac{3kT}{2m_{\rm n}}+ f\frac{aT^4}{\rho_0},
}
where $m_{\rm n}$ is the mass of a nucleon, $k$ is the Boltzmann constant and $a$ is the radiation constant.
The first term represents the approximate thermal energy of the nucleons, and the second term accounts for the thermal energy due to relativistic particles. 
The factor $f$ reflects the number of species of ultrarelativistic particles that contribute to the thermal energy. When $T << 2m_{\rm e}/k \sim 10^{10}\rm K$,
where $m_{\rm e}$ is the electron mass, thermal radiation is dominated by photons and $f = 1$. 
When $T >> 2m_{\rm e}/k$, electrons and positrons become ultrarelativistic and also contribute to radiation, and $f = 1 + 2 \times (7/8) = 11/4$. 
At sufficiently high temperatures and densities ($T \geqslant 10^{11} {\rm K}, \rho_0 \geqslant 10^{12} \rm  g\ cm^{-3}$), neutrinos are generated
copiously and become trapped, so, taking into account three flavors of neutrinos and antineutrinos, $f = 11/4+6\times(7/8) = 8$. 

In Fig.~\ref{fig:temp} we show the temperature profiles of the remnants in cases B, A2 and C, where it is clear that typical temperatures 
of our TZlOs are of order $10^{11}\ {}^o\rm K$. This is expected as the total energy available for shock heating should be of order
$M_{\rm NS}M_{\rm WD}/R_{\rm WD}$, i.e., the gravitational interaction energy when the two stars first make contact.  The total thermal energy, $E_{th}$, is then 
\labeq{}{
E_{th} \sim \frac{(M_{\rm NS} +M_{\rm WD})}{m_{\rm n}} k T \sim \frac{M_{\rm NS}M_{\rm WD}}{R_{\rm WD}}.
}
From this last equation we can estimate the characteristic temperature as
\labeq{Temperature}{
T \sim \frac{C_{\rm WD}m_{\rm n}}{(1+q)k}.
}
Thus, all things being equal (no mass loss, same kinetic energy at collision, etc.) characteristic TZlO temperatures should be directly 
proportional to the pWD compaction. Taking case A2 as an example, $C \simeq 0.01$ and $ q \simeq 2/3$, we find $T\simeq 6.5\times 10^{10}\ {}^o\rm K$,
in rough agreement with our simulations.

Using this scaling argument we can extrapolate our results to realistic WDNS head-on collisions. 
For a solar-mass WD obeying the Chandrasekhar EOS $C_{\rm WD} \simeq 3\times 10^{-4}$. Hence, we
predict that typical temperatures in realistic head-on collisions of massive WDNS systems would be of order $10^9\ {}^o\rm K$.

%
%

\section{Summary and Discussion}
\label{sec:summary}

In this work we studied the dynamics of the head-on collision of WDNS binaries
in full general relativity,
 aided by simulations that employ the Illinois  AMR
relativistic hydrodynamics code \cite{Etienne08,Illinois_new_mhd}. 
This study serves as a prelude to the circular binary WDNS
 problem which will be the subject of a forthcoming paper.

Our primary focus is on compact objects whose total mass exceeds
the maximum mass that a cold EOS can support and our goal
 is to determine whether a WDNS collision leads either to prompt
collapse to a black hole or the formation of a
 quasi-equilibrium configuration that resembles a 
Thorne-Zytkow object (TZO) \cite{ThorneZytkow77}, which we call Thorne-Zytkow-like object (TZlO). By a TZlO
 we mean a NS sitting at the center
of a hot gaseous mantle composed of WD debris. 

Due to the vast range of dynamical time scales and length scales involved 
in this problem, realistic WDNS simulations (head-on or otherwise) are
computationally prohibitive, if one employs current numerical
relativity techniques and available computational resources.
For this reason, we tackle the problem using a different approach. 

In particular, we constructed a piecewise polytropic 
EOS which captures the main physical features of NSs and, 
at the same time, scales down the size of a WD. 
We call these scaled-down models of WDs pseudo-WDs (pWDs). 
Using these pWDs, we can reduce the range of length scales and time scales 
involved, rendering the computations feasible. 

A pWD is not a realistic model of a WD. However, with
 our proposed parametrized EOS we can construct a
sequence of pWD models with gradually increasing size and perform
simulations that approach the realistic case.  Then we can make predictions 
about the realistic case by extrapolating the results from this
sequence of simulations.

Using pWDs, we performed two sets of simulations. One set of our simulations
 studied the effects of the NS mass on the final outcome, 
when the pWD is kept fixed at a mass of $0.98M_\odot$ and its size
fixed at $146$km. We choose three masses for the NS, namely
$1.4M_\odot, 1.5M_\odot, 1.6M_\odot$ (cases A1, A2 and A3, respectively). 
The other set of simulations studied the effect of the pWD 
compaction on the final outcome, 
when the NS is kept fixed at a mass of $1.5M_\odot$. 
In the latter set of calculations, 
we choose three values for the ratio of the pWD to the NS radius, 
namely $5:1$, $10:1$ and $20:1$ (cases B, A2 and C, respectively). 

In general, the head-on collision of pWDNS systems
 can be decomposed in three phases: i) acceleration, 
ii) plunge and iii) quasiequilibrium. 

During the acceleration phase the two stars accelerate towards
one another starting from rest.  As the separation decreases
 the NS tidal field becomes so strong that the pWD becomes
 highly distorted, while the NS interior is almost unchanged. 
This phase ends when the NS and pWD first make contact. 

During the plunge phase the NS penetrates the pWD, 
launching strong shocks that sweep 
through and heat the interior of the pWD. The NS outermost layers are stripped
after encountering the dense central parts of the pWD (see Fig.~\ref{fig:A2xy}), but the NS core 
is mostly unaffected, except when the compaction of the
 pWD is high (see Sec. \ref{sec:pWD_compaction_study}). 
We find that the strong shocks sweeping the pWD transfer
 linear momentum to the outer pWD layers, 
causing a large amount of pWD matter to escape to infinity. In all
calculations, we find the rest mass loss to be between $14\%-18\%$ of
the initial total rest mass. Material that did not escape to infinity
accretes onto the underlying NS and pWD matter.

Finally, during the quasiequilibrium phase, the remnant
 settles into a spherical quasiequilibrium object whose outermost
layers undergo damped oscillations. 

Although a large fraction of the initial mass escapes, the 
final total rest mass still exceeds the maximum rest mass of $1.92M_\odot$ that our
 cold EOS can support (see Fig.~\ref{fig:rest_mass_evol}).  
However, the pWDNS remnant cannot collapse promptly to form a black hole, 
because it is hot. This result is the same in all our simulations. 
However, we point out that delayed collapse 
to a black hole is almost certain after the pWDNS remnant has cooled, but this will 
occur on a timescale much larger than a hydrodynamical timescale.  

The final object consists of a cold NS core surrounded by a 
hot mantle. We quantified the results of shock heating
 by the ratio of the total pressure to the cold pressure $K=P/P_{\rm cold}$.  
In all cases $K \simeq 1$ at the center of the remnant, 
and becomes larger than unity away from the center. 
We refer to this nearly-spherical configuration as a
 Thorne-Zytkow-like object, and find this object at 
the end of all simulations, regardless of NS mass and
pWD compaction. We find that within a radius of $100$ km from the centers of mass of the remnants, 
$K$ lies in the range $[1,15]$ in cases A1, A2 and A3, $[1,35]$ in
 case B and $[1,10]$ in case C. Using a simple model for the temperature dependence of the specific
thermal energy we estimate the characteristic temperature of these objects to be of order $10^{11}\ {}^o$K.
Using a simple scaling argument (see Eq.~\eqref{Temperature}) we find that TZlO temperatures should be proportional to the compaction 
of the original pWD, so that in realistic WDNS head-on collisions
typical remnant temperatures would be of order $10^9\ {}^o\rm K$.

Furthermore, we find that the smaller the initial pWD compaction the smaller
the core densities of the pWDNS remnant. This is supported by the rest-mass density contours 
shown in Fig.~\ref{fig:xyxzCasesA} and by the values of the final central rest-mass density. 
In particular, we find that the final central rest-mass density decreases by
$9.2\%$ from case B to case A2, and $9.5\%$ from case A2 to case C. 
In addition, the final minimum value of the lapse function, which is a good indicator of collapse, 
increases with increasing initial pWD size, too. Specifically, we find this
value to be $0.57$ in case B, $0.595$ 
in case A2, and $0.609$ in case C (see Table~\ref{tab:results} for 
a summary of physical parameters of the final configurations in cases B, A2, C).  
All these facts seem to indicate that as the pWD size increases towards realistic 
WD sizes the less likely it is for the pWDNS remnant to collapse.  

An important concern regards the invariance of these results with
respect to larger initial separations.
To fully resolve this, one would need to extend the simulations to wider
  separations, but this extension is outside the scope of the current work.  
However, this work gives us some
qualitative idea about what might happen with larger initial separations. 
Larger separations imply larger kinetic energies during 
the plunge phase, which in turn imply stronger shocks. Stronger
shocks likely lead to larger mass loss
 and more intense shock heating. Therefore, our expectation is 
that head-on collisions of pWDNS systems starting at larger separations will also 
result in the formation of TZlOs 
and that such collisions would not lead to prompt formation of a black hole.

Given the consistency in the results of cases B, A2 and C, we expect
that as the parameters of our EOS are adjusted such that pWDs more closely resemble
realistic WDs, WDNS head-on collisions are likely to form a quasiequilibrium TZlO. 
If the initial total mass of the system well exceeds
the maximum mass that a cold EOS can support, then the TZlO 
will most likely have mass
exceeding the maximum mass supportable by a cold EOS, eventually
collapsing to a black hole after the remnant has cooled.


We conclude by stressing that we cannot
 use the results of this work to make definite predictions about
 either the pWDNS or the realistic WDNS circular binary problem. 
One might speculate
 that shock heating will be minimized in such a scenario,
and hence it may result in prompt collapse to a black hole. 
However, sufficient angular momentum must be shed in the circular binary
 case in order for the object to promptly form a black hole. 
To resolve these issues, hydrodynamic simulations in full general
relativity must be performed and will be the focus of a forthcoming paper.

\acknowledgments

We would like to thank Alexei M. Khokhlov and Thomas W. Baumgarte 
for helpful discussions. This paper was supported in part
by NSF Grants PHY06-50377 and PHY09-63136 as well
as NASA Grants NNX07AG96G and NNX10AI73G to
the University of Illinois at Urbana-Champaign. 
Z. Etienne gratefully acknowledges support
 from NSF Astronomy and Astrophysics Postodoctoral Fellowship AST-1002667.

\appendix

\section{Initial data Code description}
\label{appA}

In this appendix we describe some details of the fixed mesh refinement (FMR), finite difference code we developed 
for generating general relativistic initial data.

The grid structure we use is a multi-level set of properly nested, 
cell-centered uniform grids. Each grid corresponds to
 one level of refinement labeled by the 
level number $il = (0,1,2,\ldots,nl-1)$, where $nl$ is the
 total number of levels. $il=0$ corresponds to the coarsest
level and $il=nl-1$ to the finest one. All levels have the
 same number of grid points 
$nx, ny, nz \in \mathbb{Z}$ in the $x$, $y$ and $z$ directions respectively. 
The coordinates on our grid are defined as follows
\labeq{coords}{
\begin{split}
x_{il,i} & = x_{il,\rm min}+ i\cdot \Delta x_{il}, \hspace{1cm} i = 0,1,\ldots, nx-1,  \\
y_{il,j} & = y_{il,\rm min}+ j\cdot \Delta y_{il}, \hspace{1cm} j = 0,1,\ldots, ny-1, \\
z_{il,k} & = z_{il,\rm min}+ k\cdot \Delta z_{il}, \hspace{1cm} k = 0,1,\ldots, nz-1, 
\end{split}
}
where $x_{il,\rm min}, y_{il,\rm min}, z_{il,\rm min}$ are the minimum
values of the coordinates in each direction on level $il$ and 
$\Delta x_{il}, \Delta y_{il}, \Delta z_{il}$ are the mesh sizes
 in each direction on level $il$.

The mesh size between two consecutive levels differs 
by a factor of two so that
\labeq{meshsize}{
\Delta x_{il+1} = \frac{\Delta x_{il}}{2},
}
and similarly in the $y$ and $z$ directions. 

In order for the grids to be properly nested we 
demand that there exists an $i \in [0,nx-1]$ such that
\labeq{nested}{
x_{il,i} = x_{il+1,\rm min}-\frac{3}{2}\Delta x_{il+1},\hspace{0.5cm} il = 0,\ldots nl-2
}
and similarly in the $y$ and $z$ directions. 
This condition ensures that two consecutive levels share a common interface.

We now borrow FMR terminology to name two types of cells that exist
 on our grid. These are the split cells and the leaf cells or leaves. 
A split cell is one within which there exist higher level cells and
 a leaf cell is one within which there are no higher level cells. 
The total number of cells $N_{\rm tot}$ on our grid is 
\labeq{Ntot}{
N_{\rm tot} = nx\cdot ny\cdot nz\cdot nl,
}
and a straightforward calculation shows that the number of leaves is
\labeq{Nleaf}{
N_{\rm leaf} = nx\cdot ny\cdot nz \frac{7nl+1}{8}. 
}
When $nl=1$, $N_{\rm leaf}=N_{\rm tot}$, i.e., all cells are leaves, as expected.

We distinguish between these two types of cells because our solutions
 are defined only on leaves. This may be more cumbersome 
to implement, but has two main advantages. 

First, there is no ambiguity as to how one should interpolate
 values of matter sources from fine levels  
on coarse levels in order to correctly calculate the gravitational
 fields. To be more specific, let us 
assume that we have one coarse cell which is 
split into 8 cells and that we know the values of the density on the
fine cells. In Newtonian physics, to ensure that the gravitational fields
are computed correctly (at least far away), 
all we have to do is set the cell averaged density on the coarse
level such that the total mass in the coarse cell is the same as the
total mass in the enclosed fine cells. In general relativity the definition 
of gravitational mass depends not only on the density, but also 
on the gravitational fields.  Hence, there is no
 straightforward way to set the density on
coarse cells in GR. The ambiguity is immediately lifted, if one
 defines all fields only on the finest cells. 

The second advantage of using only leaves is that the memory 
requirements are minimized and the calculations are carried
 out faster because
\labeq{}{
\frac{N_{\rm leaf}}{N_{\rm tot}} = \frac{7nl+1}{8nl} \leq 1,
}
where the last inequality holds because $nl \geq 1$.

For a general second-order nonlinear elliptic equation of the form
\labeq{elleq}{
\nabla^2 u = f(u)\chi, 
}
where $f(u)$ is a nonlinear function of the variable $u$ and $\chi$
 a known scalar independent of $u$, 
our code employs a standard second-order finite difference scheme
\labeq{finite_diff}{\begin{split}
\frac{u_{il,i+1,j,k}+u_{il,i-1,j,k}-2u_{il,i,j,k}}{\Delta x_{il}^2} & + \\
\frac{u_{il,i,j+1,k}+u_{il,i,j-1,k}-2u_{il,i,j,k}}{\Delta y_{il}^2} & + \\
\frac{u_{il,i,j,k+1}+u_{il,i,j,k-1}-2u_{il,i,j,k}}{\Delta z_{il}^2} & 
 = f(u_{il,i,j,k})\chi_{il,i,j,k}. 
\end{split}
}

The finite difference stencil changes only across grid-level 
boundaries where we perform first order interpolation.

To address the nonlinearity of Eq.~\eqref{elleq} we perform 
Newton-Raphson iterations as follows. Let us assume that $u_n$ is a 
guess at step $n$. We first calculate the residual $R_n$ from Eq.~\eqref{elleq}
\labeq{Res}{
R_n = \nabla^2 u_n - f(u_n)\chi, 
}
and then solve the linearized equation for the correction $\delta u_n$ on $u_n$
\labeq{linelleq}{
\nabla^2 \delta u_n = f'(u_n)\chi \delta u_n - R_n,
}
where 
\labeq{}{
f'(u_n)=\bigg(\frac{df(u)}{du}\bigg)_{u=u_n}.
}
Once a solution to \eqref{linelleq} is found, we correct $u_n$ as
\labeq{correct}{
  u_{n+1} = u_n +\delta u_n,
}
and iterate until this procedure converges
 and a solution to Eq.~\eqref{elleq} is obtained.

We solve the equations using the PETSc linear
 solver Krylov Space (KSP) methods. KSP methods are matrix methods and hence 
we have to set up the matrix of the linear system. 


To do this we define a global index that counts all cells
 (both leaves and split cells) on our grid as
\labeq{Iglob}{
I = il + nl (i+nx\cdot j+nx\cdot ny \cdot k),
}
In this way, every leaf cell corresponds to a unique
 index $I$. However, $I$ takes values $0,1,\ldots, N_{\rm tot}-1$,  
but there 
are $N_{\rm leaf}$ leaves 
on the 
grid with $N_{\rm leaf}\leq N_{\rm tot}$. Hence $I$ cannot 
be used to count leaves, unless
$nl=1$. For this reason, we define another index $ic$, which counts
 only the leaves on our grid, as well as two mappings; 
from I to ic, $ic(I)$ and from ic to I, $I(ic)$. Since for every cell 
on our grid we can find $I$ from Eq.~\eqref{Iglob} we set
 up these mappings by defining two arrays ic\_of\_I, I\_of\_ic, 
of length $N_{\rm tot}$ and $N_{\rm leaf}$, respectively. 
Looping over $il, i,j,k$, we store $ic$ in the array ic\_of\_I
 assigning a value of $-1$ for split cells, whereas we store $I$ in the 
array I\_of\_ic. The index $I$ is used when we need the index $ic$ of
 a neighbor leaf cell in order to calculate derivatives or enter matrix elements. 

For example, let us assume that we are at a leaf cell of index $ic$
which is not near a grid-level boundary, and we want to
 enter the element of matrix $A$ that corresponds 
to the right-x neighbor of this cell (where $A$ represents the
Laplacian). From Eq.~\eqref{finite_diff} this matrix element must be
$1/\Delta x_{il}^2$. If $ic$ represents the $ic$-th row of $A$ we must
find which column of $A$ the neighbor corresponds to. We find this as
follows.

First, using the mapping from ic to I, we find the index $I=I(ic)$ 
of the leaf. Next by use of Eq.~\eqref{Iglob} 
we determine $il, i, j,k$ that correspond to $I$ using the following sequence of operations 
\labeq{}{
\begin{split}
k = &\ int(I/nl\cdot nx\cdot ny) \\
I_1 = &\ I - nl\cdot nx\cdot ny\cdot k \\
j = &\ int(I_1/nl\cdot nx) \\
I_2 = &\ I_1 - nl\cdot nx\cdot j \\
i = &\ int(I_2/nl\cdot nx) \\
il = &\ I_2 -nl\cdot i,
\end{split}
}
where $int$ means the integer part of the division.

In the next step the global index ($I_{\rm p1}$) of the right
x-neighbor is found, as $I_{\rm p1} = il + nl (i+1+nx\cdot j+nx\cdot ny \cdot k)$. 
Finally, using the mapping from I to ic we find the leaf
 (or desired column) number $ic_{\rm p1} = ic(I_{\rm p1})$. Knowing the
 column number of the neighbor, it is straightforward to 
 assign $A_{ic, ic_{\rm p1}}=1/\Delta x^2_{il}$. We use 
the same approach to set up all the matrix elements of the linear 
systems and calculate derivatives. The algorithm becomes slightly more complicated
when the neighbor cell is a fictitious cell that resides
 on a different level. In such cases we
perform first order interpolation and use the same method as outlined
above to find which matrix elements must be
filled with non-zero values. 

\section{Validation of HRSC method for a piecewise polytropic EOS}
\label{appB}

In this appendix we analyze the effect our numerical schemes have on solutions obtained with
our adopted non-smooth EOS \eqref{EOS} and a smooth counterpart of this EOS. We show that there is no essential difference. 
This result is expected because an algorithm with finite resolution cannot distinguish a smooth EOS from a non-smooth EOS,
if the smoothing operation is performed below the resolution level of the computations.

For simplicity we consider a non-smooth EOS with two branches as follows
\labeq{EOS_nonsmooth}{
P = 
\left\{
\begin{array}{ll}
\kappa_1 \rho_0^{1+1/n_1},  &  \rho_0 \leqslant  \rho_1  \\
& \\
\kappa_2 \rho_0^{1+1/n_2},  &  \rho_0 > \rho_1,
\end{array}
\right.
}
and perform a smoothing operation over a density interval $[\rho_i(1-\epsilon), \rho_i(1+\epsilon)]$ as follows
\labeq{EOS_smooth}{
P = 
\left\{
\begin{array}{ll}
\kappa_1 \rho_0^{1+1/n_1},  &  \rho_0 \leqslant  \rho_1(1-\epsilon)  \\
& \\
f(\rho_0),  &  \rho_1(1-\epsilon)<\rho_0 \leqslant  \rho_1(1+\epsilon) \ ,\\
& \\
\kappa_2 \rho_0^{1+1/n_2},  &  \rho_0 > \rho_1(1+\epsilon),
\end{array}
\right.
}
\\
where $f(\rho_0)$ is a smooth spline fit such that the EOS is continuous and has continuous first or second derivative,
depending on the order of the spline.
Our particular choice for the smoothing function is either a cubic spline or a quintic spline. In the former
case the EOS becomes $C^1$, while in the latter case the EOS becomes $C^2$. 

\begin{figure}[t]
\centering
\includegraphics[width=0.495\textwidth,angle=0]{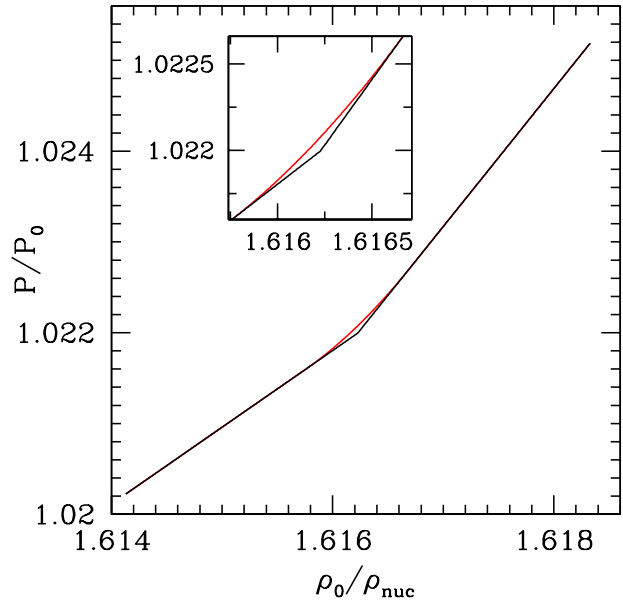}
\caption{Pressure vs rest-mass density for EOSs \eqref{EOS_nonsmooth} (black) and
\eqref{EOS_smooth} (red). Here $P_{0} = 10^{-5} \rm km^{-2}$ and $\rho_{\rm nuc} = 1.48494\times 10^{-4} {\rm km}^{-2}$.
The inset zooms in the region, where EOS \eqref{EOS_nonsmooth} is non-differentiable,
and shows that the cubic spline fit smooths out the discontinuity.
\label{fig:eos_smooth}
}
\centering
\end{figure}

We chose
$\epsilon$ to be sufficiently small so that the smoothed EOS mimics as closely as possible EOS \eqref{EOS_nonsmooth}, 
but large enough to avoid round off errors due to very large gradients. For the cubic spline we set $\epsilon=10^{-4}$, 
while for the quintic spline we set $\epsilon=10^{-2}$. In all our numerical tests we choose $k_1, k_2, n_1, n_2, \rho_1$ to correspond
to $k_2, k_3, n_2, n_3,\rho_2$ of the 10:1 EOS (see Table~\ref{tab:eosparams}), respectively. In Fig.~\ref{fig:eos_smooth} 
we show a plot of EOSs \eqref{EOS_nonsmooth} and \eqref{EOS_smooth}, where $f(\rho_0)$ is a cubic spline.

\begin{figure*}
\centering
\subfigure{\includegraphics[width=0.45\textwidth]{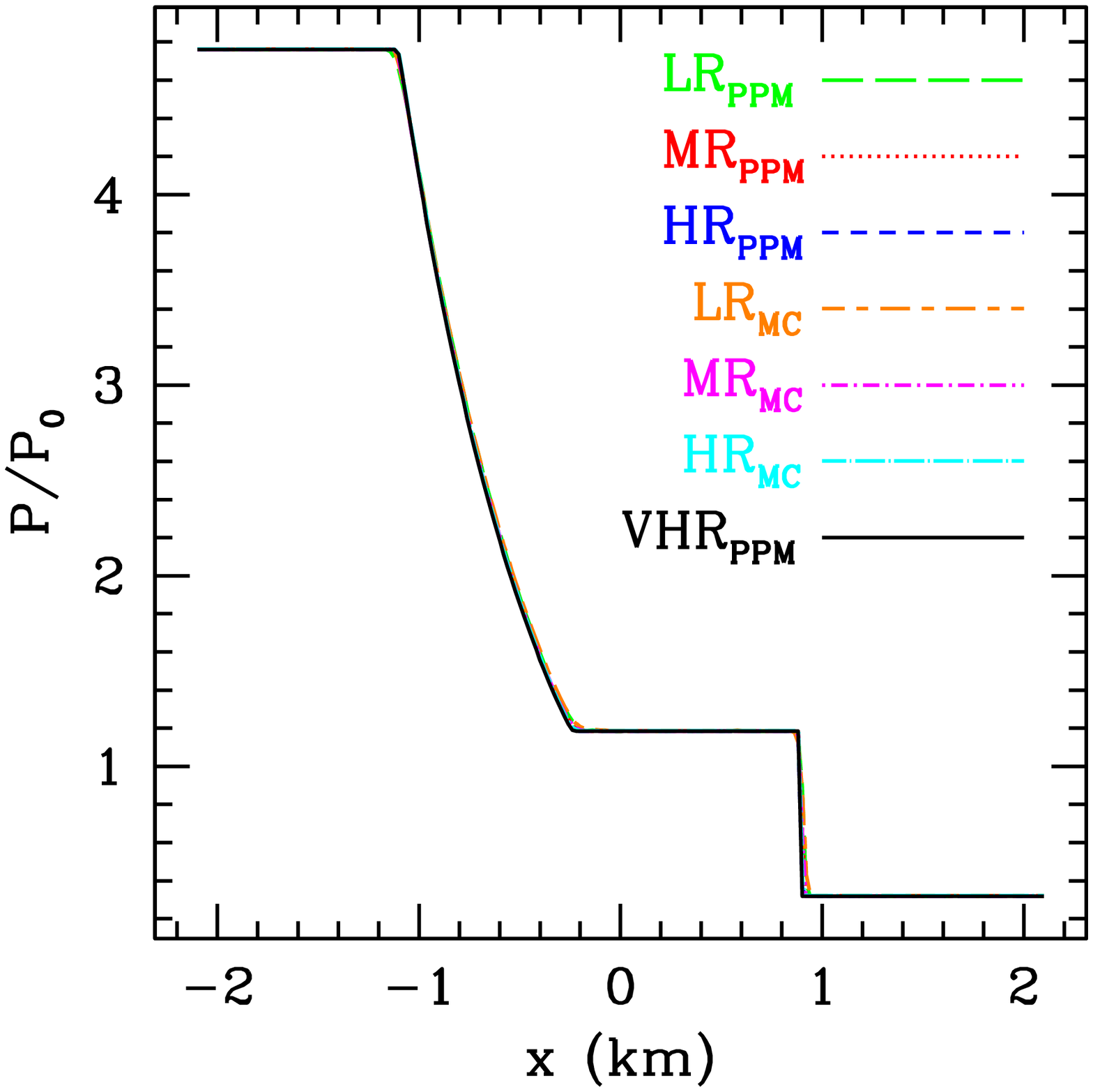}}
\subfigure{\includegraphics[width=0.45\textwidth]{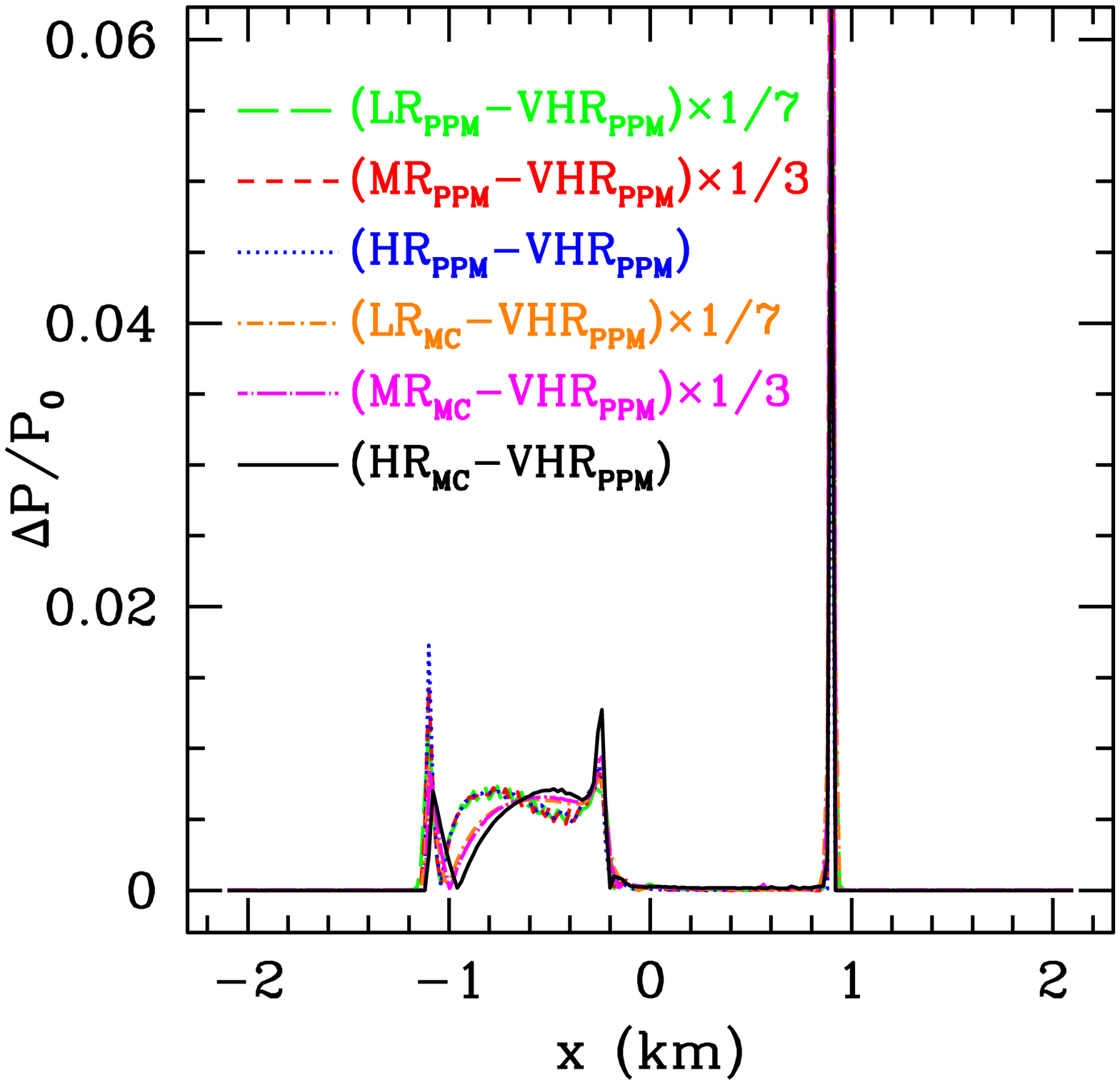}}
\caption{
      Left: Snapshot of pressure profile at $t = 2.56$ km, corresponding to a third of a sound speed crossing timescale 
      across the domain. 
      Right: Convergence plot (at $t = 2.56$ km) using as reference solution 
      the very high resolution
      solution obtained in conjunction with the smooth EOS \eqref{EOS_smooth} with a cubic spline smoothing function. 
      The labels in the plots denote the resolution (LR, MR, HR, or VHR) and the reconstruction method (PPM or MC as subscripts).
      The resolutions used are: LR = 210 , MR = 420, HR = 840, VHR = 1680 grid points. PPM stands for the piecewise parabolic reconstruction, 
      and MC stands for the monotonized central reconstruction. The plots demonstrate that all solutions overlap (left panel), regardless of
      the reconstruction
      method and the EOS used (smooth or non-smooth), and first-order convergence to the VHR run with the smooth EOS, as expected.  
      Here $P_{0} = 10^{-5} \rm km^{-2}$.
      \label{fig:pressure_prof}
}
\centering
\end{figure*}

With the smooth EOS \eqref{EOS_smooth} at our disposal we set up several 1D Riemman (or shock tube)
problems in a spatial domain of length $L = 4.2 \rm km$ and resolutions 210, 420, 840 and 1680 grid points. We set $\Gamma_{\rm th} = 1.66$ 
and use Eq.~\eqref{Ptot} with $P_{\rm cold}$ given by either EOS \eqref{EOS_nonsmooth} or EOS \eqref{EOS_smooth}. 

We have explored the parameter space of initial data $(\rho_L, P_L, u^x_L)$, $(\rho_R, P_R, u^x_R)$ for the left and right states,
and in all cases we found that the solutions obtained with EOS \eqref{EOS_nonsmooth}
almost overlap those obtained with the smooth EOS \eqref{EOS_smooth}. These results hold for both piecewise parabolic (PPM) 
and monotonized central (MC) reconstruction, regardless of resolution and the spline fit choice. Furtermore, we verified 
that all our simulations with the smooth EOS \eqref{EOS_smooth} had high enough resolution so that
data points would sample the smoothing interval $[\rho_i(1-\epsilon), \rho_i(1+\epsilon)]$ every few timesteps. 

In Fig.~\ref{fig:pressure_prof} we plot 
a snapshot of the pressure profile and do a convergence test for one of the cases we simulated with
$(\rho_R = 10^{-4},\ P_R = P_{\rm cold}(\rho_R),\ u^x_R = 0)$, $(\rho_L = 5\times 10^{-4},\ P_L = 10 P_{R},\ u^x_L=0)$. The figure shows
that all solutions overlap (left panel) and converge at the expected order (right panel) to the very high resolution solution obtained with 
PPM in conjunction with the smooth EOS.  

The solutions obtained with the smooth EOS \eqref{EOS_smooth} with quintic spline smoothing, which results in a $C^2$ 
and a convex EOS, also overlap with those of the non-smooth EOS solutions even though the smoothing interval we chose 
was much larger than in the cubic spline case, and hence the data points would sample it more frequently. Note also that the
coarsest resolution used in the shock tube tests is at least 20 times higher than the resolution used in our WDNS simulations. 
Therefore, all these results demonstrate that our numerical methods capture the correct solution and indicate that no
unphysical solutions are present in our simulations of the WDNS head-on collisions.

\bibliography{paper}

\end{document}